\documentstyle[12pt,psfig]{article}
\newtheorem{defin}{Definition}
\newtheorem{lemma}{Lemma}

\newtheorem{theor}{\large\bf Theorem}

\newcommand{\resetequ}{\setcounter{equation}{0}}
           
\newcommand{\Az}{{\cal S}}           %%%%% S call
\newcommand{\fr}{{\cal F}}           %%%%% F call
\newcommand{\tree}{{\cal T}}           %%%%% T call
\newcommand{\R}{{\bf R}}           %%%%% Reali
\newcommand{\be}{\begin{equation}}
\newcommand{\ee}{\end{equation}}
\newcommand{\bqa}{\begin{eqnarray}}
\newcommand{\eqa}{\end{eqnarray}}
\newcommand{\ba}{\begin{array}}
\newcommand{\ea}{\end{array}}
\newcommand{\p}[1]{{\partial\over \partial{#1}}}
\newcommand{\no}{\nonumber}

\newcommand{\lp}{\left (}
\newcommand{\rp}{\right )}
\newcommand{\al}{\alpha}
\newcommand{\bt}{\beta}
\newcommand{\ga}{\gamma}
\newcommand{\de}{\delta}
\newcommand{\e}{\epsilon}
\newcommand{\ze}{\zeta}

\newcommand{\la}{\lambda}

\newcommand{\si}{\sigma}

\newcommand{\om}{\omega}
\newcommand{\Om}{\Omega}

\newcommand{\La}{\Lambda}
\newcommand{\Lazero}{\Lambda_{0}}
\newcommand{\Lainv}{{\Lambda^{-2}}}
\newcommand{\Lazeroinv}{{\Lambda^{-2}_0}}
\newcommand{\Ga}{\Gamma}
\newcommand{\De}{\Delta}

\newcommand{\bpsi}{\bar{\psi}}
\newcommand{\psla}{\not\! p}
\newcommand{\xsla}{\not\! x}

\begin{document}

\centerline{\bf Continuous Constructive Fermionic Renormalization}
\vskip 2cm

\centerline{M. Disertori and V. Rivasseau}

\centerline{Centre de Physique Th{\'e}orique, CNRS UPR 14}
\centerline{Ecole Polytechnique}
\centerline{91128 Palaiseau Cedex, FRANCE}

\vskip 1cm
\medskip
\noindent{\bf Abstract}
We build the two dimensional Gross-Neveu model by a new method which requires
neither cluster expansion nor discretization of phase-space. It simply
reorganizes the perturbative series in terms of trees.
With this method we can for the first time define non perturbatively
the renormalization group differential equations of the model and
at the same time construct explicitly their solution.

\section{Introduction}
\resetequ

The popular versions of renormalization and the
renormalization group in field theory are based
on differential equations (among which the most famous one is the
Callan-Symanzik equation).
However no non-perturbative version
of these differential equations has been given until now.

On the other hand the renormalization group in statistical mechanics, for
instance for spin systems after the works of Kadanoff and Wilson, relies
 on closely related but discretized equations. When block spinning
or other discretization of momentum space is used, the result is
a discretized evolution of the effective action step by step. 
This point of view,
in contrast with the first one, has led to rigorous non perturbative
constructions for various models which have renormalizable power counting [R], 
but the methods always involved some discretization of phase space and
the outcome is a discrete (not differential) flow equation. Furthermore, 
the rigorous discretization of phase space came with a price,
namely the use of some technical tools such as cluster or Mayer expansions
which are neither popular among theoretical physicists nor among
mathematicians.

The proposal of Manfred Salmhofer to build a continuous version of
the renormalization group for Fermionic theories [S] is therefore
very interesting and welcome.
Indeed Fermionic series with cutoffs are convergent
(in contrast with Bosonic ones, which are Borel summable at best),   
and the continuous version of renormalization group which works so well
at the perturbative level should therefore apply to them.

In this paper we realize the Salmhofer proposal on the particular
example of the two-dimensional Gross-Neveu model. We rearrange Fermionic
perturbation theory according to trees, an idea first developed
in [AR2], perform subtractions only when necessary according
to the relative scales of the subgraphs,
and obtain (to our own surprise, quite easily) an explicit convergent 
representation of
the model without any discretization or cluster or Mayer expansion.
To prove the convergence requires
only some well-known perturbative techniques of
parametric representations (``Hepp's sectors''), Gram's bound
on determinants and a crucial but rather natural
concatenation of some intervals of integration for loop lines.

Therefore we can now consider that constructive theory for Fermions
has been ``reduced'' to perturbation theory. 
Remark also that since the representation we use
is an ``effective'' representation in the sense of [R],
hence with subtractions performed only when necessary according
to the relative scales of the subgraphs, we never meet the
so called problem of ``overlapping divergences'' or classification
of Zimmermann's forests. In this sense constructive renormalization 
is easier than ordinary perturbative renormalization 
(which, from the constructive
point of view, is flawed anyway because it generates renormalons).

Having an explicit convergent representation of the theory with a
continuously moving cutoff, it is trivial both
to define the continuous renormalization group equations
which correspond to the variation of this cutoff
and, at the same time,  to check that our explicit 
representation is a solution
of these equations. 

Remark however we have not yet found the way to
short-circuit our representation and to prove
that the equations and their solutions exist by a purely inductive argument
{\`a} la Polchinski [P] which would avoid an explicit formula for the solution.
This is presumably possible but this question
as well as the extension to other models,
in particular to interacting Fermions models of condensed matter physics,
is left for future investigation.
It is also important to recall that we do not see at the moment how to extend
this method to Bosons, since there are no determinant and Gram's bound
for them.

\section{Model and Main Result}
\resetequ
We consider the massive Gross-Neveu model $GN_2$, which describes $N$
types of Fermions. These Fermions interact through a quartic term.
Actually, the $GN_2$ action also requires a quadratic mass
counterterm and a wave function counterterm
in order for the ultraviolet limit to be finite.
Therefore the bare action in a finite volume $V$ is (using
the notations of [FMRS]): 
\bqa
\Az_V &=& {\la\over N} \int_V d^2x\; 
[\sum_a \bpsi_a(x)\psi_a(x)]^2\label{blic}\\
&&+ \de m  \int_V d^2x\; [\sum_a \bpsi_a(x)\psi_a(x)] +  
\de \zeta \int_V d^2x\; [\sum_a \bpsi_a(x) i \not\!\partial  \psi_a(x)] \no
\eqa
where $\la$ is the bare coupling constant, $\de m$
and $\de\ze$ are the bare mass and wave function counterterms,
and $a$ is the color index: $a=1,...,N$. 
The action (\ref{blic}) and the power counting of the $GN_2$
model are like the ones for  the Bosonic
$\phi_4^4$ theory, except that, unlike the latter, the $GN_2$ theory  
is asymptotically free
for $N\ge 2$, a condition which we assume from now on.
The free covariance in momentum space is
\be
C_{ab}^{\ga\de}(p)=\de_{a,b}\lp\frac{1}{-\psla +m}\rp_{\ga,\de}
=  \de_{a,b} \lp\frac{\psla +m}{p^2+m^2}\rp_{\ga,\de}
\ee
where $\ga,\de$ are the spin indices, and $a,b$ are the color indices.
Most of the time we skip the inessential
spin indices to simplify notation.
The mass $m$ is the renormalized mass.
To avoid divergences, according to the notations of
[KKS] we introduce an {\em ultraviolet cut-off $\La_0$}
and (for later study of the renormalization group
flow) a {\em scale parameter $\La$} which plays the role of 
an infrared cutoff:
\bqa
C_\La^{\La_0}(p) &=&
C(p) \left [ \eta\lp{\scriptstyle (p^2+m^2)\over
\scriptstyle \Lazero^2 }\rp -\eta\lp{\scriptstyle (p^2+m^2)\over 
\scriptstyle\La^2}\rp \right ].
\eqa
 The cutoff
function $\eta$ might be any function which
satisfies $\eta(0)  = 1$, which is smooth, monotone and rapidly decreasing
at infinity (this means faster that any fixed power).
For simplicity in this paper we restrict ourselves
to the most standard case $\eta(x)=e^{-x}$. In this case both  $C_\La^{\La_0}$
and its Fourier transform have explicit so called parametric representations:
\bqa
C_\La^{\La_0}(p) &=& \int_{\La_0^{-2}}^{\La^{-2}}  (  \not\! p + m ) 
\; e^{-\al (p^{2}+m^{2})}  d\alpha
\no\\
C_\La^{\La_0}(x-y) &=& \pi \int_{\La_0^{-2}}^{\La^{-2}}  
\lp i {(\not\! x -\not\! y)\over 2\al^{2}} + {m\over \al } \rp
 e^{-\al m^{2} -|x-y|^{2}/4\al}
d\alpha 
\label{param}\eqa

%In some applications it might be useful to require a cut-off function with
%compact support:
%\centerline{\hbox{\psfig{figure=eta.eps,height=2cm,width=3cm}}}

We define now the connected truncated Green functions, 
also called vertex functions,
which are the coefficients of the {\em effective action}.
The partition function with external fields $\xi,\bar\xi $ is
\bqa
Z_V^{\La\Lazero}(\xi,\bar{\xi})& =& \int
d\mu_{C^{\Lazero}_\La}(\psi,\bpsi)  
e^{-\Az_V(\psi,\bpsi)+<\psi,\xi >+ <\xi,\psi>}\no\\
<\psi,\xi > &:=& \int_{V} d^2x\; \bar{\psi}(x)\xi(x).
\eqa 
The vertex function with $2p$ external points is:
\bqa
\Gamma^{\La\Lazero}_{2p}(\{y\}, \{z\}) &:=& 
\Gamma^{\La\Lazero}_{2p}(y_1,...,y_p,z_1,...,z_p)\label{connected-function}\\
 &=&   \lim_{V\rightarrow \infty} 
{\scriptstyle \de^{2p} \over 
\scriptstyle\de\xi(z_1)..\de\xi(z_p)\de\bar{\xi}(y_1)..
\de\bar{\xi}(y_p)} 
\left .\lp (\ln Z_V^{\La\Lazero} - F )(C^{\Lazero}_\La)^{-1}(\xi) \rp
\right |_{\xi=0}
\no\eqa
where $F (\xi)= <\xi, C^{\Lazero}_\La \xi>$ is the bare propagator, and 
color indices are implicit. 
These functions (in fact distributions)
form the coefficients of the effective action 
(expanded in powers of the external fields)
at energy $\La$ with UV cutoff $\Lazero$. 
Developing the exponential in $Z$ and attributing prime and double prime
indices respectively to the mass and wave function counterterms we have:
\bqa 
 &&Z_V^{\La\Lazero}(\xi) =
\sum_{p=0}^\infty \frac{1}{p!^2} \sum_{n,n',n''=0}^\infty 
\frac{(-1)^{n+n'+n''}}{n!n'!n''!} 
\sum_{a_i b_i c_i d_i}   
{\scriptstyle \lp{\scriptstyle \la\over
\scriptstyle N}\rp^{n}}\lp\de m\rp^{n'}
\lp\de\ze\rp^{n''}\label{blublu} \\
&&\int_{V} d^2y_1...d^2y_p d^2z_1...d^2z_p d^2x_1...d^2x_n d^2x'_1...
d^2x'_{n'} 
d^2x''_1...d^2x''_{n''} \prod_{i=1}^p \xi_{d_i}(z_i)
\bar{\xi}_{c_i}(y_i)\no\\
 &&\left \{\ba{ccccccccccc}
y_{1,c_1} &...& y_{p, c_p}& x_{1, a_1}&
x_{1,b_1}&...&x_{n,a_n}&x_{n,b_n}& x'_{1, a'_1}& ...&x''_{n'', a''_{n''}}\\
z_{1,d_1} &...& z_{p, d_p}& x_{1, a_1}&
x_{1,b_1}&...&x_{n,a_n}&x_{n,b_n}&x'_{1, b'_1}& ...&x''_{n'', b''_{n''}}\\
\ea
\right \}\no
\eqa
where we used Cayley's notation for the determinants:
\be
\left \{\ba{c}
u_{i,a}\\ v_{j,b}\\ \ea \right \} = 
\det (D_{ab}(u_i-v_j))
\ee
and $a_i,b_i,a'_i,b'_i, a''_i,b''_i, c_i, d_i$ are the color indices.
By convention 
\be
D_{ab}(u_i-v_j):= 
C_{ab}(u_i-v_j)
\ee
except when the second index is the one of
a $\psi$ field hooked to a $\de\ze$ vertex. In this particular case
the vertex has a so called derivative coupling, and therefore
%in that case one should not forget that 
the propagator
$D$ bears a derivation, namely $D_{ab''}(u_i-v_j):=$ \mbox{$
i\not\!\partial_{v_{j}}C_{ab''}(u_i-v_j)$}$:=C'_{ab''}(u_i-v_j)$.
This derived propagator is explicitly
\be
C_\La^{'\,\La_0}(x-y) = \pi \int_{\La_0^{-2}}^{\La^{-2}}  
\lp {|x-y|^{2}\over 4\al^{3}}  + {im(\not\! x-\not\! y)\over 2\al^{2}}
 - {1\over \al^{2}}  \rp 
 e^{-\al m^{2} -|x-y|^{2}/4\al}
d\alpha 
 \ee
Expanding the determinant in (\ref{blublu}) one obtains the usual
perturbation theory in terms
of Feynman graphs with the three types of vertices
corresponding to the three terms of the action (\ref{blic}), 
and the logarithm is simply the sum
over connected graphs. To see if a graph is connected,  
it is not necessary to know its whole structure but only a tree in it. Based
on this remark the logarithm
of (\ref{blublu}) was computed in [AR2] using an expansion which is 
intermediate
between the determinant form (\ref{blublu}) and the fully expanded
Feynman graphs. This expansion is based on a {\em forest formula}.
Such formulas, discussed in [AR1], are Taylor expansions with integral
remainders. They test the coupling or links (here the propagators) 
between $n\geq 1$ points (here the vertices)
and stop as soon as the final connected components are built. 
The result is therefore a sum over forests, which are 
simply defined as union of disjoint trees. A forest is therefore
a (pedantic, but poetic) word for a
Feynman graph without loops, and our point of view is that these
are the natural objects to express Fermionic perturbation theory.

Here we use the most symmetric forest formula,
the {\em ordered Brydges-Kennedy Taylor formula}, which states [AR1]
that for any smooth function $H$ of the $n(n-1)/2$ variables
$u_{l}$,  $l \in P_n = \{(i,j)| i,j\in \{1,..,n\}, i\neq j\}$,
\be
H |_{u_{l}=1} = \sum_{o-\fr} \lp \int_{0\le w_{1} \le ...\le w_k 
\le 1} 
\prod_{q=1 }^{k} dw_{q}
\rp
\lp \prod_{q=1 }^{k}\p{u_{l_{q}}}  H \rp ( w^{\fr}_{l}({w_{q}}), l \in P_n)
\label{bloblo}
\ee 
where $o-\fr$ is any ordered forest, made of  $0\le k\le n-1$ 
links $l_{1},...,l_{k}$ over the $n$ points. To each link $l_{q}$  $q=1,...,k$
of $\fr$ is associated the parameter $w_{q}$, and to each pair $l=(i,j)$
is associated the weakening factor
$ w^{\fr}_{l}({w_{q}})$. These factors replace the variables
$u_{l}$ as arguments of the derived function $\prod_{q=1 }^{k}\p{u_{l_{q}}} H$
in (\ref{bloblo}).
These weakening factors $ w^{\fr}_{l}({w})$ are themselves functions
of the parameters $w_{q}$, $q=1,...,k$ through the formulas 
\bqa
w^{\fr}_{i,i}(w)&=&1\no\\ 
w^{\fr}_{i,j}(w)&=&\inf_{l_{q}\in P^{\fr}_{i,j}}w_{q}, \quad\quad
\hbox{if $i$ and $j$ are 
connected by $\fr$}\no\\&&
\hbox{where $P^{\fr}_{i,j}$ is the unique path in the forest 
$\fr$ connecting $i$ to $j$}\no\\
w^{\fr}_{i,j}(w)&=&0 \quad \quad\hbox{if $i$ and $j$ are not
connected by $\fr$}.
\label{w-factor}\eqa

We apply this formula to the determinant in (\ref{blublu}), inserting
the interpolation parameter $u_{l}$ in the cut-off (but only between distinct
vertices, so not for the ``tadpole'' lines):
\bqa
C_{\La}^{\La_{0}}
(x,y,u) &=& \de (x-y)C_{\La}^{\La_{0}}(x,x)+ 
[1-\de (x-y)]
C_{\La}^{\La_{0}} (x,y,u)\no\\ 
&:=& C^{\La_0(u)}_\La (x,y)\no\\ &:=&
\pi \int_{\La_0^{-2}(u)}^{\La^{-2}}  
\lp  i {(\not\! x -\not\! y)\over 2\al^{2}} + {m\over \al } \rp
 e^{-\al m^{2} -|x-y|^{2}/4\al}
d\alpha  
\label{interpol}\eqa
where 
\be
\Lazero^{-2}(u)= 
\La
^{-2} + u(\Lazero^{-2}-\La^{-2}).
\ee
We use similar interpolation for the $C'$ propagators.
When $u$ grows from 0 to 1, the ultraviolet cut-off of the interpolated
propagator (between distinct vertices) grows therefore from $\La$ to
$\Lazero$.

\centerline{\hbox{\psfig{figure=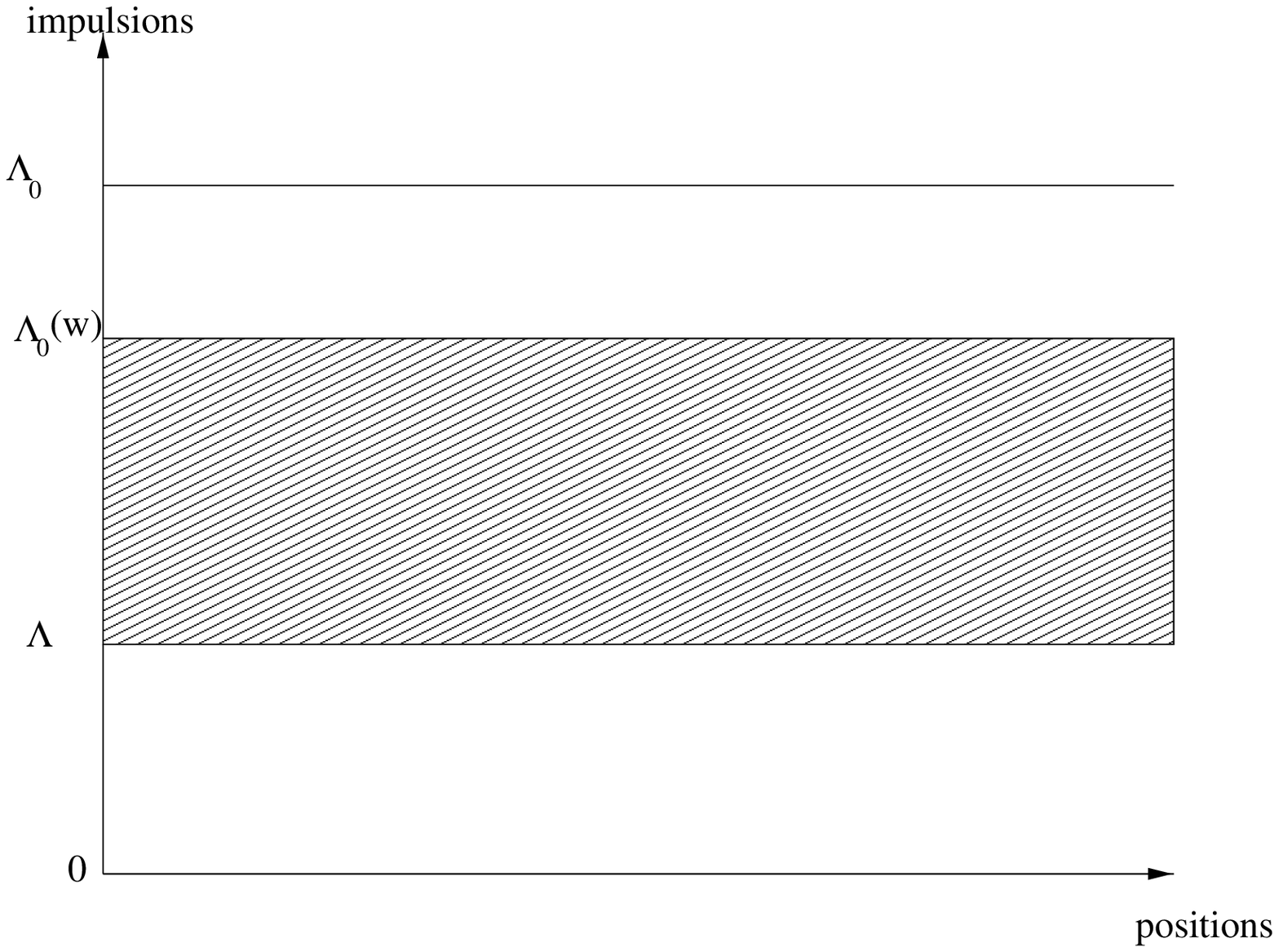,height=7cm,width=10cm}}}
\centerline{Figure 1}

We define 
\bqa
&& C_{\La}^{\La_{0}, u}(x,y) :=  \p{u}C^{\Lazero}_\La(x,y,u)=
\pi \lp i {(\not \! x - \not \! y )\Lazero ^{4}(u)\over 2 } 
+ m \Lazero ^{2}(u)
\rp \cdot \no\\
&& \quad  \cdot (\Lainv -\Lazeroinv) 
e^{-m^{2}\La_{0}^{-2} (u)-|x-y|^{2}\Lazero^{2} (u)/4}
\label{covar}\\
&& C_{\La}^{'\, \La_{0}, u}(x,y) :=  \p{u}C^{'\, \Lazero}_\La(x,y,u)=
 \pi \bigg( {|x-y|^{2}\Lazero ^{6}(u)\over 4}+\no\\
&&\quad + { i m (\not \! x - \not \! y )\Lazero ^{4}(u)\over 2 } 
  - \Lazero ^{4}(u)\bigg) (\Lainv -\Lazeroinv)
e^{-m^{2}\La_{0}^{-2} (u)-|x-y|^{2}\Lazero^{2} (u)/4}\no
\eqa
The derivative of $\eta$ fixes $C^u$ at an energy near $\Lazero(u)$. 
We observe that for any fixed $\epsilon'$ we have the scaled decay:
\be
|C^u(x,y)| \leq K \La_0^3(u)  (\Lainv -\Lazeroinv) 
e^{-|x-y|(1-\epsilon')\Lazero^m (u)
-\epsilon' m^{2}\La_{0}^{-2} (u)/2}\label{hello1}
\ee
\be
|C^{'\, u}(x,y)| \leq K \La_0^4(u)  (\Lainv -\Lazeroinv) 
e^{-|x-y|(1-\epsilon')\Lazero^m (u)-\epsilon' m^{2}\La_{0}^{-2} (u)/2}
\label{hello2}\ee
where $K$ is a constant 
depending only on $\epsilon '$ and
\be\Lazero^m (u) := \sup[m,\Lazero (u)]\label{hello3}
\ee.

Applying this interpolation and the ordered forest formula (\ref{bloblo})
to the propagators in the determinant of  (\ref{blublu})
we obtain

\bqa 
&&Z_V^{\La\Lazero}(\xi)  =
 \sum_{p=0}^\infty \frac{1}{p!^2} \sum_{n,n',n''=0}^\infty \frac{1}{n!n'!n''!} 
\sum_{o-\fr}\sum_{{\cal C}ol}\sum_{\Om}   
 {\scriptstyle \lp{\scriptstyle \la\over
\scriptstyle N}\rp^{n}} \lp\de m \rp^{n'}
\lp\de\ze\rp^{n''}
\e(\fr, \Om)\no\\
&&\int_{V} d^2y_1...d^2y_p d^2z_1...d^2z_p d^2x_1...d^2x_{\bar n} 
 \prod_{r=1}^p \xi_{d_r}(z_r)
\bar{\xi}_{c_r}(y_r)\no\\
&& \int_{0\le w_{1} \le ...\le w_{k}\le 1} 
\biggl[\prod_{q=1}^{k}  D_{\La}^{\La_{0},  w_q}(\bar x_{l_{q}}, x_{l_{q}})
dw_q \biggr]
 [\det]_{\hbox{left}}(w^{\fr}(w))\no\\
&&
\label{sviluppo1}\eqa
where for simplicity the position of any vertex
is simply denoted by the letter $x$ and $\bar n := n + n' + n''$.
$\bar x_{l_{q}}$ and $ x_{l_{q}}$ are the ends of line
$l_{q}$. $[\det]_{\hbox{left}}$ is the remaining
determinant. Its entries correspond to the remaining fields
necessary to complete each vertex of the forest into a quartic 
or quadratic vertex, according to its type (interaction or counterterm).
For this model, remark that the sum 
$\sum_{o-\fr}$ is performed only over the ordered forests that have, for
each point $x_i$ coordination number $n(i)\leq 4$
or $n(i)\leq 2$ depending of the type of the vertex (all other terms
being zero). 
The additional sums over  ${\cal C}ol$ and $\Om$ correspond to
coloring choices at each vertex and ``fields versus antifields'' choices
at each line and vertex [AR2]. The sign $\e(\fr, \Om)$ comes in from
the antisymmetry of Fermions and is computed in [AR2]: here we only
need to know that it factorizes over the connected components of $\fr$. 
To find the expression for $\ln Z$ we write $Z$ as an exponential.
In  equation (\ref{sviluppo1}), the determinant factorizes
over the ordered trees $\tree_1...\tree_j$ forming the forest. Indeed
one can resum all orderings of the ordered forest $\fr$ compatible
with fixed orderings of its connected components, the trees  
$\tree_1...\tree_j$. Furthermore the
``weakening factor'' $w^{\fr}$ vanishes between vertices belonging to
different connected components. Hence:
\bqa
 Z_V^{\La\Lazero}(\xi) & =& 
\sum_{p=0}^\infty \frac{1}{p!^2} \sum_{n,n',n''=0}^\infty \frac{1}{n!n'!n''!} 
\sum_{j=0}^n \frac{1}{j!}\sum_{n_1,...n_j \atop n_1+...+n_j=n}
\sum_{n'_1,...n'_j \atop n'_1+...+n'_j=n'}
\sum_{ n''_1,...n''_j \atop n''_1+...+n''_j=n''}\no\\
&&\sum_{\stackrel{p_1,...p_j,p'}{p_1+...+p_j+p'=p}} \frac{n!n'!n''!}{
n_1!...n_j!n'_1!...n'_j!n''_1!...n''_j!}
\frac{p!^2}{p_1!^2...p_j!^2p'!^2}
p'!\no\\
&&(\xi,C^{\Lazero}_\La\xi)^{p'}
\prod_{i=1}^j  \left [{\scriptstyle 
\lp{\scriptstyle \la\over \scriptstyle N}\rp^{n_i}} \lp\de m \rp^{n'_{i}}
\lp\de\ze\rp^{n''_{i}}
A(n_i,n'_i,n''_i,p_i)\right ]
\eqa
where
\bqa 
&&A(n_i,n'_i,n''_i,p_i) = 
\sum_{\tree_i}\sum_{{\cal C}ol_{i}, \Om_{i}}  \e(\tree_i, \Om_{i})
\int d^2y_1...d^2y_{p_i} d^2z_1...d^2z_{p_i}
d^2x_1...d^2x_{\bar n_i}\no\\
&& \prod_{r=1}^p \xi_{d_r}(z_r)
\bar{\xi}_{c_r}(y_r)
\int_{0\le w_{1} \le ...\le w_{\bar n_{i}-1}\le 1} 
\biggl[\prod_{q=1}^{\bar n_{i}-1}  
D_{\La}^{\La_{0},  w_q}(\bar x_{l_{q}}, x_{l_{q}})
dw_q \biggr]\no\\
&&[\det]_{\hbox{left,i}}(w^{\tree_{i}}(w))
\eqa
where $\bar n_i$  is the number of vertices in the 
ordered tree $\tree_i$, which has
therefore $\bar n_{i}-1$ lines,
$p_i$ is the number of external fields of type $y$ 
(and $z$) attached to the $\tree_i$, and $p'$ is the number of free
external propagators (not connected to any vertex) in the forest.  
%\newpage
This can be written as an exponential, hence
\bqa
&&\ln Z_{V}^{\La\Lazero} (\xi)=(\xi,C^{\Lazero}_\La\xi) +  
\sum_{p=0}^\infty \frac{1}{p!^2}
\sum_{n,n',n''=0}^\infty \frac{1}{n!n'!n''!}  
 {\scriptstyle \lp{\scriptstyle \la\over
\scriptstyle N}\rp^{n}}\lp\de m \rp^{n'}
\lp\de\ze\rp^{n''}\no\\
&&\sum_{o-\tree} \sum_{{\cal C}ol, \Om} \e(\tree, \Om)
\int_{V} d^2y_1...d^2y_{p} d^2z_1...d^2z_{p}
d^2x_1...d^2x_{\bar n}\prod_{r=1}^p \xi_{d_r}(z_r)
\bar{\xi}_{c_r}(y_r)\no\\
&& 
\int_{0\le w_{1} \le ...\le w_{\bar n-1}\le 1} 
\biggl[\prod_{q=1}^{\bar n-1}  D_{\La}^{\La_{0},  w_q}
(\bar x_{l_{q}}, x_{l_{q}})dw_q \biggr]
[\det]_{\hbox{left}}(w^{\tree}(w))   
\eqa 
where $\tree$ is an ordered tree over $\bar n$ points, and the external
points are all connected to the tree. 
Now, applying the definition (\ref{connected-function}),
we obtain the vertex functions, for which the limit
$V\to\infty$ can be performed (because the external points hooked to the tree 
ensure convergence). 
The set
\bqa
E&=& \{(i_1,...i_p,j_1,...,j_p)| i_1,..i_p,j_1,...j_p\in \{1,...,\bar n\}\}
\label{ext-lines}
\eqa
fixes the internal points to which the $2p$ external lines hook.

We recall the well-known fact
that the vertex functions in $x$-space are in fact distributions. For instance
it is easy to see that
when some of the external points $ {i_{k}} $, $ {j_{k}} $ in the previous
sum  coincide, one has to factor out the product of
the corresponding delta functions of the external arguments
to obtain smooth functions.
This little difficulty can be treated either by considering
the vertex functions in momentum space (they are then ordinary functions of
external momenta, after factorization of global momentum conservation),
or by smearing the vertex functions with test functions. Here we adopt this
last point of view. 
The quantity under study is then 
$\Ga_{2p}^{\La\La_0}$ 
smeared with smooth test functions $\phi_1(y_1)$, ..., $\phi_p(y_p), $ $
\phi_{p+1}(z_1), ...,\phi_{2p}(z_p)$:
\bqa
\lefteqn{\Ga_{2p}^{\La\Lazero}(\phi_1,...\phi_{2p})= 
\int d^2y_1...d^2y_p d^2z_1...d^2z_p }\no\\
&&
\Ga_{2p}^{\La\Lazero}(y_1,...,y_p,z_1,...,z_p) \phi_1(y_1)...\phi_p(y_p)
\phi_{p+1}(z_1)...\phi_{2p}(z_p).
\eqa
where we asked the test functions to have compact support: 
$\phi \in {\cal D}(\R^{2})$. 

Remark that when some external antifield hooks to a $\de \ze$ vertex, the
amputation by $C$ instead of $C'$ leaves a $\de '$ distribution, which means
a derivative acting on the corresponding test function. 
 
We obtain the formula:
\bqa
&&\Ga_{2p}^{\La\Lazero}(\phi_1,...\phi_{2p})= 
\sum_{n,n',n''=0}^\infty 
 {\scriptstyle \lp{\scriptstyle \la\over
\scriptstyle N}\rp^{n}} \lp\de m \rp^{n'}
\lp\de\ze\rp^{n''}\frac{1}{n!n'!n''!}
\label{sviluppo2}\\
&&\sum_{o-\tree}\sum_{E}\sum_{{\cal C}ol,\Om}  \e(\tree, \Om)\int 
d^2x_1...d^2x_{\bar n}
\phi_1(x_{i_{1}})...\phi_{2p}(x_{j_{p}})
\no
\\
&&\int_{0\le w_{1} \le ...\le w_{\bar n-1}\le 1} 
\biggl[\prod_{q=1}^{\bar n-1}  D_{\La}^{\La_{0},  w_q}
(\bar x_{l_{q}}, x_{l_{q}})dw_q \biggr][\det]_{\hbox{left}}(w^{\tree}(w),E)
\no 
\eqa
where the propagator $D$ is now $C$ or $C'$ according to 
the discussion above.
%, and in the case $p=1$ the trivial term 
%with $n=n'=0$  should be ${i \de a  \not \! \de ' (y-z) \over 1
%+\de a C' (y-z)}$.
\medskip

When renormalization is introduced, it will be convenient to
use the BPHZ 
subtraction prescription at 0 external momenta, which corresponds to
integrate the vertex functions over all arguments except one.
In this prescription 
one defines the renormalized coupling constant as the 4-vertex 
function of the full theory at zero external momenta:

\bqa
\frac{\la_{ren}}{N} &:=& \widehat\Ga_{4}^{\La\Lazero} (0,0,0,0) = 
\int d^2x_{2}d^2x_{3}d^2x_{4}\;\Ga_{4}^{\La\Lazero} (0,x_{2},x_{3},x_{4})
\eqa
Moreover we want the renormalized mass and wave function constant
to be respectively $m$ and 1. This means that we impose the additional
renormalization conditions:
\bqa
\de m_{ren} &:=& \widehat\Ga_{2}^{\La\Lazero} (0,0) = 
\int d^2x_{2}\Ga_{2}^{\La\Lazero} (0,x_{2}) = 0
\eqa
\bqa
\de \ze_{ren} &:=&  \not\!\partial\widehat\Ga_{2}
^{\La\Lazero} (0,0) = 
 \int d^2x_{2} i \not \! x_{2} \Ga_{2}^{\La\Lazero} (0,x_{2}) = 0
\eqa
With these conditions the whole theory (at fixed renormalized mass
$m$) becomes parametrized only by $\la_{ren}$,
hence not only $\la$ but also $\de m$ and $\de \ze$ in (\ref{blic})
become functions of $\la_{ren}$. This of course
has a precise meaning only if we can construct the theory
and solve the renormalization group flows, which is precisely what we
are going to do.
We can express the main result of this paper as a theorem on the
existence of the ultraviolet limit of the vertex
functions and of the renormalization group flows.
Recall that the theory is not directly the sum but the Borel sum
of the renormalized perturbation theory. In summary 

\begin{theor} 
The limit $\Lazero\to\infty$ of  
$\Ga_{2p}^{\La\Lazero}(\phi_1,...\phi_{2p})$  exists and
is Borel summable in the renormalized coupling constant
 $\la_{ren}$, uniformly in $N$ (where $N$ is the number of colors).
Since the parameter $\La$ varies continuously, the 
continuous renormalization group equations and in particular the $\bt$ 
function  are also  well defined in the limit $\Lazero\to\infty$. 
\end{theor}
\medskip
The first part of the theorem is similar to [FMRS], but the
second part (the existence of the
continuous renormalization group equations) is new. 
The rest of the paper is devoted to the proof of this theorem.

The precise bounds on the smeared vertex functions are given in
theorem 3 below. They are uniform in $N$ 
(and in fact proportional to $N^{1-p}$). Let us discuss also briefly
the dependence in $m$, the renormalized mass. 
For $m \ne 0$ fixed, we can define the physical scale of the
system by putting $m=1$. The theorem is then uniform
in the infrared cutoff $\La$, including the point $\Lambda = 0$.
In the case $m=0$ our method requires a nonzero 
infrared cutoff $\Lambda \ne 0$. Since this cutoff is the only scale
of the problem, we can then put it to 1: $\Lambda = 1$.
In this last case, improperly called the ``massless theory'', 
we know that there should be
a non-perturbative mass generation [GN]. This mass generation has been
proved rigorously for the model with fixed ultraviolet cutoff and
large number $N$ of components in [KMR], using the Matthews-Salam formalism
of an intermediate Bosonic field and a cluster expansion 
with a small/large field expansion. Our result in the massless
case $m=0$ with a finite infrared cutoff $\Lambda$ should 
therefore glue with the method and results of [KMR]
to obtain at large $N$
the mass generation of the full model without ultra-violet cutoff.

\section{The Expansion}
\resetequ

\subsection{The continuous band structure}

\centerline{\hbox{\psfig{figure=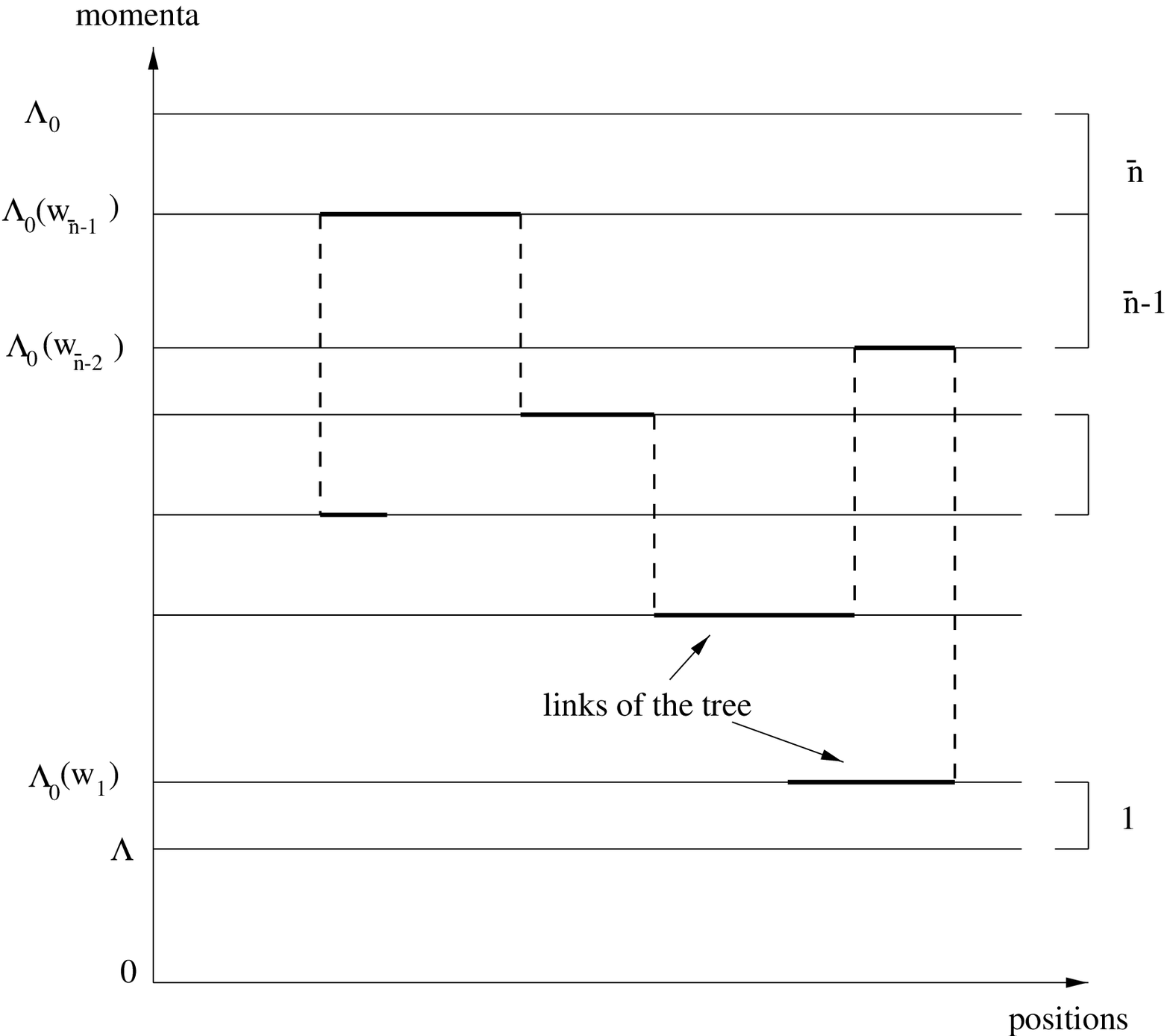,height=7cm,width=10cm}}}
\centerline{Figure 2}

Remark that in (\ref{sviluppo2}) 
\be
w_1\leq w_2\leq...\leq w_{\bar n-1} \Longrightarrow 
\Lazero(w_1)\leq\Lazero(w_2)..\leq\Lazero(w_{\bar n-1}).
\label{old-diff}\ee
This naturally cuts the  space of momenta into $\bar n$ bands
$B=\{1,...,\bar n\}$
(see Figure 2).

Looking at equation (\ref{param}), we see that the
covariance can be
written as a sum of propagators restricted to single bands:
\be
C_\La^{\Lazero}(p) =  \sum_{k=1}^{\bar n}
C^{\Lazero({w}_k)}_{\Lazero({w}_{k-1})}(p)
=  
\sum_{k=1}^{\bar n} 
\int_{\La_0^{-2}(w_k)}^{\La_0^{-2}(w_{k-1})}  (  \not \! p + m ) 
e^{-\al (p^{2}+m^{2})}  d\alpha
 = C(p)  \sum_{k=1}^{\bar n} \eta^k
\label{eta-band0}\ee
where we defined
\bqa
\eta^k &:=& e^{-\La_0^{-2}(w_{k}) (p^{2}+m^{2})} - 
e^{-\La_0^{-2}(w_{k-1}) (p^{2}+m^{2})}
\label{eta-band}\eqa
and we adopted the convention
\be
{w}_0=0 \Rightarrow \Lazeroinv({w}_0)=\Lainv
\quad {w}_{\bar n}=1 \Rightarrow
\Lazeroinv({w}_{\bar n})=\Lazeroinv.
\ee
Similar formulas hold for $C'$ but with an additional $\not \! p$.

To cure the ultraviolet divergences we have to combine
the divergent local parts of some subgraphs with counterterms
and reexpress the series for $G$ as an effective series
in the sense of [R]. For that purpose we use the band structure 
to distinguish the convergent and divergent subgraphs hence decide
where renormalization is necessary.

\subsection{Notations}

Now we fix some notations. It is convenient to
give indices to the fields variables
or the half-lines which correspond to these fields
after Grassmann integration. We observe that there are several types 
of such variables, the half-lines which form the lines of the tree, 
the external variables (which correspond to amputated lines)  
and the entries (rows
or columns) in the determinant ${\rm det_{left}}$. 
These entries will be called
``loop fields'' or ``loop half-lines'' since they form the usual
loop lines of the Feynman graphs if one expands the determinant.
We define $E$ and $L$ as the set of all external and loop half-lines.
For each level $i$ there 
is a tree-line $l_i$ with two ends corresponding
to two half-lines called $f_i$ and $g_i$. 
The loop fields $\psi(x_g)$ 
and $\bar{\psi}(x_f)$ are called $h_f$ and $h_g$, and (when expanded)
the loop line $\bar{\psi}(x_f)\psi(x_g)$ is called $l_{fg}$ (it corresponds
to a particular coefficient in the determinant ${\rm det_{left}}$). 
Each tree half-line $f_i$ or $g_i$, each loop field 
$h_f$ or $h_g$ is hooked to a vertex 
called $v_{f_{i}}$ or $v_{g_{i}}$ or $v_f$ or $v_g$. We need also to care
about the set $S$ of special fields (or antifields) which are hooked to the
$\de \ze$ vertices and correspond to propagators 
$C'$ which have different ``power counting'' than $C$.  
Finally the index of the highest tree-line hooked 
to a vertex $v$ is 
called $i_v$.

Now  $[\det]_{\hbox{left}}$ is the determinant of a matrix
$(n+1-p)\times(n+1-p)$. The corresponding loop fields can 
be labeled by an index $a =1,...,2n+2-2p$. The matrix elements are 
$D(x_f,x_{g},w_{v_f,v_{g}}^{\tree}( w))$. Therefore in terms of bands
the line $l_{fg}$ is restricted by the 
weakening factor $w_{v_f,v_{g}}^{\tree}( w)$ to belong to the bands
from 1 to the lowest index in the path $P^{\tree}_{f,g}$ (this path 
$P^{\tree}_{f,g}$ is defined in equation (\ref{w-factor})).
We call $i^{\tree}_{f,g}$ this index:
\bqa
i^{\tree}_{f,g}=   \inf\; \{ q \  | \  l_{q} \in P^{\tree}_{f,g}  \}
\eqa
\bqa
D(x_f,x_{g},w_{v_f,v_{g}}^{\tree}( w))&=& 
D(p)\ \sum_{k= 1} ^{i^{\tree}_{f,g}}\; \eta^k (p)
\label{band1}\eqa
By multilinearity one can expand the determinant in  (\ref{sviluppo2}) 
according to the different bands in the sum (\ref{band1}) for each  
row and column.
\be
[\det]_{\hbox{left}}( w^{\tree}( w ),E) = 
\sum_\mu \det{\cal M}(\mu)
\ee
where we define the 
attribution $\mu$ as a collection of band indices for each  
loop field $a$:
\be
 \mu= \{ \mu(f_1), ..\mu(f_{n+1-p}),\mu(g_1), ..\mu(g_{n+1-p})\} \ ,\ 
\mu(a)\in B  \ \ {\rm for}\   a=1...2n+2-2p. 
\label{half-lines}
\ee
Now, for each attribution $\mu$ we need to exploit power counting. This 
requires notations for the various types of fields or half-lines
which form the analogs of the quasi local subgraphs of [R] in our formalism. 
We define:
\bqa
&& T_k =   \{ l \in \tree |\; i_{v_{l}} \geq k\}\no\\
&& IT_k =   \{ f_{i}, g_{i} \in \tree |\; i\geq k\}\no\\
&& IL_k= \{ a \in L |\; \mu(a)\geq k \} \no\\ 
&& EE_k = \{ f, g \in E | i_{v_f}, i_{v_g}\geq k\}\no\\
&& ET_k = \{f_{i} | i_{v_{f_{i}}} \geq k, i<k \}
\cup \{g_{i} | i_{v_{g_{i}}} \geq k, i<k \}\no\\ 
&& EL_k = \{a \in L| i_{v_a} \geq k,
\mu(a) < k\}\no\\ 
&& N_k =  \{ v {\rm \ of \ type \ } \lambda \  | i_{v} \geq k  \}\no\\
&& N'_k =  \{ v {\rm \ of \ type \ } \de m\  | i_{v} \geq k  \}\no\\
&& N''_k =  \{ v {\rm \ of \ type \ } \de \ze\  | i_{v} \geq k  \}\no\\
&& \bar N_{k} =   N_k\cup N'_k \cup N''_k \ ,\ 
 G_k =   IT_k \cup IL_k \ ,\ E_k = EE_k \cup ET_k\cup EL_k\no\\
&& E''_k = E_{k} \cap S \ ,\ T_k =   \{ l_{i} |\; i\geq k\}
\label{definizioni}
\eqa
where we recall that $S$ in the last definition is the set of those
fields hooked to a vertex of type $\de\ze$ which bear a derivation.
We note $|A |$ the number of elements in the set $A$. For instance
the reader can check that $|IT_1| = 2\bar n -2$ and that $|T_1| = \bar n -1$.
Each $G_k$ has $c(k)$ connected components $G_k^j$, $j=1,..,c(k)$.
All the definitions in (\ref{definizioni})
can be restricted to each
connected component.
 Applying power counting, the convergence degree for the subgraph $G_i^k$ is
\be
\om (G_i^k) =\frac{1}{2} (|E_i^k| +2|N_i^{'k}| -4)
\label{conv-degree}\ee
where we assumed that no external half-line hooked to a vertex of type
$\de \ze$ bears a $i\not \!\!\partial$. To assure this for any $G_i^k$, 
we apply, for each vertex $v''$,  the operator  $i\not\!\!\partial$ (or  
$-i \not\!\!\partial$)  
to the highest tree half-line hooked to $v''$ (there is always at least one). 
In this way for all $k$ $|E''_k|=0$, and no loop line bears a gradient. Then
${\cal M}(\mu)$ is a matrix whose coefficients are
\be
{\cal M}_{fg}(\mu) (x_f,x_g)= \de_{\mu(f),\mu(g)}
\int {d^2p \over (2\pi)^2 }\; e^{-ip(x_f-x_g)}C(p)\;\eta^{\mu(f)}(p) W^{\mu(f)}_{v_f,v_g}
\ee
where 
\bqa
W^k_{v,v'}&=& 1 \quad \hbox{if $v$ and $v'$ are connected by $T_k$}\no\\
        &=& 0  \quad \hbox{otherwise}
\label{defw}\eqa
since we always have $D=C$ in the matrix ${\cal M}(\mu)$.

From (\ref{conv-degree}) we see that there are three types of 
divergent subgraphs:

- for $|E_i^k|=4$, $|N_i^{'k}|=0$ 
we have logarithmic divergence ($\om(G_i^k)=0$);

- for $|E_i^k|=2$, $|N_i^{'k}|=0$  we have linear divergence 
($\om(G_i^k)=-1$);

- for $|E_i^k|=2$, $|N_i^{'k}|=1$  
we have logarithmic divergence ($\om(G_i^k)=0$).

In fact the divergent graphs are only those for which the algebraic 
structure of the external legs is of one of the three types in (\ref{blic}).
For instance not all four-point subgraphs are divergent, but only those for
which the flow of spin indices follows the flow of color indices [GK][FMRS].
Using the invariance of ${\cal L}$ under parity and charge conjugation one
finds that all counterterms which are not of the three types in (\ref{blic})
are zero (this means that the corresponding subgraphs have 0 local part). 
Then renormalizing these subgraphs we improve power counting without 
generating new counterterms.
 In what follows, for simplicity, ``divergent subgraph'' always  means
subgraph with two or four external legs (this means we will renormalize some
subgraph which does not need it but this does not affect the convergence of the
series). 
Also for simplicity we change the definition of convergence degree 
(\ref{conv-degree}) in 
\be
\om' (G_i^k) =\frac{1}{2} (|E_i^k|-4) \ .
\label{conv-degree1}\ee

To cure divergences, we apply to the amplitude of each divergent subgraph
$g$ the operator $(1-\tau_g)+\tau_g$. In the momentum space $\tau_g$ is the 
Taylor expansion at order $-\om(g)$
of the amplitude $\hat g (p)$ at $p=0$. 
The operator $1-\tau_g$ makes the amplitude convergent when the 
UV cut-off is sent to infinity. The remaining term $\tau_g \hat g $ gives a
local counterterm for the coupling constant
that depends on the energy of the 
external lines of $g$. At each vertex $v$, we can resum the series of all 
counterterms obtained applying $\tau_g$ to all divergent subgraphs 
(for different
attributions $\mu$) that have the same set of external lines
as $v$ itself. In this way we
obtain an effective coupling constant which depends on the energy 
$\La_0(w_{i_{v}})$ 
of the highest tree line hooked to the vertex $v$. This is true
because {\em after applying the $1-\tau_g$ operators}, for each 
graph with nonzero amplitude the highest index at each vertex coincides
with the highest {\em tree index $i_{v}$} at each vertex! 
Indeed at vertices $v$
for which this is not true, there are loop fields with attribution $\mu$ higher
than $i_{v}$. By  (\ref{defw}) they must contract together forming
tadpoles, which are set to zero by the  $1-\tau_g$ operators. The
corresponding graphs therefore disappear from the expansion.

For each attribution $\mu$ we define the set of divergent subgraphs as
\be
D_\mu := \{\  G_i^k | \om'(G_i^k)\leq 0\}.
\label{diverg}\ee
 The action of $\tau_g$ is 
\be
\tau_g \hat{g}(p_1,..,p_k) = \sum_{j=0}^{-\om'(g)} \frac{1}{j!} 
\frac{d^j}{dt^j} \hat{g}(tp_1,...tp_k)_{|p=0}\quad k=2,4.
\ee
With this definition the effective constants $\la_w$, $\de m_w$, $\de\ze_w$ 
 turn out to be 
the vertex functions $\Ga_4$, $\Ga_2$ and $\not\!\!\partial\Ga_2$ for an 
effective theory
with infrared parameter $\La=\La_0(w)$:

\subsection{Effective constants}

In the space of positions, the operator $\tau_g$ is applied 
by partial integration 
to the product of  external propagators $a(x_1,...x_{v_e})$, as in [R]:
\be
\tau^*_g(v_e)a(x_1,...x_{v_e}) = \sum_{j=0}^{-\om'(g)} \frac{1}{j!}
\frac{d^j}{dt^j} a(x_1(t),...,x_{v_e}(t))_{|t=0} 
\ee
where $x_i(t)=x_{v_e}+t(x_i-x_{v_e})$, and $v_e$ is an external vertex of $g$
chosen as `reference vertex'. The choice of this reference vertex is
given in section IV.3.1. 
As announced we find the three possible counterterms of (\ref{blic}).
For $|E_i|=4$ we have
\be
 \tau^*(x_1) \prod_{i=1}^4 C(x_i,y_i)=  \prod_{i=1}^4 C(x_1,y_i), 
\ee
so the counterterm is
\bqa
&&\int d^2x_1 \prod_{i=1}^4C_{\al_i\al'_i}^{a_ia_i}(x_1,y_i) 
\int d^2x_2 d^2x_3 d^2x_4\;
g(0,x_2,x_3,x_4)_{\al_1\al_2\al_3\al_4}^{a_1 a_2 a_3 a_4}\no\\
&&= \int d^2x_1 \prod_{i=1}^4C_{\al_i\al'_i}^{a_ia_i}(x_1,y_i) 
\;\hat{g}(0,..,0)
\eqa
This gives a coupling constant counterterm.
For $|E_i|=2$ we have
\be
\tau^*(x_1) 
[C(x_1,y_1) C(x_2,y_2)]= 
C(x_1,y_1) \left [ C(x_1,y_2) + (x_2-x_1)^\mu \p{x_1^\mu} C(x_1,y_2)
\right ].
\ee
Integrating over internal points, we obtain a mass
counterterm from the first  term in the sum:
\bqa
&&\int d^2x_1 \prod_{i=1}^2 C_{\al_i\al'_i}^{a_ia_i}(x_1,y_i) 
\int d^2x_2\;  g_{\al_1\al_2}^{a_1 a_2}
(0,x_2)\no\\
&& =  \int d^2x_1 \prod_{i=1}^2 C_{\al_i\al'_i}^{a_ia_i}(x_1,y_i) 
\; \hat g_{\al_1\al_2}(0)\de_{a_1,a_2}\no\\
&&=\int d^2x_1 \prod_{i=1}^2 C_{\al_i\al'_i}^{a_ia_i}(x_1,y_i)
\;\de_{a_1,a_2}\;  f_1(0) 
\eqa
where we applied the development 
\be
\hat g(p) = f_1(p^2) + \ga_5 f_2(p^2)+ \psla f_3(p^2) + \ga_5\psla f_4(p^2)
\ee 
and we adopted for the gamma matrices the conventions in [FMRS]. 
By invariance under charge conjugation and parity $f_2(0)=f_4(0)=0$. 
For  the   second term in the sum we obtain a wave function counterterm:
\bqa
&&\int d^2x_1 C_{\al_1\al'_1}^{a_1a_1}(x_1,y_1)\p{x_1^\mu}  
C_{\al_2\al'_2}^{a_2a_2}(x_1,y_2) 
\int d^2x_2 (x_2-x_1)^\mu g_{\al_1\al_2}^{a_1 a_2}(x_1-x_2)=\no\\
&& = \int d^2x_1 C_{\al_1\al'_1}^{a_1a_1}(x_1,y_1)\p{x_1^\mu}  
C_{\al_2\al'_2}^{a_2a_2}(x_1,y_2) i 
\p{p^\mu} \hat g(p)_{|p=0} \de_{a_1a_2}\no\\
&&= \int d^2x_1 C_{\al_1\al'_1}^{a_1a_1}(x_1,y_1)i \not\!\partial   
C_{\al_2\al'_2}^{a_1a_1}(x_1,y_2) f_3(0).
\eqa

\begin{theor}
If we apply to each divergent subgraph $g\in D_\mu$, 
for any  attribution $\mu$,  the operator 
$(1-\tau_{g})+\tau_{g}= R_g+\tau_g$, the function (\ref{sviluppo2}) 
can be written as 
\bqa
\lefteqn{\Ga_{2p}^{\La\Lazero}(\phi_1,...\phi_{2p})= 
\sum_{n,n',n''=0}^\infty  \frac{1}{n!n'!n''!}
\sum_{o-\tree}\sum_{E,\mu}\sum_{{\cal C}ol,\Om}  \e(\tree, \Om)\int  
d^2x_1...d^2x_{\bar n} }\no\\ 
&& 
\int_{0\le w_{1} \le ...\le w_{\bar n-1}\le 1}\prod_{q=1}^{\bar n-1} 
dw_q \;  
\left [\prod_{v} \lp \frac{\la_{w(v)}}{N}\rp \right ] 
\left [\prod_{v'}\de m_{w(v')}\right ] 
\left [\prod_{v''}\de \ze_{w(v'')}\right ] \no\\
&& \prod_{G_i^k\in D_\mu} R_{G_i^k} \;  
\biggr[\prod_{q=1}^{\bar n-1}  D_{\La}^{\La_{0},  w_q}
(\bar x_{l_{q}}, x_{l_{q}}) \, \det {\cal M}(\mu)
\phi_1(x_{i_{1}})...\phi_{2p}(x_{j_{p}})\biggr] 
\label{sviluppo4}
\eqa
where the constants $\la_w$, $\de m_w$, $\de\ze_w$ are the 
`effective constants', defined as:
\bqa
\frac{\la_w}{N} &=& \hat \Ga_{4}^{\Lazero(w), \Lazero}(0,0,0,0) = 
\int d^2x_2 d^2x_3 d^2x_4\; \Ga_{4}^{\Lazero(w), \Lazero }(0,x_2,x_3,x_4)\no\\
\de m_w &=&   \hat\Ga_{2}^{\Lazero(w), \Lazero }(0,0) = 
\int d^2x_2 \; \Ga_{2}^{\Lazero(w),\Lazero }(0,x_2)\no\\
\de \ze_w &=&   \not\!\partial 
\hat\Ga_{2}^{\Lazero(w),\Lazero }(p)_{|p=0} =  
 \int d^2x_2\; i\xsla_2   \Ga_{2}^{\Lazero(w),\Lazero }(0,x_2)
\label{eff-const}\eqa
The effective constants are 
the vertex functions $\Ga_4$, $\Ga_2$ and $\not\!\partial\Ga_2$ for an 
effective theory
with infrared parameter $\La_0(w)$, 
and the renormalized constants correspond to the effective ones at the 
energy $\La$. (For the massive theory recall that we can use $\La=0$).
\bqa
\la_{w=0} &=& \la_r\no\\
\de m_{w=0} &=&  \de m_r = 0 \no\\
\de \ze_{w=0} &=&  \de \ze_r =0
\eqa
\end{theor}
The reshuffling of perturbation theory performed by Theorem II can be proved 
by standard combinatorial arguments as in [R] (the only difficulty
was discussed above, when we remarked that the 
parameter $w$ of the effective constants always
corresponds to the highest tree line
of the vertex. Otherwise the effective vertex generates
a tadpole graph whose later renormalization gives 0). 

\section{Convergence of the series.}
\resetequ

\begin{theor}
Let $\epsilon > 0$ be fixed. Suppose $\La^m$  (defined below)  
belongs  to 
some fixed compact $X$ of $]0,+\infty )$. The series (\ref{sviluppo4}) 
is absolutely convergent for $|\la_{w}|$, $|\de m_w|$, 
$|\de\ze_w|\leq c$, $c$ small enough. 
This convergence is uniform in $\La_{0}$ and
$N$ (actually $\Ga_{2p}$ is proportional to $N^{1-p}$). 
The ultraviolet limit 
$\Ga^{\La}_{2p}=\lim_{\Lazero\rightarrow\infty} 
\Ga^{\La\Lazero}_{2p}$ 
exists and satisfies the bound:
\bqa\lefteqn{
 |\Ga_{2p}^\La(\phi_1,...\phi_{2p})|
 \leq (p!)^{5/2}[ K(c,\e, X)]^p
\; (\La^m)^{2-p}\;N^{1-p}\;
}\\&&  
||\phi_1||_1 \;
\prod_{i=2}^{2p} ||\phi_i||_{\infty,2}\  e^{-(1-\epsilon )
\La^m d_T(\Om_1,...\Om_{2p})}  
\no\eqa
where 
\be \La^m :=\sup{[\La, m]} \ ,
\ee
\be  ||\phi_i||_{\infty,2}:= \biggr (||\phi_i||_{\infty}+
||\phi'_i||_{\infty}
+||\phi''_i||_{\infty}\biggr ) \ ,
\ee
$\Om_i$ is the compact support of $\phi_i$, $K(c,\e,X)$ is some 
function of $c$ $\e$ and $X$, which tends to zero
when $c$ tends to 0,    
$||\phi_i||_{\infty}=\sup_{x\in \Om_i}|\phi_i(x)|$,  
$||\phi_1||_1=\int d^2x |\phi_1(x)|$, and  
\bqa
d_T(\Om_1,...\Om_{2p}) &:=& \inf_{x_i\in \Om_i} 
d_T(x_1,...x_{2p})\no\\
d_T(x_1,...x_{2p})&:=& \inf_{u-\tree} 
\sum_{l\in \tree} |\bar{x}_l-x_l|.
\eqa 
where in the definition of $d_T(x_1,...x_{2p})$, called the ``tree
distance of $x_1,...x_{2p}$'', the infimum over $u-\tree$ is taken 
over all unordered trees (with any number
of internal vertices) connecting $x_1,...x_{2p}$. 
\end{theor}
This bound means that one can construct in a non perturbative sense
the ultraviolet limit of either the massive theory with any infrared
cutoff $\Lambda$ including  $\Lambda=0$, or the weakly coupled
massless theory with nonzero infrared
cutoff $\Lambda$. To complete Theorem 1 from Theorem 3,
one needs only to check Borel summability by expanding explicitly at finite
order $n$ in $\la_{ren}$ and controlling the Taylor remainder.
This additional expansion generates a finite number of 
Taylor operators $\tau_{g}$ for a finite number of non quasi-local subgraphs,
which are responsible for the $n!$ of Borel summability [R].
Since this is rather standard we will not include this additional argument 
here. Finally the renormalization group equations are discussed
in section V. The rest of the section is 
devoted to the proof of this theorem.
\subsection{Plan of the proof}
\resetequ
To prove the theorem we show that the absolute value of the term $(n,n',n'')$
in the sum (excluding the effective constants)
is bounded by $K^{\bar n}$. The strategy for the proof 
consists in moving the absolute value inside all sums and integrals, 
bounding the product of effective constants, 
\be
\left [\prod_v \left |\la_{w(v)}\right | \right ] 
\left [\prod_{v'}|\de m_{w(v')}|\right ] 
\left [\prod_{v''}|\de \ze_{w(v'')}|\right ]\leq c^{\bar n} \ ,
\ee 
then taking $c < K^{-1}$.

The loop determinant will be bounded  by a Gram inequality, and we shall use 
the tree lines decay to bound the spatial integrals. 
Actually, we cannot move the absolute value directly 
inside the sum over attributions  
because  $\#\{\mu\}\simeq\bar n!$. In other words  fixing the 
band index for each single half-line
develops too much the determinant. The way to overcome this difficulty is to
remark that the attributions contain much more
information than necessary. We can in fact 
group the attributions into packets to reduce the number of
determinants to bound. 
We  observe that, if for the level $i$ a connected component $G_i^k$ has 
$|EE_i^k|+|ET_i^k|\geq 5$, 
the subgraph is convergent and we do not need to 
know the band indices for the loop lines in that connected
component. So for each convergent  $G_i^k$:
\begin{itemize}
\item{} if $|EE_i^k|+|ET_i^k|\geq 5$, we do not want to know anything on loop
lines;
\item{} if $|EE_i^k|+|ET_i^k|< 5$,  we just want to fix 
$5-|EE_i^k|-|ET_i^k|$
half-lines with energy lower than $i$, but
we are not interested on the energy  of the other half-lines; 
\end{itemize}
Instead of expanding the loop determinant over lines and columns
as a sum over all attributions 
\be
\det {\cal M} = \sum_{\mu} \det{\cal M}(\mu,E)
\ee
we write it as a sum over a smaller 
set $\cal P$ (called the set of packets). These packets
are defined by means of the function
\bqa
\phi: \{\mu\} & \longrightarrow & {\cal P}\no\\
      \mu  &\mapsto & {\cal C} = \phi(\mu)
\eqa
but this function must respect some constraints related to
the future use of Gram's inequality. This motivates the following definition:
\begin{defin}
The pair $({\cal P},\phi)$ is called a ``Gram-compatible pair'' if
\be
\forall \;{\cal C}\in {\cal P}, \forall a,\; \exists\; J_a({\cal C})
 \subset B
\ee
with the property
$\phi^{-1}({\cal C}) = \{\mu | \mu(a)\in J_a({\cal C})\  \forall\; a\}$.
\end{defin}
This definition ensures that there exists a matrix ${\cal M'}$ such that
\be
\sum_{\mu\in \phi^{-1}({\cal C})} \det{\cal M}(\mu) 
= \det {\cal M'}({\cal C})
\ee
and that Gram's inequality can be applied to $\det {\cal M'}({\cal C})$, 
as shown in Lemma 4.

\subsection{Construction of {\cal P}}
We build first the {\em partition} ${\cal P}$ of the set of attributions
into packets. These packets should contain the informations
we need over $|E_k^j|$. In contrast with attributions
there should be few of them; more precisely they
should satisfy $\#{\cal P}\leq K^{\bar n}$. Finally, together  
 with the function
$\phi$,  they should form a Gram-compatible pair.
To define ${\cal P}$ we introduce some preliminary
definitions and notations.

To each  ordered tree $o-\tree$ we can associate a  rooted tree $R_\tree$, 
which pictures the inclusion relation of the $G^j_k$ [R]. 
We can picture this tree with two types of vertices: crosses and dots.
We recall that the leaves
of a rooted tree are the vertices of the rooted tree with coordination number
one. The leaves in our case are the dot-vertices and 
correspond exactly to the vertices $v$, $v'$ or  $v''$ of the initial
ordered tree $\tree$.
The other vertices of $R_\tree$ are crosses.
Each  cross $i$ corresponds to a line $l_i$ of the initial
ordered tree $\tree$, 
and has coordination number three, except the root which has 
coordination number two.
To build $R_\tree$ we take  the lowest line in $\tree$,  $l_1$, as root 1.

\medskip
\centerline{\hbox{\psfig{figure=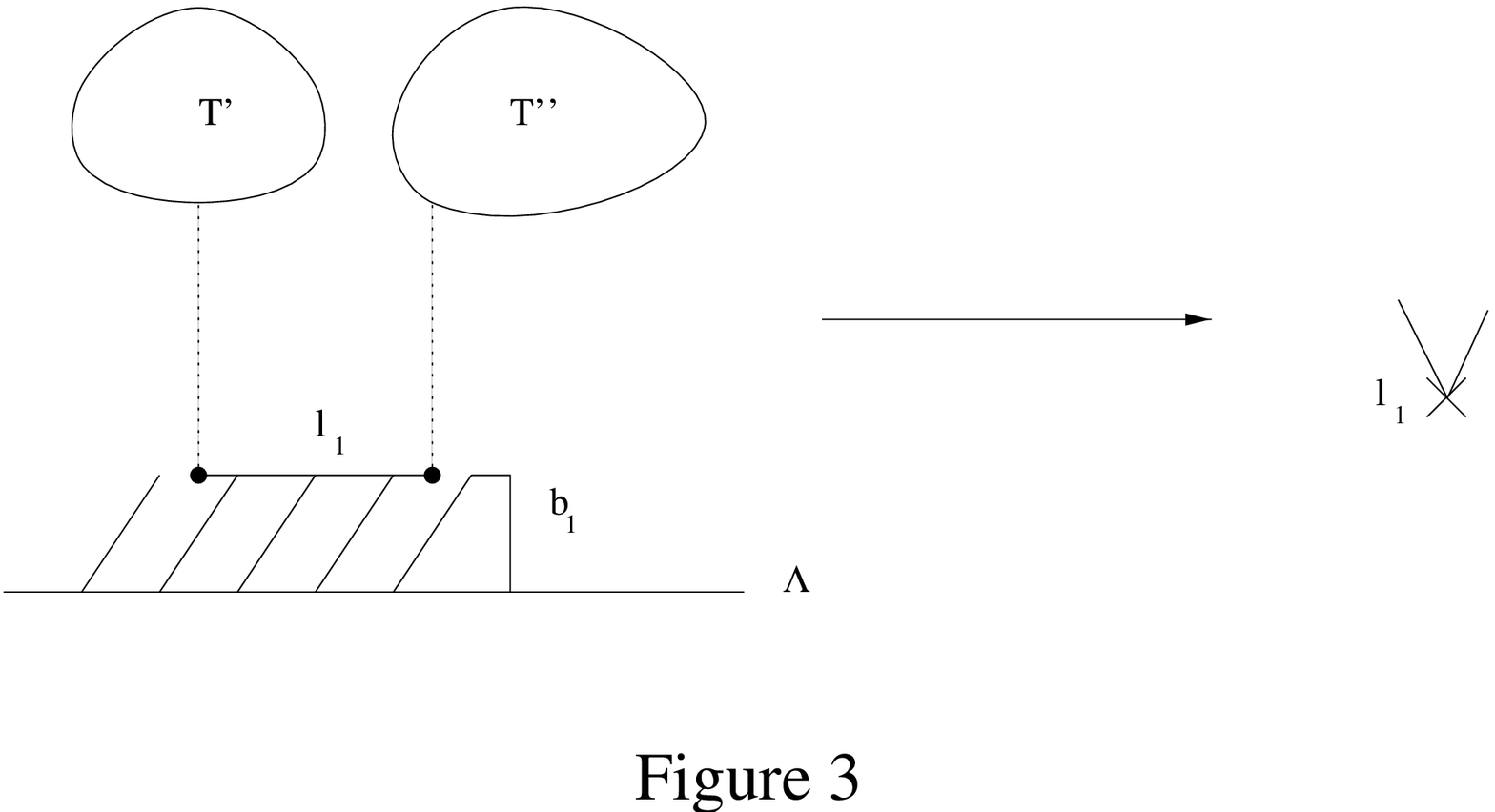,height=4cm,width=9cm}}}

This line $l_{1}$, or root, separates $\tree$ into two connected components  
$\tree'$ and $\tree''$ possibly reduced to a single vertex. 
When $\tree'$ or $\tree''$ is a single vertex, it gives
a dot connected to 1. Otherwise it gives a cross, which is
the lowest line of $\tree$  
in it. This procedure is repeated at each cross-vertex 
obtained, and generates $R_\tree$.

\centerline{\hbox{\psfig{figure=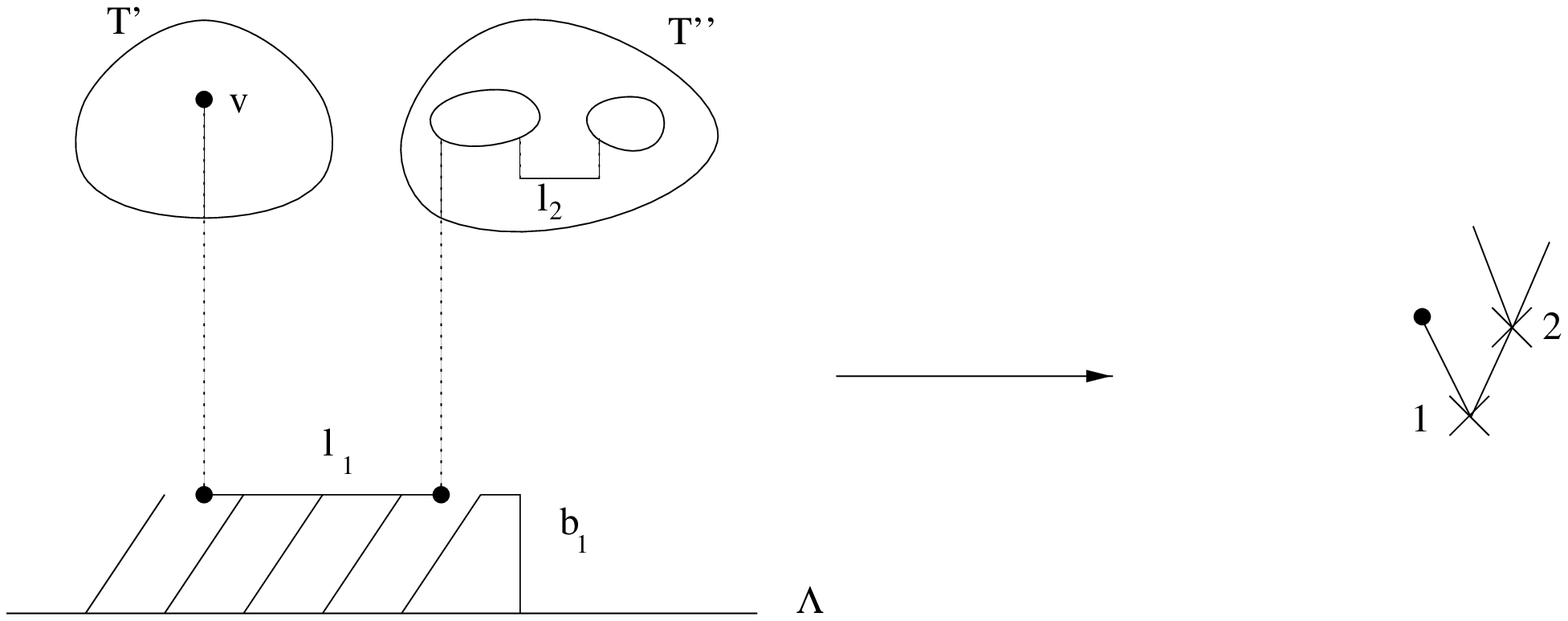,height=4cm,width=10cm}}}
\centerline{Figure 4}
\medskip

Finally to complete the picture to each dot
of $R_{\tree}$ we hook all loop half-lines hooked to the 
corresponding vertex (there could be none).
We define the {\em ancestor of i} ${{\cal A}(i)}$ 
as the cross-vertex just under $i$ 
in $R_\tree$ and we call $v_a$, the dot-vertex 
to which the half-line $a$ 
is hooked and $i_a$ the cross-vertex connected to $v_a$ (which represents
a line of the initial tree!). 
For each cross-vertex $i$ we define
\be
t_i := \{ l_j\in \tree | j\geq i, l_j\; \hbox{connected to} \; l_i
\; \hbox{by}\; T_i\}
\ee  

\medskip
\centerline{\hbox{\psfig{figure=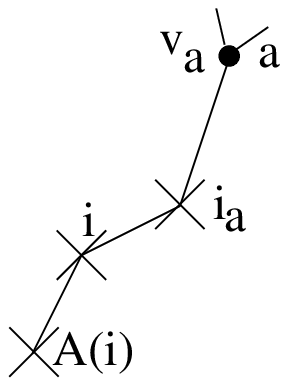,height=2cm,width=2cm}}}
\centerline{Figure 5}
\medskip

An example of a tree with its associated $R_\tree$ is given in Figure 6:

\medskip
\centerline{\hbox{\psfig{figure=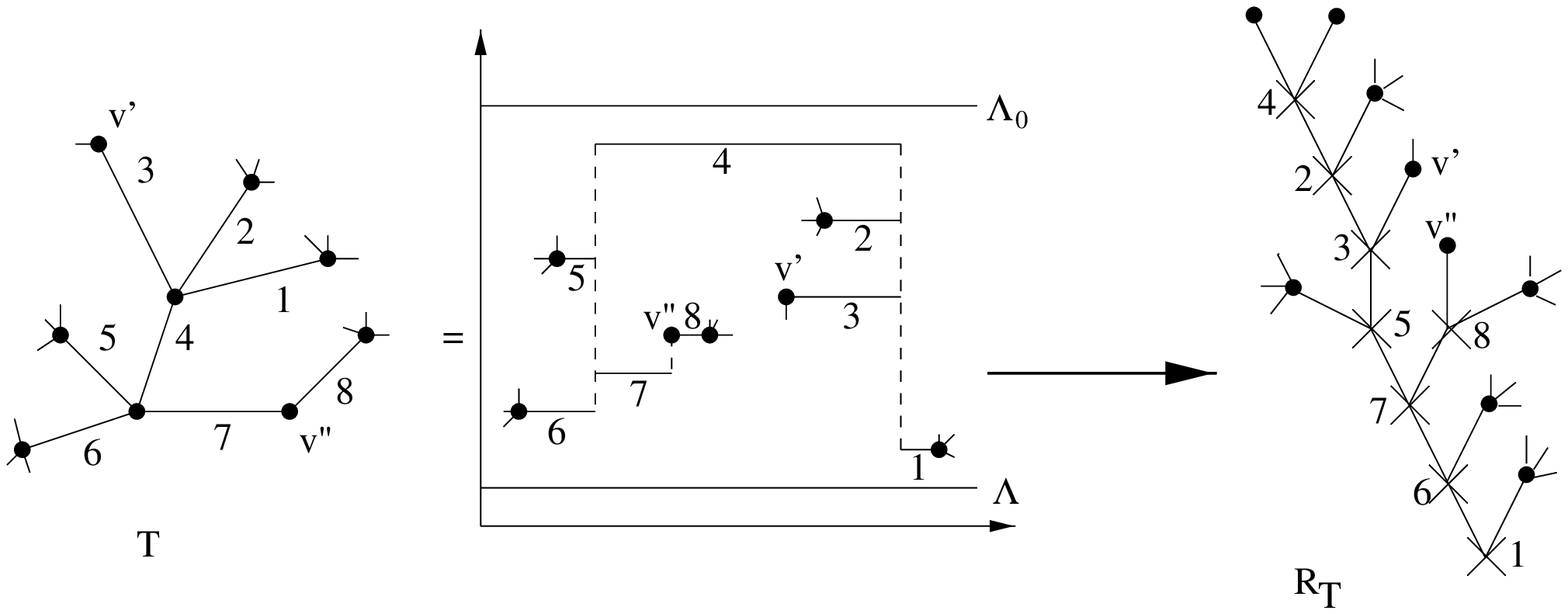,height=4cm,width=10cm}}}
\nobreak\centerline{Figure 6}
\medskip
\goodbreak

For each tree line (cross-vertex) $i$ 
and each connected component $G_{i}^k$, 
no new line connects to $t_i$ in the interval
between $i$ and ${\cal A}(i)$.  Hence 
$\om'(G_{i'}^{k'})\geq \om'(G_{{\cal A}(i)+1}^{k''})$ 
$\forall i\geq i'>{\cal A}(i)$  and $T_{i'}^{k'}\subset G_{i'}^{k'}
\subset G_{i}^k$.
Therefore we 
can neglect what happens in this interval and 
generalize the definitions of (\ref{definizioni}) for  
the internal lines of a subgraph $G_i^k$. We define:
\bqa
g_i & :=& t_i \cup \{a\in L| i_a\geq i, v_a\in t_i, \mu(a)\geq {\cal A}(i)+1\}
\no \\
eg_i & :=&  et_i \cup ee_i \cup el_i \no\\
 et_i &:= &  \{ f_{i'} | i_{v_f}\geq i, v_f\in t_i, i'<i\}
\cup  \{ g_{i'} | i_{v_g}\geq i, v_g\in t_i, i'<i\}\no\\
 ee_i &:=& \{ f,g \in E | i_{v_f}, i_{v_g}\geq i, v_f, v_g\in t_i\}\no\\
 el_i &:=& 
\{ a\in L |i_a\geq i,  v_a\in t_i, \mu(a)\leq {\cal A}(i)\}
\label{definiblou}
\eqa
This set of definitions (\ref{definiblou}) concerns
the connected component $g_{i}$ above line $i$.
Remark that we defined as loop internal lines of $g_{i}$, 
all loop lines higher than ${\cal A}(i)$. We also need
some additional definitions concerning the other connected
components:
\bqa
i(k)&:= & \inf_{\{j\geq i, v_j\in T_i^k\}} j\no\\
g_i^k & :=& g_{i(k)}
\no \\
eg_i^k & :=&  et_i^k \cup ee_i^k \cup el_i^k \no\\
 et_i^{k} &:= & et_{i(k)} \quad 
ee_i^k := ee_{i(k)} \quad el_i^k := el_{i(k)}
\label{definibla}
\eqa
This second set of definitions is used only much 
later in the bounds when all connected components are considered at once.
\begin{defin}
A \underline{chain} $C_{a,i}$ is the unique path in $R_\tree$ 
from the half-line $a$ to the cross-vertex  $i$ with 
 $i_a\geq_T i$:
\be
C_{ai}:= \{i'|i\leq_T i'\leq_T i_a\}\cup \{a\}
\ee 
\end{defin}
In the following, we write $i_a\geq_T i$ to specify that $v_a$ and
$v_i$ are connected by $t_i$. 

\centerline{\hbox{\psfig{figure=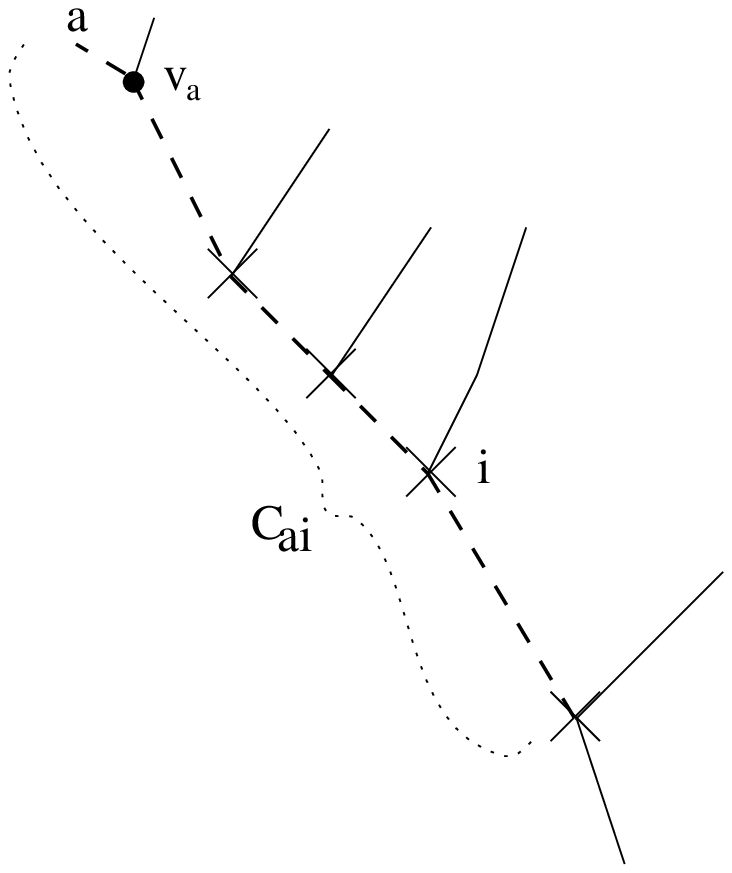,height=4cm,width=4cm}}}
\centerline{Figure 7}

\begin{defin}
a \underline{class} ${\cal C}$ is a set of chains over $R_\tree$ with the
properties:
\bqa
&& \forall C_{ai}\in {\cal C}, \ 
\forall C_{a'i'}\in {\cal C} \ {\rm one \ has}\ a\neq a' \no\\
&&\forall\; i \quad c_i \leq \max[0;5-|ee_i|-|et_i|- c'_i]
\label{chain0}\eqa
where we defined:
\bqa
c_i &=& \#\{ C_{ai}\in {\cal C}| \; i \; \hbox{fixed}\}\no\\
c'_i& =& \#\{C_{a'i'}\in {\cal C}|\; i_{a'}\geq_T i, \; i'<i\}
\label{chain1}
\eqa
\end{defin}
So $c_{i}$ is the total number of chains arriving at $i$ and  
$c'_{i}$ is the total number of chains passing through $i$ and
continuing further below.  
This definition ensures therefore that there  are at 
most five chains passing through each cross $i$. 
\begin{defin}
The partition ${\cal P}$ is 
the set of all possible classes ${\cal C}$ over $R_{\tree}$.
\end{defin}
To verify that this is a good definition, we have to prove three lemmas.
\begin{lemma}
The cardinal of ${\cal P}$ is bounded by  $K^{\bar n}$.
\end{lemma}
\paragraph{Proof:} we prove that 
${\cal P}\subseteq {\cal P'}$ and  $\#{\cal P}'\leq K^{\bar n}$. 
We define ${\cal P'}$ as the set of all sets of chains ${\cal D}$, that are
unions of five subsets  (possibly empty) $Y_j$, where $Y_j$ is a set of
 completely disjoint chains (this means they have no cross and no dot 
in common).
\be
{\cal P'}:= \{{\cal D}\}    \qquad {\cal D} := \cup_{j=1}^5 Y_j.
\ee

To build a set of disjoint chains $Y_j$, we have at most
three possible choices for each  vertex: at each cross-vertex we can have
no chain passing, a chain going right or left; at each dot-vertex
touched by a chain, we have to choose among three (at most) loop half-lines.
Putting all this together we have:
\be
\# {\cal P'} \leq (3^5)^{{\bar n}-1}(3^5)^{2n+2-2p}\leq K^{\bar n}
\ee
where the number 5 comes because each element of ${\cal C}$ is made of five 
sets $Y_j$.

Figure 8 shows an example of disjoint sets built in this way.

\centerline{\hbox{\psfig{figure=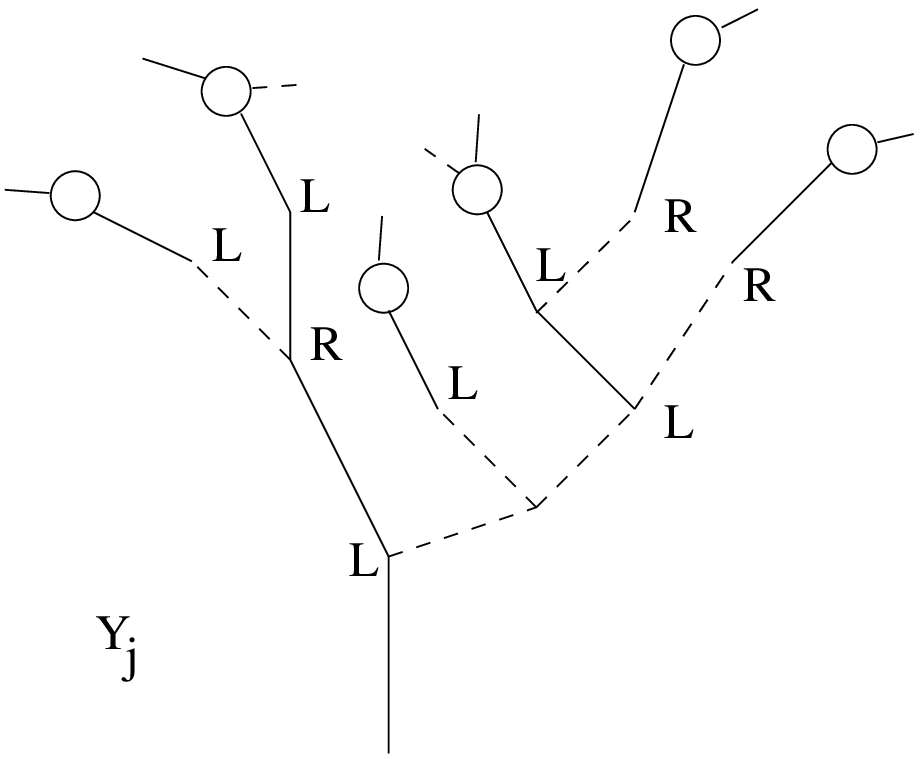,height=4cm,width=5cm}}}

\medskip
\centerline{Figure 8}
\medskip
  
Now we prove that ${\cal P}\subseteq {\cal P'}$ by induction on $i$. 
For each ${\cal C}\in {\cal P}$ we define ${\cal C}(i)$ as the 
subset of ${\cal C}$ that contains only chains ending in some point
(cross-vertex) of 
the unique path connecting $i$ to the root. 
\be
{\cal C}(i) := \{C_{ai'}\in {\cal C}|\; i'\leq_T i\}
\ee
This set satisfies the following induction law:
 if, for ${\cal A}(i)$ there are five sets (eventually empty) 
of disjoint chains 
$Y_1({\cal A}(i))$...
$Y_5({\cal A}(i))$ with 
\be
{\cal C}({\cal A}(i)) = \cup_{j=1}^5 Y_j({\cal A}(i)), 
\label{hypoth}\ee
then there are five sets $Y_1(i)$,...,$Y_5(i)$ with 
${\cal C}(i)=\cup_j Y_j(i)$.
This can be seen observing that ${{\cal C}(i)}$ can be written as 
\be
{\cal C}(i) = {\cal C}({\cal A}(i)) \cup \{C_{ai'}\in {\cal C}| i'=i\}. 
\ee
Among the five sets $Y_j$ forming  ${\cal C}({\cal A}(i))$ there are $c_i'$ 
ones containing chains passing through $i$: 
 $Y_1({\cal A}(i))$,..., $Y_{c'_i}({\cal A}(i))$.
If $c'_i+|ee_i|+|et_i|\geq 5$, there are no chains ending at $i$ so
${\cal C}(i)={\cal C}({\cal A}(i))\subset {\cal P'}$.
If  $c'_i+|ee_i|+|et_i|< 5$ there are $c_i$ chains ending at $i$
$C_{a_1,i},...C_{a_{c_i},i}$,  with 
 $c_i\leq 5-c'_i$,
so we can define 
\bqa
Y_j(i) &=& Y_j({\cal A}(i)) \quad {\rm for} \ j\leq c'_i, \no\\
Y_{c'_i+j}(i) &=& Y_{c'_i+j}({\cal A}(i))\cup \{C_{a_j i}\}\quad 
j=1,...,c_i,\\
Y_j(i) &=& Y_j({\cal A}(i)) 
\quad  {\rm  \ for\ } j>c'_i+c_i.\no
\eqa
With these definitions we have
\be
{\cal C}(i)= \cup_{j=1}^5 Y_j \subset {\cal P'}
\ee
Now, the hypothesis (\ref{hypoth})  is true for the root $r$. 
In fact, by construction, we
have at most five chains ending at  $r$: $C_{a_1,r},...C_{a_5,r}$. 
If we define:
\be
Y_1(r) = \{C_{a_1r}\},...,Y_5(r)=\{C_{a_5r}\}
\ee 
we have
 ${\cal C}(r)= Y_1\cup..\cup Y_5 \subset {\cal P'}$. Working the induction
up to the leaves of $R_{\tree}$ completes the proof of the lemma.
\hfill $\Box$

\begin{lemma}
There exists a function 
$\phi: \{\mu\} \longrightarrow {\cal P}$
which associates to each attribution $\mu=(\mu(1), \mu(2),...)$ a
class ${\cal C}$ in ${\cal P}$.
\end{lemma}
To define $\phi$ we fix an \underline{order} over the half-lines
and the lines of $R_\tree$. We do it
turning around $R_\tree$ clockwise and we call $n(a)$
the index of $a$ in the ordering and $s_i$ the index of the line in
$R_\tree$ connecting $i$ to ${\cal A}(i)$. 

\medskip

\centerline{\hbox{\psfig{figure=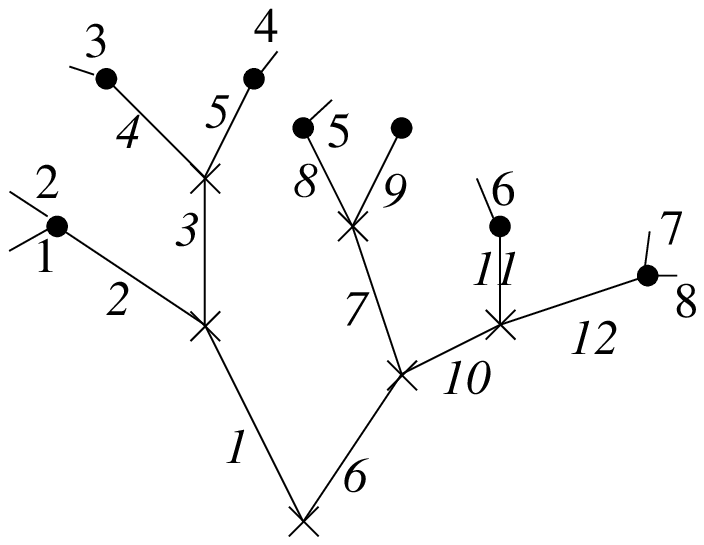,height=4cm,width=5cm}}}

\medskip
\centerline{Figure 9}  

\medskip

We build the class $\phi(\mu)$ as a union of chains
by induction, defining first the chains in $\phi(\mu)$ ending
at the root, then the ones ending at the 
cross connected to the root by the line 1,
and so on, following the
ordering $s_i$. Therefore for each $i$ we consider the set
$A_i = \{ a\in el_i|  \not\!\exists C_{ai'}\in \phi(\mu) \ \hbox{with }\ 
i'<i\}$
which is the set of loop half-lines that are external lines for 
$g_i$ and are not
connected to a chain in $\phi(\mu)$ ending lower than $i$. 
%\item{}  
\paragraph{-} If  $[5-|ee_i|-|et_i|-c'_i]>0$ and 
$\# A_i < [5-|ee_i|-|et_i|-c'_i]$ we have a divergent subgraph, and
we add to the part already built of $\phi(\mu)$  
all the chains starting at an element of $A_i$ and ending at $i$, so 
\be
c_i= \#A_i.
\ee 
%\item{} 
\paragraph{-} If  
$\# A_i \geq \max [0, 5-|ee_i|-|et_i|-c'_i]$,  
we have a convergent subgraph, so we put 
\be
c_i = \max [0, 5-|ee_i|-|et_i|-c'_i] 
\ee
and we add to the part already built of $\phi(\mu)$  the
$c_i$ chains ${\cal C}_{a',i}$, 
with $a'=a_i^j, \  j=1...,c_i$, which start at the 
$c_i$ elements in $A_i$ that have the lowest values of $n(a)$, and end at $i$. 

In this way we obtain a set of chains with the  two properties  
(\ref{chain0}).
For each $\mu$,   $\phi(\mu)$ is an element of ${\cal P}$ 
and  $\{\phi^{-1}({\cal C})\}_{{\cal C}\in {\cal P}}$ 
is a partition of the set of attributions.\hfill $\Box$

\medskip

\centerline{\hbox{\psfig{figure=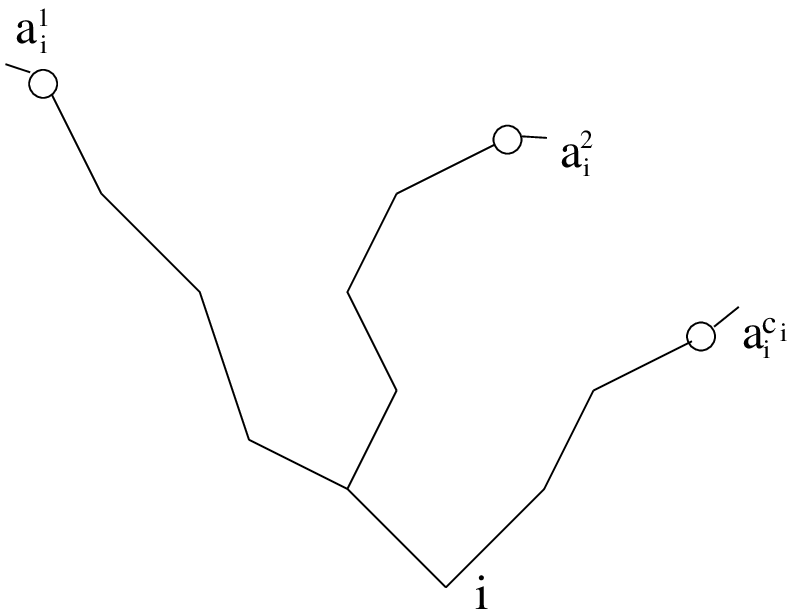,height=4cm,width=5cm}}}
\medskip
\centerline{Figure 10}
\medskip
We call $B_{i}$ the set of half-lines in $A_{i}$ which are the starting points
of chains in  $\phi(\mu)$ ending at $i$ (see Figure 10). 
Therefore in the divergent case
$B_{i}=A_{i}$ and in the convergent case $B_{i} = \{a_i^j, \  j=1...,c_i\}$.
We also define 
\be  eg_{i}({\cal C}) := et_i  \cup ee_i  \cup \{  a | i_{a} \ge_{T} i 
\ {\rm and} \ 
a \in B_{i'} \ {\rm for\ some\ } i'\le_{T} i\} 
\label{externalbla} 
\ee
With this definition we have
$|eg_{i}({\cal C})| =c_i+c'_i+|et_i|+|ee_i| $. Remark that
in the divergent case $|eg_{i}| \le 4$, one has 
$|eg_{i}|=|eg_{i}({\cal C})|$,
and in the convergent case one has 
$|eg_{i}|\geq |eg_{i}({\cal C})| \ge 5$.
The next lemma describes the structure of the classes ${\cal C}$.
\begin{lemma} For each class ${\cal C}\in {\cal P}$ and each
half-line $a=1,...,2n+2-2p$ there exists a subset of band indices
$J_a({\cal C})\subseteq B$ such that
\be
\phi^{-1}({\cal C}) = \{\mu |\mu(a)\in J_a({\cal C})\;\forall\;a\}.
\ee
\end{lemma}
\paragraph{Proof}
The existence of the $c_i$ chains  $C_{a i}$ for $a\in B_{i}$
ending at $i$ implies a certain 
set of constraints on attributions. We distinguish two situations.
\paragraph{1)}  If $|eg_{i}({\cal C})| \le 4$ (divergent case)
\begin{itemize}
\item{} $\forall a \in B_{i}, \ \underline{\mu(a)\leq {\cal A}(i)}$; 
\item{} $\forall$ $a\not\in B_{i} $ with $i_a\geq_T i$, 
$\underline{\mu(a)> {\cal A}(i)}$.
\end{itemize}
\paragraph{2)} If  $|eg_{i}({\cal C})| \ge 5$ (convergent case)
\begin{itemize}
\item{} $\forall a\in B_{i} , \  \underline{\mu(a)\leq {\cal A}(i)}$; 
\item{} $\forall 
a \not\in B_{i}$ with $i_a\geq_T i$, and $n(a)< \max_{a'\in B_{i'}} n(a')$,
 $\underline{\mu(a)>{\cal A}(i)}$;
\end{itemize}
In any other case, there is no particular constraint.
We observe that the underlined constraints for $\mu(a)$ are therefore 
determined by the chain
structure and the ordering, but the crucial point is that
they are independent from each other.
Hence $J_a({\cal C})$ is an interval
in terms of band indices.
Remark that if some chain in ${\cal C}$ starts from $a$, 
it must end at some unique
$i$, called $i'_{a}$. In this case we define $ M(a,{\cal C}) = 
{\cal A}(i'_{a}) $. Otherwise we define  $ M(a,{\cal C}) = i_{a}$. 
Moreover for each $i'$ such that $a\not\in B_{i'}$ we have two 
different lower bounds on $\mu(a)$, 
 depending whether $g_{i'}$ is divergent or convergent.
So the constraints in cases 1 and 2 simply mean  
$ m(a,{\cal C}) \le \mu(a)\le M(a,{\cal C}) $, where 
\bqa
M(a,{\cal C}) &=& {\cal A}(i'_{a})   \ {\rm if}\ a \in B_{i'_{a}}
\quad , \quad
M(a,{\cal C})  = i_{a}    \ {\rm otherwise} \no\\
 m(a,{\cal C}) &=& \sup_{i'\in I(a,{\cal C})} [{\cal A}(i')+1] \no\\
 m(a,{\cal C}) &=& 1 \quad  {\rm if}\ I(a,{\cal C}) = \emptyset
\label{bounds}\eqa
and
\be
I(a,{\cal C}) := \{ i'\;| i_{a} \ge_{T} i', \  a \not\in B_{i'}, 
\ {\rm and, \ if }\ |eg_{i'}({\cal C})| \ge 5,
\ n(a)< \max_{a'\in B_{i'}} n(a') \}
\ee

In summary the constraints are expressed by
\bqa
\phi^{-1}({\cal C}) &=& \{\mu | \;\mu(a)\in J_a({\cal C})\; \forall\; a\}\no\\
J_a({\cal C}) &= &[m(a,{\cal C}), M(a,{\cal C})]
\eqa   
\hfill $\Box$
\paragraph{Example:}
As an example, in Figure 11 we have three chains 
$C_{a,3}$, $C_{a',2}$ and $C_{a'',1}$. The
bands are: $i_1+1 \leq \mu(a)\leq i_2$, 
$i_0+1\leq\mu(a')\leq i_1$, $\mu(a'')\leq i_0$.

\centerline{\hbox{\psfig{figure=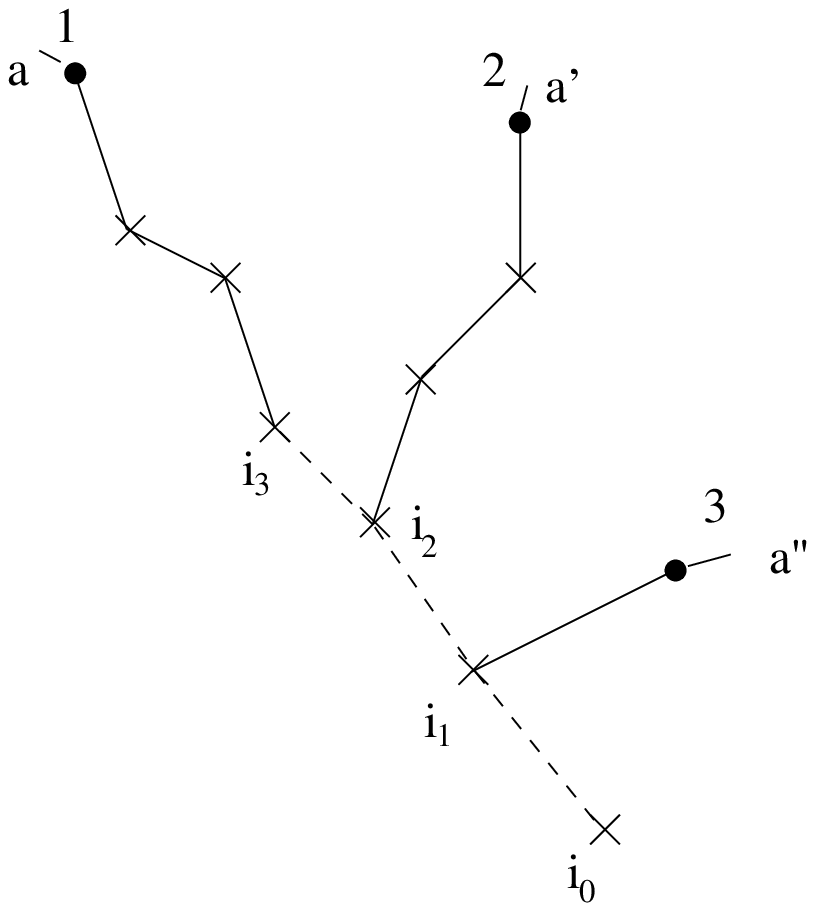,height=5cm,width=5cm}}}

\medskip
\centerline{Figure 11}
\medskip

We observe that, after packing the attributions into classes, the 
sets $T_i$, $t_i$,  $ee_i$, $et_i$ 
are still well defined, but we can no longer define
$g_i$ and $el_i$. We already defined $eg_{i}({\cal C})$
in (\ref{externalbla}). We add further definitions
\bqa
g_i({\cal C})& :=& t_i \cup il_i({\cal C})\no\\
il_i({\cal C}) &:=& \{ a\in L| 
i_a\geq_T i, M(a, {\cal C})\geq {\cal A}(i)+1\}\no\\
el_i({\cal C}) &:=& \{ a\in L| i_a\geq_T i, M(a, {\cal C})\leq {\cal A}(i)\}
\label{externalblou}
\eqa
which generalize the notions of internal and external loop lines. 
Remark that $eg_i({\cal C})=  et_i  \cup ee_i \cup el_i({\cal C})$,
and $|el_i({\cal C})|= c_{i} + c'_{i}$.
In the same way we extend these definitions
to the other connected components 
\be g_i^k({\cal C}):= g_{i(k)}({\cal C})  \ ,\ 
il_i^k({\cal C}) := il_{i(k)}({\cal C})   \ ,\  
el_i^k({\cal C}):= el_{i(k)}({\cal C}) 
\label{externalbli}
\ee 
Furthermore the generalized definitions for the convergence degree and the 
set of divergent subgraphs after packing the attributions into  classes
become:
\bqa
\om(g_i({\cal C}))& :=& (|eg_i({\cal C})|-4) / 2 .\no\\
D_{\cal C}&:=& \{g_i({\cal C}) \; | \ 
\om(g_i({\cal C}))\leq 0\}.
\label{externalblu}
\eqa

We return now to the loop determinant in (\ref{sviluppo4}).
Lemma 3  ensures that 
\be
\sum_{\mu\in \phi^{-1}({\cal C})} det {\cal M}(\mu) = det {\cal M'}({\cal
C})
\ee
and that for each loop half-line $a$ there exists a characteristic function 
\be
\chi_a({\cal C}):k\in B\rightarrow \{0,1\} \quad
\chi_a^k({\cal C}) = 
\left\{\ba{cc}
0 & \hbox{if}\; k\not\in J_a({\cal C})\\
1 &  \hbox{if}\; k\in J_a({\cal C})\ .\\
\ea\right.
\ee
Therefore the matrix elements for ${\cal M'}({\cal C})$ can be written 
\bqa
{\cal M'}_{fg}(x_f,x_g)&=&
\int {d^2p\over (2\pi)^2}\; e^{-ip(x_f-x_g)} C(p) 
\sum_{k\in B} \chi_{a(f)}^k\chi_{a(g)}^k \eta^k(p) 
W^k_{v_f,v_g}\no\\
&=&\int {d^2p\over (2\pi)^2}\; F^*_f(p) G_g(p) 
\sum_k \chi_{a(f)}^k\chi_{a(g)}^k \eta^k(p) W^k_{v_f,v_g}\no
\eqa
where we omit for simplicity to write the dependence 
in ${\cal C}$, and we defined:
\be
F_f(p)= e^{ix_f p} \frac{1}{(p^2+m^2)^{\frac{1}{4}}}\qquad 
G_g(p)= e^{ix_g p} \frac{(-\psla+m)}{(p^2+m^2)^{\frac{3}{4}}}.
\label{FG}\ee
$v_a$ is the vertex to which the half-line $a$ is hooked and
 $\eta^k$ is the cutoff restricted to the band $k$ (see equation 
(\ref{eta-band})). 
Finally $W^k$ is the $\bar n\times \bar n$ matrix defined in 
equation (\ref{defw}). Our next lemma is crucial since it bounds the loop
determinant without generating any factorial.
\begin{lemma}
The matrix
${\cal M'}({\cal C})$ satisfies the following Gram inequality:
\be
|\det {\cal M'}({\cal C})| \leq \prod_f \left [
\int {d^2p \over (2\pi)^2}\; \eta_{\cal C}^f(p)  |F_f(p)|^2
\right ]^{\frac{1}{2}}
 \prod_g \left [
\int {d^2p\over (2\pi)^2}\; \eta_{\cal C}^g(p)  |G_g(p)|^2
\right ]^{\frac{1}{2}}
\label{gram}\ee
where the cutoff functions $\eta_{\cal C}^f(p)$ and $\eta_{\cal C}^g(p)$
corresponding to fields $f$ and $g$ are defined in equation 
(\ref{new-cutoff}) below.
\end{lemma}
\paragraph{Proof}
The Gram inequality states:

{\em If $M$ is a $n\times n$ matrix with elements $M_{ij}= <f_i,g_j>$
and $f_i$, $g_j$ are vectors in a Hilbert  space, we have 
$|\det M|\leq \prod_{i=1}^n ||f_i||\;  \prod_{j=1}^n ||g_j||$.}

To apply Gram's inequality, the matrix elements must be written as
scalar products.
We introduce the $q\times q$  matrix $1_q$ which is not the identity, but the
matrix with all coefficients equal to 1. It is obviously a positive symmetric
matrix. We observe that the matrix $W^k_{v,v'}$ is block diagonal 
with diagonal blocks of type $1_{q_j}$, and $\sum q_j=\bar n$.  
Each block corresponds to 
all the vertices in a given connected component of $T_k$.
Therefore $W$ itself is positive symmetric.
We can define the symmetric matrix $(2n+2-2p)\times (2n+2-2p)$:
\be
R_{ab}^k:= \chi_a^k \chi_b^k 
\ee
where $a$ and $b$ are the indices for the loop half-lines. 
By a permutation of field indices, we can list first the $q$
half-lines for which $\chi^k_a({\cal C})=1$. 
In this way the matrix $R$ becomes 
$2\times 2$ block diagonal positive of the type
\be
\lp\ba{cc}
1_q &0\\
0&0\\
\ea\rp.
\ee
Now we can group  
$W$ and $R$ in a unique matrix (tensor product)
\be
{\cal W}^{k}_{v,a;v',b}:= \chi_a^k \chi_b^k W_{v,v'}^k
\ee
that is still  positive  as we can diagonalize separately $W$ and
$R$.
Hence the matrix 
\be
\sum_k \eta^k {\cal W}^{k}_{v,a;v',b}=T_{v,a;v',b}
\ee
is symmetric positive, as is a linear combination
(with positive coefficients $\eta^k$) of symmetric
positive matrices; therefore we 
can take its square root (which is also positive symmetric):
\be
T_{v,a;v',b} = \sum_{w,c}U_{v,a;w,c} U_{w,c;v',b}.
\ee
Now, we can write ${\cal M}'_{fg}$  as 
\bqa
{\cal M}'_{fg}
&=&\int  {d^2p\over (2\pi)^2}\; F^*_f(p) G_g(p) 
T_{v_f,a(f);v_g,a(g)}\no\\
&=& \int  {d^2p\over (2\pi)^2}    \; F^*_f(p) G_g(p) \sum_{v's}
U_{v_f,a(f);v',s} U_{v',s;v_g,a(g)}
\eqa
If we introduce the vectors
\be
{\cal F}^f_{v's}(p) = F_f(p) U_{v',s;v(f),a(f)}\qquad
{\cal G}^g_{v's}(p) = G_g(p) U_{v',s;v(g),a(g)}
\ee
we can write ${\cal M}'_{fg}$ as
\be
 {\cal M}'_{fg}= \int {d^2p\over (2\pi)^2} \;\sum_{v', s}  {\cal F}^{f*}_{v's}\;
\; {\cal G}^g_{v's} \;
= <\vec{{\cal F}}^f,\vec{{\cal G}}^g>.
\ee
Now we can apply Gram's inequality:
\be
|\det {\cal M}'_{fg}| \leq \prod_{f=1}^{n+1-p} ||\vec{{\cal F}}^f||\; 
 \prod_{g=1}^{n+1-p} ||\vec{{\cal G}}^g||
\ee
where
\[
||\vec{{\cal F}}^f||^2 = \int {d^2p\over (2\pi)^2} \; \sum_{v', s} 
({\cal F}^f_{v's})^t ({\cal F}^f_{v's}) 
  =  \int  {d^2p\over (2\pi)^2} \;\sum_{v's} U_{v(f),a(f);v',s}  
U_{v',s;v(f),a(f)} |F_f|^2  
\]
\vskip-.7cm
\bqa       &&       = \int  
{d^2p\over (2\pi)^2}\; T_{v(f),a(f); v(f),a(f)} |F_f|^2
               = \int  {d^2p\over (2\pi)^2} \sum_k 
\chi^k_{a(f)} \chi^k_{a(f)} W^k_{v(f),v(f)} \eta^k |F_f|^2\no\\
&&= \int  {d^2p\over (2\pi)^2}\; (\sum_k \eta^k(p)\chi^k_{a(f)}) |F_f(p)|^2
= \int  {d^2p\over (2\pi)^2}\; \eta^{a(f)}(p) \; |F_f(p)|^2
\eqa
as $(\chi^k_{a(f)})^2 = \chi^k_{a(f)}$, and,
as the bands in $\chi_a$ are adjacents, 
the cut-offs sum up
(using equations (\ref{eta-band0}-\ref{eta-band})
to give 
\be
\eta_{\cal C}^a (p) := \left [  \eta\lp \frac{p^2+m^2}
{\Lazero({w}_{M(a,{\cal C})})}\rp -
 \eta\lp \frac{p^2+m^2}
{\Lazero({w}_{m(a,{\cal C})-1})}\rp \right ]
\label{new-cutoff}\ee
We can treat in the same way $G$ and this achieves the proof of (\ref{gram}).
\hfill $\Box$

\subsection{Bound on the series}

We are now in the position to  bound the series (\ref{sviluppo4}).
After packing the attributions into packets we can put the absolute value
inside the integrals and the sums and  bound the product of effective
constants by $c^{\bar n}$.  
Moreover, we observe that
the two sums $\sum_{{\cal C}ol, \Om}$ in (\ref{sviluppo4})
are bounded by taking the supremum over ${\cal C}ol$ and $\Om$ and
multiplying by the number of elements. We have
\bqa
&&\#\{\Om\} \leq 2^{2n+n'+ n''-p}< 4^{\bar n}2^{-p}\no\\
&& \#\{{\cal C}ol\} \leq N^{n+1-p}
\eqa
Indeed to estimate $\#\{{\cal C}ol\}$ remark that,  
once $\tree$ and $\Om$ are known, the circulation of color indices 
is determined. If there are no external color indices fixed (vacuum graph), 
the attribution of color indices costs $N^2$ at the first
four-point vertex (taken as root)
and climbing inductively into the tree layer by
layer a factor $N$ for each of the remaining 
four-point vertices of the tree (see [AR2]). 
The two-point vertices do not contribute
as color is conserved at them. When we have fixed the $p$ independent
external colors for the $2p$ external fields only
$N^{n+1-p}$ choices remain. 
 
We introduce some notations. Recalling the definitions (\ref{externalblou})
and (\ref{externalblu})
we say that a divergent subgraph $g_i({\cal C})\in D_{\cal C}$ is `D1PR' 
(`dangerous one particle reducible')  if, by cutting a single tree line, 
we can cut it into two
subgraphs $g_j({\cal C})$ and  $g_{j'}({\cal C})$, one of them, say 
$g_j({\cal C})$, being a two point subgraph. 
The line to cut is then the tree line $l_{{\cal A}(j)}$. 
In Figure~12 we show some examples of 
D1PR subgraphs, where tree lines are solid lines and loop 
half-lines are wavy.

\medskip

\centerline{\hbox{\psfig{figure=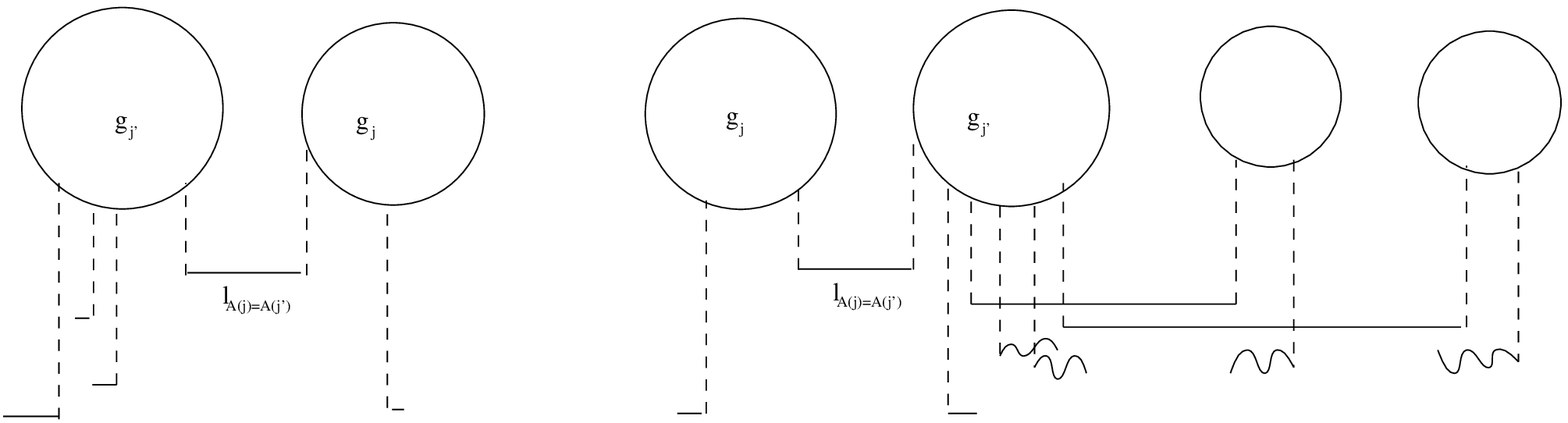,height=3cm,width=11cm}}}
\medskip
\centerline{Figure 12}
\medskip

All subgraphs that cannot be cut in this way are called D1PI 
(`dangerous one particle irreducible'). We say that a 
four-point D1PI  subgraph $g_i({\cal C})$ is `open' (as in [R]) if 
there exists a 
two-point subgraph $g_j({\cal C})\in D_{\cal C}$ (called its closure) 
with $j\leq_T i$ (then $g_i({\cal C})\subset 
g_j({\cal C})$) and they have two external lines in common (see Figure~13).

\medskip

\centerline{\hbox{\psfig{figure=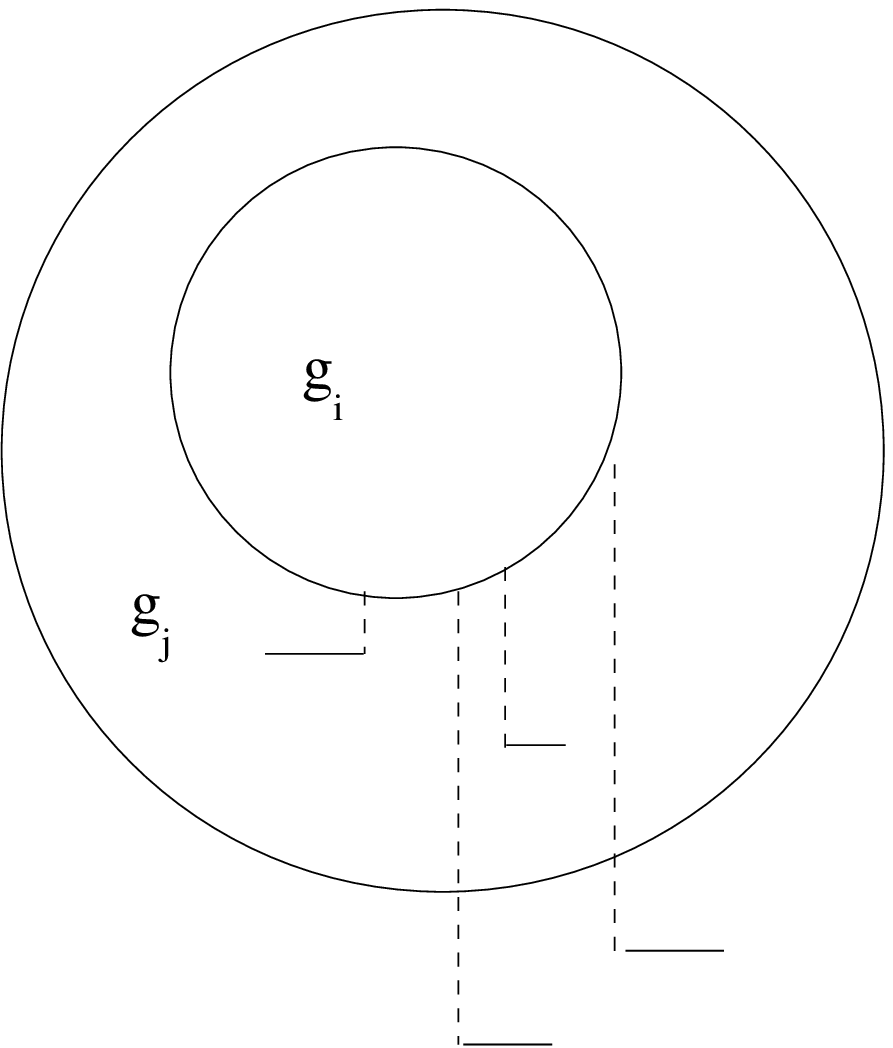,height=3cm,width=3cm}}}
\medskip
\centerline{Figure 13}
\medskip
     
 A four-point
subgraph is called `closed' if it is D1PI but not open. A two-point
D1PI subgraph is always closed
by definition. This classification of subgraphs is useful, as  
only closed subgraphs contribute in the product 
$\prod_{g\in D_{\cal C}} (1-\tau^*_g)$.
Applying the definition of $\tau_g$ in the momentum space one can see that:

- if $g_i({\cal C})$ is D1PR and $g_j({\cal C})$ 
is the corresponding divergent subgraph, then
\be
\tau_{g_i({\cal C})} (1-\tau_{g_j({\cal C})})=0
\ee
so the renormalization of $g_i({\cal C})$ is ensured by that of 
$g_{j}({\cal C})$;

- if $g_{i}({\cal C})$ is four-point and open, 
and   $g_{j}({\cal C})$ is the associated two-point subgraph
containing it, then
\be
(1-\tau_{g_{j}({\cal C})})(\tau_{g_{i}({\cal C})})=0. 
\ee

For any $g_{i}({\cal C})\in D_{\cal C}$ 
we know exactly which loop half-lines  are external lines, therefore 
we can still apply the operator $1-\tau^*_g=R_g^*$ to the external 
propagators, and distinguish closed subgraphs. Hence we 
define 
\be
D_{\cal C}^c:= \{g_{i}({\cal C}) \in D_{\cal C}| g_{i}({\cal C}) \; 
\hbox{closed} \}
\ee
and we apply $R_g^*$ only to $g\in D_{\cal C}^c$.   
By the relation of partial order in $R_\tree$ we see that 
for each pair $g_{i}({\cal C})$,$g_{i'}({\cal C})\in D_{\cal C}^c$ 
we can only have that
$g_{i}({\cal C}) \cap g_{i'}({\cal C}) =\emptyset$, 
or $g_{i}({\cal C}) \subseteq  g_{i'}({\cal C})$ (if 
$i'\leq i$).  Hence  $D_{\cal C}^c$ has a forest structure. 
Following [R] we define the `ancestor' of 
$g_{i}({\cal C}) \in D_{\cal C}^c$, called $B(g_{i}({\cal C}))$, as 
the smallest  subgraph in $D_{\cal C}^c$ containing $g_{i}({\cal C})$:
\be
B(g_i({\cal C})):= g_{i'}({\cal C}), \quad i'= \max_{ g_{i''}({\cal C}) 
\in D_{\cal C}^c, \ g_{i}({\cal C})\subseteq  g_{i''}({\cal C})} i''.  
\ee
With all these bounds and definitions, the sum (\ref{sviluppo4}) becomes:
\bqa
&&|\Ga_{2p}^{\La\Lazero}(\phi_1,..,\phi_{2p})|\leq 
\sum_{n,n',n''=0}^\infty N^{1-p}\; (cK)^{\bar n}\; \frac{1}{n!n'!n''!}
\sum_{o-\tree}\sum_{E,{\cal C} }  \label{sviluppo5}\\                       
&&\int  
d^2x_1...d^2x_{\bar n}\;
\int_{0\le w_{1} \le ...\le w_{\bar n-1}\le 1}\prod_{q=1}^{\bar n-1} 
dw_q \; \no\\
&& \left | \prod_{g\in D^c_{\cal C}} R^*_{g} \;  
\biggr[\prod_{q=1}^{\bar n-1}  D_{\La}^{\La_{0},  w_q}
(\bar x_{l_{q}}, x_{l_{q}}) \;\det {\cal M'}({\cal C})\; 
\phi_1(x_{i_{1}})...\phi_{2p}(x_{j_{p}})\biggr]\right | \no
\eqa
Before performing any bound we must study the action of 
 $\prod_{g\in D_{\cal C}^c} R^*_{g}$ on the tree propagators, the loop 
determinant and the external test functions. As the external half-lines for any
subgraph cannot be of type $C'$ we will write $C$ instead of $D$ in the
formulas. We distinguish two situations.
\paragraph{1)} If $|eg({\cal C})|=4$  then 
$\om(g_i({\cal C}))=0$ and the action of $R^*_g$ is:
\bqa
\lefteqn{R^*_g(x_1) \prod_{i=1}^4 C(x_i,y_i) :=
   \sum_{i=2}^4 R^0_{gi}(x_1) [\prod_{i=1}^4 C(x_i,y_i)] }
\label{process}\\
&& = C(x_1,y_1)\left [
\sum_{i=2}^4 \prod_{2\leq j<i} C(x_j,y_j) [C(x_i,y_i)-C(x_1,y_i)] 
\prod_{i<j\leq 4} C(x_1,y_j)\right ]\no\\
&& = C(x_1,y_1)\left [
\sum_{i=2}^4 \prod_{2\leq j<i} C(x_j,y_j) [\de_0C(x_i,x_1,y_i)] 
\prod_{i<j\leq 4} C(x_1,y_j)\right ]\no
\eqa
where we took as reference vertex $x_1$ and we defined $R^0_{gi}$ as
the operator that moves the
external line with $i$ on the reference vertex $x_1$, 
and applies a difference $\de_0C(x_i,x_1,y_i)$ between
two covariances on the line $i$. 
 
\paragraph{2)} If  $|eg_i({\cal C})|=2$ then  
$\om(g_i({\cal C}))=-1$ and the action of $R_g^*$  is:
\bqa
\lefteqn{R^*_g(x_1) C(x_1,y_1) C(x_2,y_2):=
R^1_g(x_1) C(x_1,y_1) C(x_2,y_2) }\no\\
&=& C(x_1,y_1)\left [ C(x_2,y_2)-C(x_1,y_2) - 
(x_2-x_1)^\mu \p{x_1^\mu} C(x_1,y_2) \right ]\no\\
&=& C(x_1,y_1)\left [\de_1C(x_2,x_1,y_2)\right ]\label{process1}
\eqa
where we took as reference vertex $x_1$.

\subsubsection{Choice of the reference vertex.}
Now, for each $g_i\in D^{c}_{\cal C}$ we call the reference vertex 
$v_e(g_i({\cal C}))$.  
In this paper the choice of this vertex is adapted to the tree
$\tree$, and is different from previous rules such as  [R], chap.II.
We adopt the following rule. We call the first external vertex of the graph,
the one with position $x_{i_1}$
the root of the tree. We define
$D_{\cal C}^{0c}$  and $D_{\cal C}^{1c}$ as the subsets of 
four-point and two-point subgraphs
in  $D_{\cal C}^c$.

For every subgraph $g \in D^{0c}_{\cal C}$ and any vertex of $g$
there is a single path in $\tree$ joining this vertex to the root. 
This path must contain a single well defined external line
of $g$. The vertex to which this external line hooks is by definition our
reference vertex for $g$ (see Figure 14).

\medskip
\centerline{\hbox{\psfig{figure=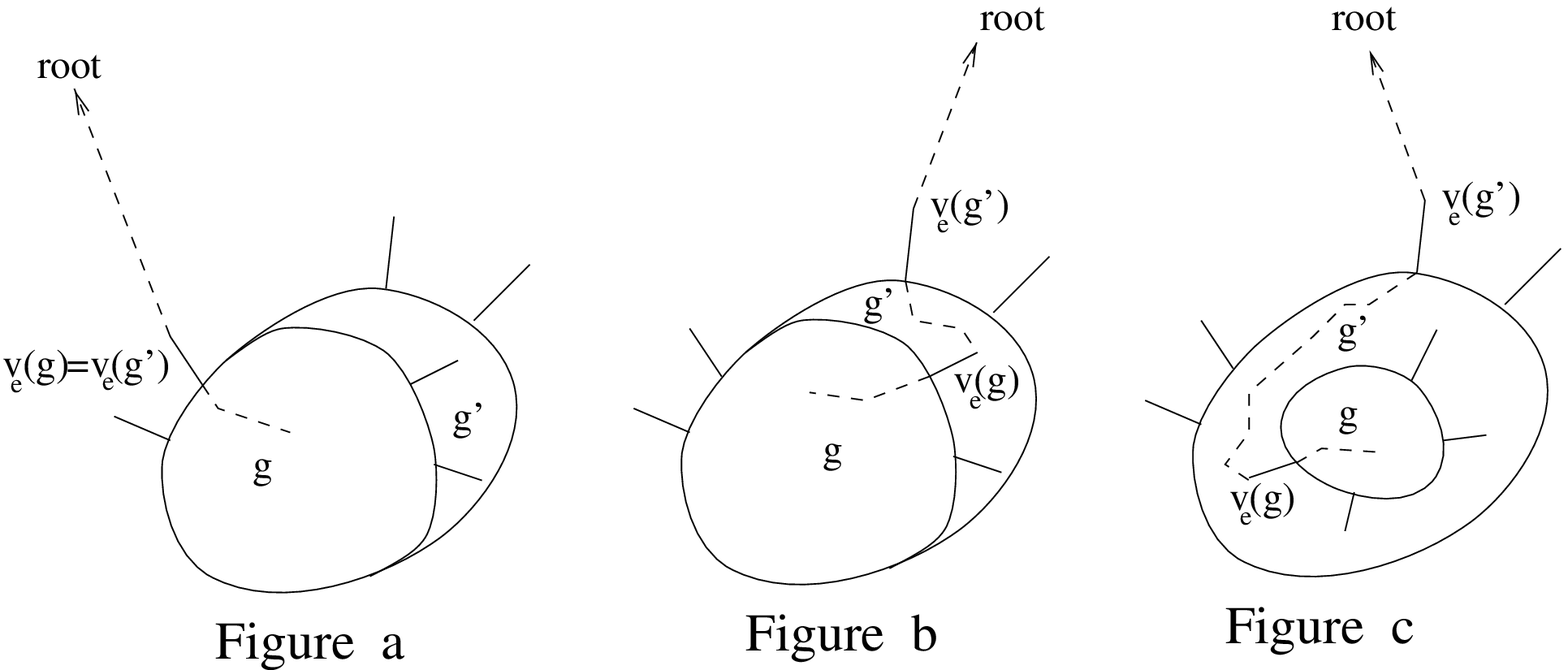,height=5cm,width=11cm}}}
\medskip
\centerline{Figure 14}
\medskip

This rule leaves us free of choosing in a different way 
the reference vertex for any two-point D1PI 
subgraph $ D^{1c}_{\cal C} $. The rule must ensure that  open subgraphs 
 and  D1PR subgraphs are automatically  
renormalized by renormalization of their closure or proper parts. 
We decide to take as border vertex of any 
subgraph $g \in D^{1c}_{\cal C} $ the one to which the highest of the two
external half-lines of $g$ hooks. Remark that this external half-line
is always a tree half-line in $\tree$, so we know its scale.
This rule, shown in Figure 15, 
fulfills the desired requirement, as will be shown below.

\medskip 
\medskip
\centerline{\hbox{\psfig{figure=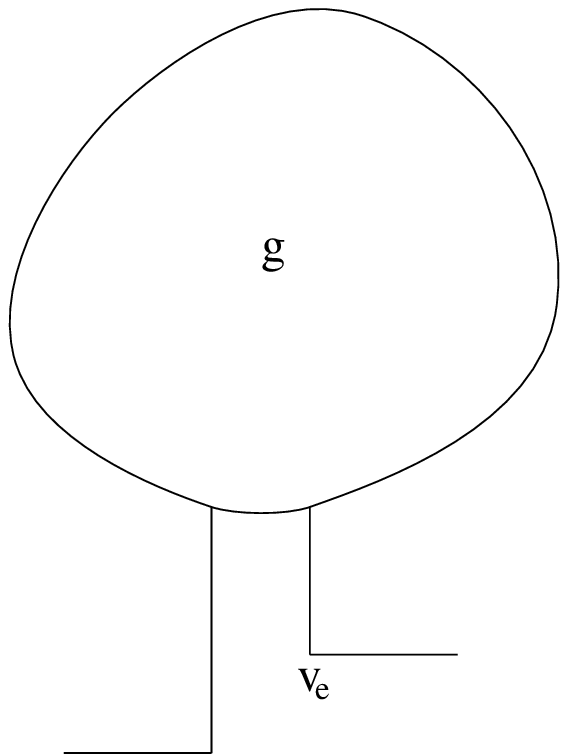,height=3cm,width=3cm}}}
\medskip
\centerline{Figure 15}
\medskip

Finally we add a rule which is not strictly speaking necessary but
simplifies the discussion: to compute the action
of the renormalization operator we perform first all operations corresponding
to two point subgraphs, then all operations  corresponding
to four point
subgraphs, starting from the smallest graphs towards the largest.
This rule ensures  that any external half-line of a subgraph $g$  
bearing one or two gradients because of the action of the Taylor
operator for $g$, cannot bear additional gradients from
the later action of another Taylor
operator for a different subgraph $g'$.

\subsubsection{Processes}

Returning to equations (\ref{process}) and (\ref{process1}), we start the
renormalization for the two point subgraphs
from the leaves of $R_\tree$ (hence from the smallest
subgraphs at highest energy) and go down. Then we perform the
renormalization of four point subgraphs. Some of them may be already
convergent due to renormalization of two point subgraphs.
Also, even after fixing the reference vertex for each closed subgraph, 
there remains an arbitrary ordering of the other external lines
for each four-point graph and a sum
over three possible terms as shown in \ref{process}. 
Again some of these terms may themselves renormalize
some lower four point subgraphs, so the outcome of the renormalization
is difficult to capture in a single formula. We index the terms finally
obtained by an index $P$, called the process,
which summarize all these choices made for the four point subgraphs. Hence  
\be
\prod_{g\in D_{\cal C}^c} (R_g) = \biggl(\sum_P 
\prod_{g\in D_{\cal C}^{0c}(P)} 
[R_g(P)] \biggr)\prod_{g\in D_{\cal C}^{1c}} R^1_g
\ee 
where
$D_{\cal C}^{0c}(P)$ is the subset of $D_{\cal C}^{0c}$ made of the subgraphs
for which $R_{g({\cal C})}^*\ne 1$, hence for which there is a non-trivial
renormalization.
\bqa
R_g(P) &=& R^0_{g\, i(P)}, \; i(P)\in \{2,3,4\} 
\quad \hbox{if} \; g\in D_{\cal C}^{0c}(P)
\eqa
Hence in equation (\ref{sviluppo5}), the absolute value inside the 
integrals can be
bounded by:
\bqa
\lefteqn{\left | \prod_{g\in D^c_{\cal C}} R_{g} \;  
\biggr[\prod_{q=1}^{\bar n-1}  D_{\La}^{\La_{0},  w_q}
(\bar x_{l_{q}}, x_{l_{q}}) \;\det {\cal M'}({\cal C})\;
\phi_1(x_{i_{1}})...\phi_{2p}(x_{j_{p}})
\biggr]\right |} 
\label{sviluppo6} \\
&\leq & \sum_{P}
\prod_{q=1}^{\bar n-1}  
|D_{\La}^{r, \La_{0},  w_q}(\bar x_{l_{q}}, x_{l_{q}})|
|\det {\cal M'}^r({\cal C})|\;
|\phi^{r}_1(x_{i_{1}})...\phi^{r}_{2p}(x_{j_{p}})|
\no
\eqa
where we defined 
$D^r$, ${\cal M'}^r$, $\phi^r$ 
as the functions obtained after the application of
$\prod_{g\in D^c_{\cal C}(P)} R_{g}(P)$. Again we bound the sum over
processes $P$ by the supremum 
times the number of possible processes. This number is bounded by $3^{n-1}$.
Indeed we recall that for
any forest ${\cal F}$ of closed four-point subgraphs,
we have $f({\cal F}) \le n-1$, where $f({\cal F})$ is the number of
four-point subgraphs
in ${\cal F}$ [CR, Lemma C1]; this maximal number is not changed
by adding $n'+n''$ two point vertices because of one particle
irreducibility of the closed subgraphs. 

From now on we work therefore with a fixed process $P$.
We introduce some notations. We define  $L^0(P)$ and $L^{1}$ 
as the set of loop half-lines which bear some single or double gradient  
respectively
by some $R^0_{gi}(P)$ or $R^1_g$ operator, $L^{r0}(P)$ as  
the set of loop half-lines moved to the reference vertex by  some  
$R^0_{gi}(P)$ and $L^u(P)$ the  loop 
half-lines left unchanged. 
In the same way we define the sets $T^0(P)$, $T^{1}$, 
 $T^{r0}(P)$ and  $T^u(P)$ for the tree half-lines, 
and  $E^0(P)$, $E^{1}$, 
$E^{r0}(P)$ and  $E^u(P)$  for the external half-lines. 
To avoid confusion, from now on 
we call $f_i$ and ${\bar f}_i$ the two half-lines forming
the tree-line $l_i$.

\subsubsection{Interpolations of the lines}

For a four-point subgraph the difference $\de_0C$ is expressed by 
\bqa
\de_0C(x,x_v,y)&=& \int_0^1 dt \frac{d}{dt} C(x(t),y)  \label{uffa1}
\eqa 
For a two-point subgraph $\de_1C$ is expressed by 
\bqa
\de_1 C(x,x_v,y)&=& 
\int_0^1 dt (1-t) \frac{d^2}{dt^2} C(x(t),y)\label{uffa2}
\eqa

This means that the external line hooked to $x$ has been hooked to the
point $x(t)$ on any differentiable path joining $x$ to $x_v$ and has now a 
propagator (see Figure 16)
\medskip
\medskip
 
\centerline{\hbox{\psfig{figure=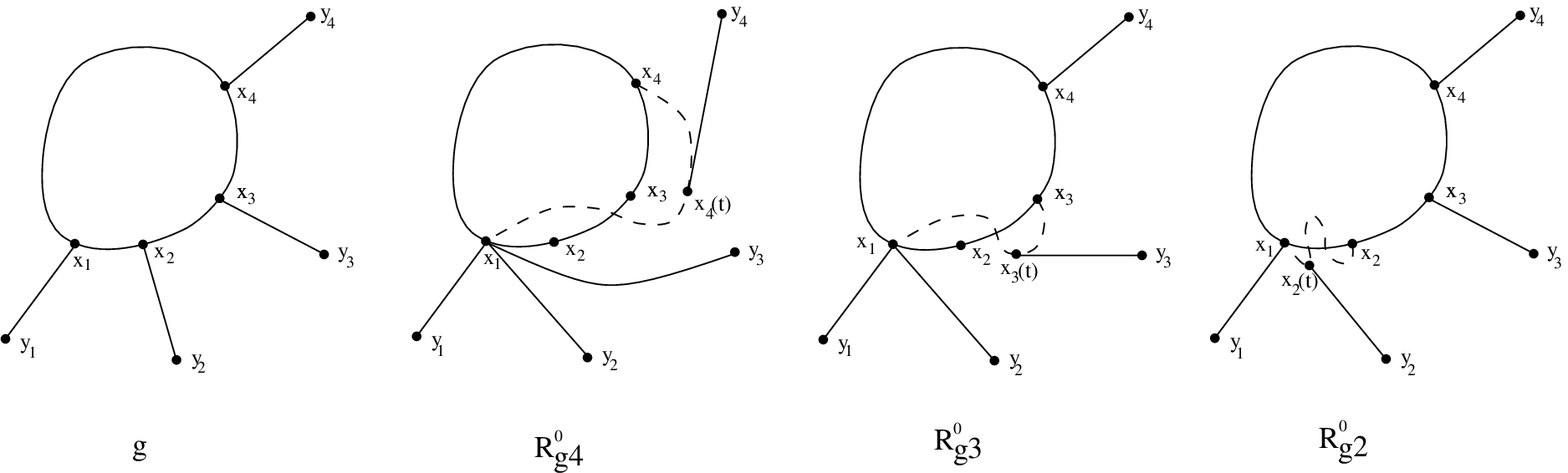,height=3.5cm,width=11cm}}}

\medskip
\centerline{Figure 16}
\medskip

\be
C^0(x(t),y) := \frac{d}{dt} C(x(t),y) 
\ee
or (see Figure 17)
\be
 C^1(x(t),y):= (1-t)  \frac{d^2}{dt^2} C(x(t),y).
\ee

\medskip
\medskip
\centerline{\hbox{\psfig{figure=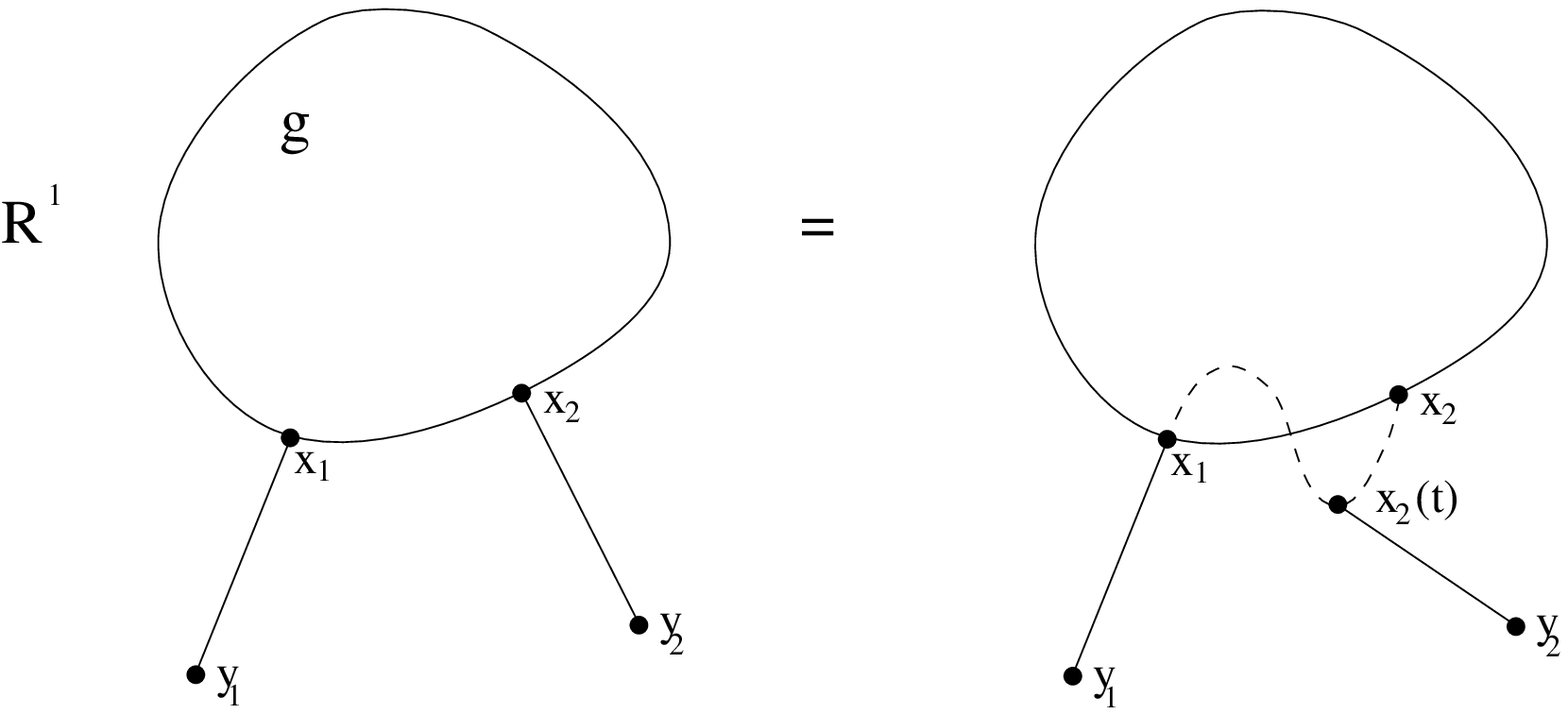,height=3.5cm,width=9cm}}}

\medskip\centerline{Figure 17}
\medskip

In previous perturbative or constructive works, 
this path $x(t)$ is always defined to be the linear
segment connecting $x$ to $x_v$ hence is parametrized by 
\be
x(t) := x_{v} +t (x-x_{v}) \quad x(0)=x_v\;\; x(1)=x \label{uffa3}
\ee
But with the continuous band structure this 
obvious choice when applied to tree half-lines
leads to difficulties. 
It is therefore more convenient to treat differently the
loop, tree and external half-lines. Loop lines and external half-lines
(except the root) do not affect spatial
integration (recall that this spatial integration is performed using the decay
of the tree lines).
So for them we can choose the  obvious linear interpolation that makes
easier to 
factorize the matrix elements of ${\cal M}'$ 
as scalar products and to apply Gram's inequality.
For 
the tree lines it will be convenient to exploit
the existence of $\tree$ to choose a different path. 

\paragraph{IV.3.2.A: Loop lines} Now for each $h_a\in L^0(P)\cup L^1$ 
we define
\be
x_a(t) :=  x_{v_e} +t (x_a-x_{v_e})
\ee
which is the obvious linear path. 

The  propagator for the line bearing the difference becomes  
respectively for a four-point
and a two-point subgraph
\be
C^0(x_a(t),y) := \frac{d}{dt} C(x_a(t),y) =  (x_{a}- x_{v})^\mu
\p{x^\mu } C(x_a(t),y)
\label{moved3}\ee
\be
 C^1(x_a(t),y):= (1-t)  \frac{d^2}{dt^2} C(x_a(t),y)=
 (x_{a}- x_{v})^\mu (x_{a}- x_{v})^\nu  \p{x^\mu }\p{x^\nu } C(x_a(t),y)  .
\label{moved4}\ee

If the second end of the line $y$ 
is also moved we just apply the same
formulas to $C^0(x_a(t),y)$ and   $C^1(x_a(t),y)$,  interpolating $y$. 
Introducing these formulas into the matrix element ${\cal M'}^r_{fg}$  
we can factorize it as the scalar product of $F^r_f$ and $G^r_g$ where
\bqa
&& F^r_f= F_f \quad\hbox{for}\quad h_f \not\in  L^0(P)\cup L^1\\
&& F^r_f= \int dt (x_f-x_v)^\mu \p{x^\mu}F_f(x_f(t))  
\qquad\hbox{for}\qquad h_f \in L^0(P)\no\\
&& F^r_f= \int dt (1-t)  
(x_f-x_v)^\mu (x_f-x_v)^\nu \p{x^\nu}\p{x^\mu}F_f(x_f(t))\no   
\; \; \hbox{for}\;  h_f\in L^1 
\eqa
where  $F_f, G_g$ are defined in 
(\ref{FG}). 
The same definitions hold for $G^r_g$.
With these definitions the determinant is bounded by
\be
|\det {\cal M'}^r({\cal C})| \leq  \prod_{f} ||F^r_f|| \ \ 
 \prod_{g} ||G^r_g|| 
\ee
Now we can bound the norms using the following lemma.

\begin{lemma}
The norms of $F_f$ and $G_g$  satisfy the bounds
\bqa
||F_f||_{\cal C}^2& \leq & 
  K \ [\Lazero({w}_{M(f,{\cal C})})
-\Lazero({w}_{m(f,{\cal C})-1})]\no\\
||G_g||_{\cal C}^2& \leq & 
  K \  [\Lazero({w}_{M(g,{\cal C})})
-\Lazero({w}_{m(g,{\cal C})-1})]
\eqa
\end{lemma}
\paragraph{Proof} 
Applying the definition 
(\ref{new-cutoff}),  $||F_f||^2_{\cal C}$ is written
\bqa
||F_f||_{\cal C}^2& =& \int \frac{d^2p}{(2\pi)^2}\; \frac{1}{(p^2+m^2)^{\frac{1}{2}}} 
\left [\eta\lp \frac{p^2+m^2}{\Lazero^2({w}_{M(f,{\cal C})})} \rp
-\eta\lp\frac{p^2+m^2}{\Lazero^2({w}_{m(f,{\cal C})-1})} \rp \right ]\no\\
 &= & \int \frac{d^2p}{(2\pi)^2}\; (p^2+m^2)^{\frac{1}{2}} 
\int_{\Lazeroinv({w}_{M(f,{\cal C})})}^{\Lazeroinv
({w}_{m(f,{\cal C})-1})} d\al 
\lp -\eta'[(p^2+m^2)\;\al]\rp \no \\
&\leq  & 
\int_{\Lazeroinv({w}_{M(f,{\cal C})})}^{\Lazeroinv(
{w}_{m(f,{\cal C})-1})} d\al \;
\pi \;\al^{-\frac{3}{2}} \int_{0}^\infty dv \sqrt{v} [-\eta'(v)] \no\\
&\le & 
K\; [\Lazero({w}_{M(f,{\cal C})})-\Lazero({w}_{m(f,{\cal C})-1})] 
\eqa
where, in the third line, we performed the change of variable 
$v= \al (p^2+m^2)$. The same result holds for $||G_g||^2_{\cal C}$.
\hfill $\Box$
\medskip

A similar argument can be performed for loop lines  
with some gradient applied. Each derivative adds a 
factor $\al^{-\frac{1}{2}}$ in the integral.
With these definitions the determinant is bounded by
\bqa
&&\prod_{a\in L^u(P)\cup L^{r0}(P)}  [\Lazero({w}_{M(a,{\cal C})})
-\Lazero({w}_{m(a,{\cal C})-1})]^{\frac{1}{2}} \no\\
&&\prod_{a\in L^0(P)} |x_a-x_{v(a)}| [\Lazero^3({w}_{M(a,{\cal C})})
-\Lazero^3({w}_{m(a,{\cal C})-1})]^{\frac{1}{2}}\no\\
&&\prod_{a\in L^1} |x_a-x_{v(a)}|^2 [\Lazero^5({w}_{M(a,{\cal C})})
-\Lazero^5({w}_{m(a,{\cal C})-1})]^{\frac{1}{2}}  
\eqa

\paragraph{IV.3.2.B: External lines}
The only external line essential in spatial integration is the root 
$y_1$, then we can choose this point as reference vertex for the whole graph
so that it is never interpolated. For the other external lines we take 
again the easiest formula, the linear one.
All gradients generated by moving the external lines 
in fact apply to the test functions. Therefore the 
product is bounded by
\bqa
\prod_{h_{i_e}\in E^{u}(P)\cup E^{r0}(P), i_e\neq 1} ||\phi_{i_e}||_{\infty}
\prod_{h_{i_e}\in E^{0}(P)}  |x_{i_e}-x_{v}| ||\phi'_{i_e}||_{\infty}\no\\
\prod_{h_{i_e}\in E^{1}}  |x_{i_e}-x_{v}|^2 ||\phi''_{i_e}||_{\infty}
\eqa

\paragraph{IV.3.2.C: Tree lines} Now we consider tree lines.

We observe that the set of $f_i,{\bar f}_i\in T^{r0}$ modifies the tree
$\tree$ 
but does not disconnect it in the sense that it simply changes
the hooking vertices of some line. On the other hand, interpolating each 
$f_i, {\bar f}_i\in T^0(P)\cup T^1$ with the linear rule 
(equation (\ref{uffa3}))
in an intuitive sense ``disconnects'' the tree, since the point $x(t)$
in general no longer hooks to some point on a segment corresponding to a tree 
line. This defect would lead to difficulties when integrating over spatial
positions. 
To avoid it we express the differences $\de_0C$ and $\de_1C$
using the connection between external vertices
of any subgraph which is provided by the tree $\tree$ itself. 
But, as the tree $\tree$ is itself
modified by renormalization, 
this process has to be inductive, starting from the smallest 
two point subgraphs
of  $D_{\cal C}^{1c}$ and proceeding towards the biggest, then again from
the smallest four point subgraphs
of  $D_{\cal C}^{0c}(P)$ and proceeding towards the biggest.

Our induction creates progressively
a new tree $\tree(P,J)$. To describe it, we number the
subgraphs to renormalize in the order the operations are performed
as $g_{1},$... $g_{r}$. At the stage $1\le p\le r$, before renormalization
of $g_{p}$, the tree is called 
$\tree(P,J_{p-1})$ (we put $\tree(P,J_{0})=\tree$). If the renormalization
of $g_{p}$ as specified by the process $P$ does not act on a tree half-line
external to $g_{p}$, we neither modify $J_{p-1}$ nor $\tree(P,J_{p-1})$.
If the renormalization
of $g_{p}$ results in some half-tree line $f_i$ or ${\bar f}_i$
belonging to $T^{r0}(P)$, as shown in Figure 18, case 1), we do not
modify $J_{p-1}$, so we put $J_{p}=J_{p-1}$, but we modify $\tree(P,J_{p-1})$,
that is we define $\tree(P,J_{p})$ as $\tree(P,J_{p-1})$ but with the half
line now hooked to the reference vertex of $g_{p}$.

Finally when the renormalization of $g_{p}$ interpolates a tree 
half-line $f_i$ or ${\bar f}_i$, we modify both $J_{p-1}$ and 
$\tree(P,J_{p-1})$. There exists a
unique path ${\cal P}^{\tree(P,J_{p-1})}_{x_f,x_v}$ joining the vertex
$x_{f}$ where the half-line hooked to the fixed vertex of $g_{p}$.
This path is made of $q$ lines and goes through $q+1$ vertices
with positions $x_0=x_v, x_{1},...,x_q=x_f$.
We interpolate the half line using this path instead of the
linear path. This means that we write, if $g_{p}$ is a four point subgraph, 

\bqa
\de_0C(x_{f},x_v,y)& =& \sum_{j=1}^q \de_0C(x_j,x_{j-1},y)\no\\
&=&
 \sum_{j=1}^q   (x_j-x_{j-1})^\mu 
\int_0^1 dt \p{x_j(t)^\mu} C(x_j(t),y)\label{diff0}
\eqa
and if $g_{p}$ is a two point subgraph we write
\bqa
&&\de_1C(x_f,x_v,y) = \de_0C(x_f,x_v,y) -(x_f-x_{v})^\mu
\p{x^\mu_{v}}C(x_{v},y)
\no\\ 
&&= 
 \sum_{j=1}^q  \de_0 C(x_j,x_{j-1},y)
-  \sum_{j=1}^q (x_j-x_{j-1})^\mu 
\left [\p{x_{j-1}^\mu} C(x_{j-1},y)-\right.\no\\
&& - \left.
\left [\sum_{k=1}^{j-1} \p{x_{k}^\mu}C(x_{k},y)
- \p{x_{k-1}^\mu}C(x_{k-1},y)\right ]\right]\no\eqa
\vskip-.4cm
\bqa
&=&  \int_0^1 dt \sum_{j=1}^q   (1-t) 
(x_j-x_{j-1})^\mu   
(x_j-x_{j-1})^\nu
\frac{\partial^2}{\partial^\mu \partial^\nu } C(x_j(t),y)
\no\\
&+& \int_0^1 dt \sum_{j=1}^q \sum_{k=1}^{j-1}    
(x_j-x_{j-1})^\mu   
(x_k-x_{k-1})^\nu
\frac{\partial^2}{\partial^\mu \partial^\nu } C(x_k(t),y)
\label{diff1}
\eqa
where for each $j$ (and $k\le j$) we defined
\be
x_j(t) := x_{j-1} +t (x_j-x_{j-1})\ ,\  x_k(t) := x_{k-1} +t (x_k-x_{k-1})\  .
\ee

Then we update $J$ and $\tree$. In the first case, where $g$ is a four
point subgraph, we define $J_{p}=J_{p-1}\cup \{j\}$, where $j$ is the index of
the line of $\tree(P,J_{p-1})$ chosen in \ref{diff0}. In the second
case, where $g$ is a two
point subgraph, we define $J_{p}=J_{p-1}\cup \{j\}\cup\{k\}$, 
where $j$ and $k$ are the indices of
the lines of $\tree(P,J_{p-1})$ chosen in \ref{diff1}. Finally
we update the tree according to Figure 18, case 2 and 3.

\medskip
\medskip
\centerline{\hbox{\psfig{figure=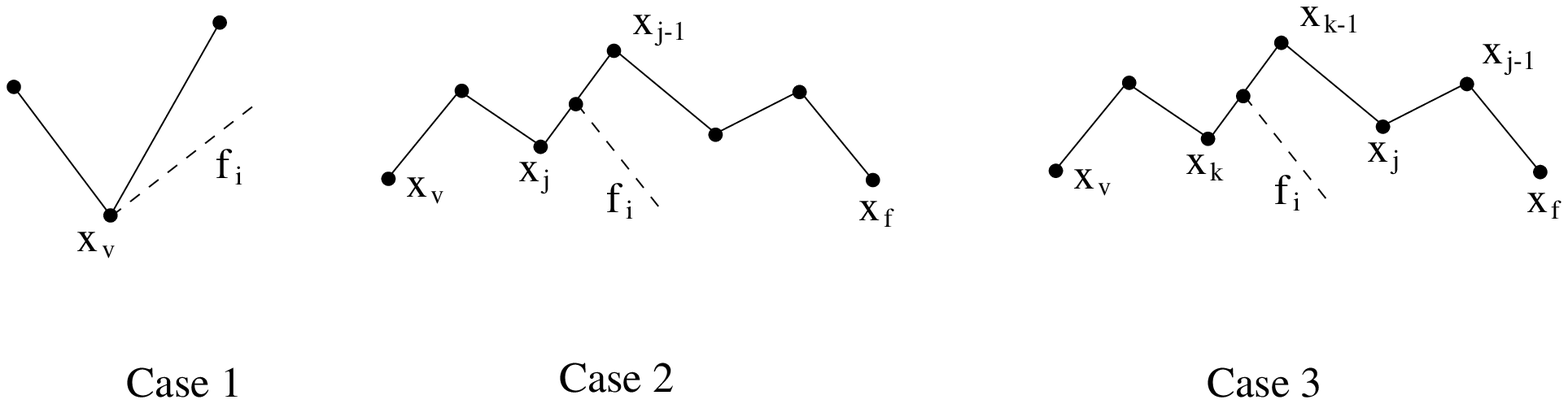,height=3cm,width=11cm}}}
\medskip
\centerline{Figure 18}
\medskip

This means that, for $g_p$ a four point subgraph,
in  $\tree(P,J_{p})$ the external line hooked to $x_f$
is now hooked to the point 
$x_j(t)$ on the  tree segment $[x_{j-1}, x_j]$ 
and has propagator
\be
 C^0(x_j(t),y):=   (x_j-x_{j-1})^\mu  \p{^\mu} C(x_j(t),y).
\label{moved1}\ee
For $g_p$ a two point subgraph, 
the external line previously hooked to $x_f$
is now hooked to the point 
$x_k(t)$ on the tree segment $[x_{k-1}, x_k]$ (with $k\le j$) 
and has propagator
\[
C^1(x_k(t),y):=  (x_j-x_{j-1})^\mu
(x_k-x_{k-1})^\nu
\frac{\partial^2}{\partial^\mu \partial^\nu } C(x_k(t),y)
\quad k\neq j
\]\vskip-.8cm
\be
C^1(x_j(t),y):=(1-t)  (x_j-x_{j-1})^\mu
(x_j-x_{j-1})^\nu
\frac{\partial^2}{\partial^\mu \partial^\nu } C(x_j(t),y)
.\label{moved2}
\ee

Remark that the new tree $\tree(P,J_{p})$
has therefore
one additional vertex and one line more than $\tree(P,J_{p-1})$.
The final tree built inductively in this way,  $\tree(P,J)$ is still a tree
connecting all initial vertices, with at most $n-1$ new vertices
and new lines (as $n-1$
is the maximal number of closed divergent subgraphs).

The treatment of two or four point subgraphs and the rules for
their fixed vertex being different, we write $J=J^0 \cup J^1$,
where $J^0$ is the set of indices $j$ for the interpolations associated to 
the renormalization of four point graphs, and $J^1$ is the set of indices 
$j_{g}$ and $k_{g}$ for the interpolations associated to 
the renormalization of two point graphs.

\subsubsection{Bound on the sum over $J^1$ and on the 
associated distance factors}

Now, before going on, in order to reproduce the
spatial decay between supports of Theorem 3,
we take out  a fraction $(1-\e )$ 
of the exponential decay of each tree line in (\ref{hello1})
and (\ref{hello2}): this factor is bounded by
\be 
\prod_{i=1}^{\bar n -1} e^{-(1-\e )|{\bar x}_i-x_i|\La^m_0(w_i)}
\leq  e^{-\La^m 
(1-\epsilon )d_T(\Om_1,...
\Om_{2p})}  
\ee
and we keep the remaining decay 
$e^{-(\e/2) |{\bar x}_i-x_i|\La^m_0(w_i)}$
(adjusting $\e' =\e/2$)
for two  purposes: a fraction of this decay is used to perform 
spatial integration and the other to bound the sum over $J^0$ and the 
distance factors generated by equations 
(\ref{moved3})-(\ref{moved4}) and (\ref{moved1})-(\ref{moved2}).
Therefore we have to bound, for a fixed process $P$
\be \sum_{J^1} \sum_{J^0}\;\int dx \;
A(x,J,P,\tree)\; B(x,J,P,\tree)
\ee
where
\bqa
&& A(x,J,P,\tree) :=
\prod_{f_i,\bar f_i\in T^{1}}  |x_j-x_{j-1}||x_k-x_{k-1}| 
 \prod_{h_a\in L^{1}}  |x_a-x_{v}|^2 
\no\\ 
&&\prod_{h_{i_e}\in E^{1}}  |x_{i_e}-x_{v}|^2 
\prod_{l\in\tree(P,J) }e^{-(\e/4) 
|{\bar x}_l-x_l|\La^m_0(w_l)}
\label{mod1}
\eqa
\bqa
&& B(x,J,P,\tree) :=
\prod_{f_i,\bar f_i\in T^{0}(P)}  |x_j-x_{j-1}|
\prod_{h_a\in L^{0}(P)}  |x_a-x_{v}|  \no\\ 
&& 
\prod_{h_{i_e}\in E^{0}(P)}  |x_{i_e}-x_{v}| 
\prod_{l\in\tree(P,J) }e^{-(\e/4) 
|{\bar x}_l-x_l|\La^m_0(w_l)}
\label{mod0}
\eqa
where $J$ specifies in particular the distance factors $ |x_j-x_{j-1}| $
and $ |x_k-x_{k-1}| $, as explained above.

The strategy of the bound is to write
\be \sum_{J^1} \sum_{J^0}\int dx \; A(x,J,P,\tree)\; B(x,J,P,\tree)
\le \sum_{J^1} \sum_{J^0} A(J,P,\tree)\int dx\; B(x,J,P,\tree)
\ee
where $ A(J,P,\tree) := \sup_{x} A(x,J,P,\tree) $. 

For each divergent subgraph $g_i\in{\cal D}_{\cal C}^{c}$
we define $t(i)$ 
as the index of the lowest tree line in the path 
of $\tree(P,J)$ joining $x_{v(g_i)}$ to the interpolated
half-line which renormalize it. The next
lemma proves that $ A(J,P,\tree) $ is bounded by something
independent of $J$:

\begin{lemma}
\be
 A(J,P,\tree) \le K^n \prod_{g_i\in{\cal D}_{\cal C}^{c1}}
[\Lazero(w_{t(i)})-\Lazero(w_{{\cal A}(i)})]^{-2}
\ee
where $K$ is some $\e$ dependent constant.
\end{lemma}

\noindent{\bf Proof}
For each loop or external line the difference $|x-x_v|$ can be bounded, 
applying several triangular inequalities, by the sum over the tree lines 
on the unique path in  $\tree(P,J)$ connecting $x$ to $x_v$.

We observe that the same tree line $l_j$ can appear in several paths connecting
different pairs of points $x_v,x$. Using the same
fraction of its exponential decay many times  might generate 
some unwanted factorials since  
$\sup_{x} x^n\exp(-x)= (n/e)^n$.
To avoid this problem we define $D_j$ as 
the set of subgraphs 
$g_i\in {\cal D}_{\cal C}^{1c}\cup{\cal D}_{\cal C}^{0c}(P)$
which use
the tree distance $|\bar x_{l_j}-x_{l_j}|$ to bound the norm 
of $|x  -x_{v(g_{i})}|$ or its 
square and we apply the relation
\bqa
 e^{-\frac{\e}{4}|\bar x_{l_j}-x_{l_j}|\Lazero^m(w_j)}
&\leq &  
e^{-\frac{\e}{4}|\bar x_{l_j}-x_{l_j}|\sum_{g_i\in D_j}[\Lazero(w_{t(i)})-
\Lazero(w_{{\cal A}(i)})]} \label{morceaux}
\eqa
With this expression a different decay factor
is used for each subgraph. 
Applying  this result  and the inequalities
$x e^{-x}\leq 1 $, $ x^2 e^{-x}\leq 1$ completes the proof of the lemma.
\hfill${\Box}$
\medskip

It is proved 
in the next section that the sum and spatial integral\\
 $\sum_{J^0}\int dx\;B(x,J,P,\tree)$ is in fact independent of 
$P$ and  $J^1$ (and of the interpolation parameters $t$ that we omitted).
Therefore the sum over $J^1$  will simply lead to the bound of
the next section multiplied by the cardinal of the set 
$J^1$, that is by $|J^1|$. This is done thanks
to the following lemma:
 
\begin{lemma}
We have $|J^1 |\le e^{2\bar n}$.
\end{lemma}

\noindent{\bf Proof} We consider the graphs $g_{1},... g_{r_{1}}$
of $ D_{\cal C}^{1c}$
in the order used for their renormalization in subsection IV.3.2.C.
We define, for each such two point subgraph $g\in D_{\cal C}^{1c}$
the set $ {\cal A}(g)$ of  maximal subgraphs $g'$, $g'\in D_{\cal C}^{1c}$, 
$g'\subset g$, and
the reduced graph $g/D_{\cal C}^{1c}$  where each
$g'\in {\cal A}(g)$ has been reduced to a single point.

We also count the number $L_{g}$ of lines on the unique path in the tree
$\tree \cap g/D_{\cal C}^{1c}$ which joins the two external vertices of $g$.
Remark that this number $L_{g}$ is independent of $J$ and that
$\sum_{g\in D_{\cal C}^{1c}}L_{g}\le \bar n-1$. Finally we define the subset
$ {\cal A}'(g)$ of  $ {\cal A}(g)$ made of those $g'$ in  $ {\cal A}(g)$
which appear as reduced points on this unique path (see Figure 19).

\medskip
\centerline{\hbox{\psfig{figure=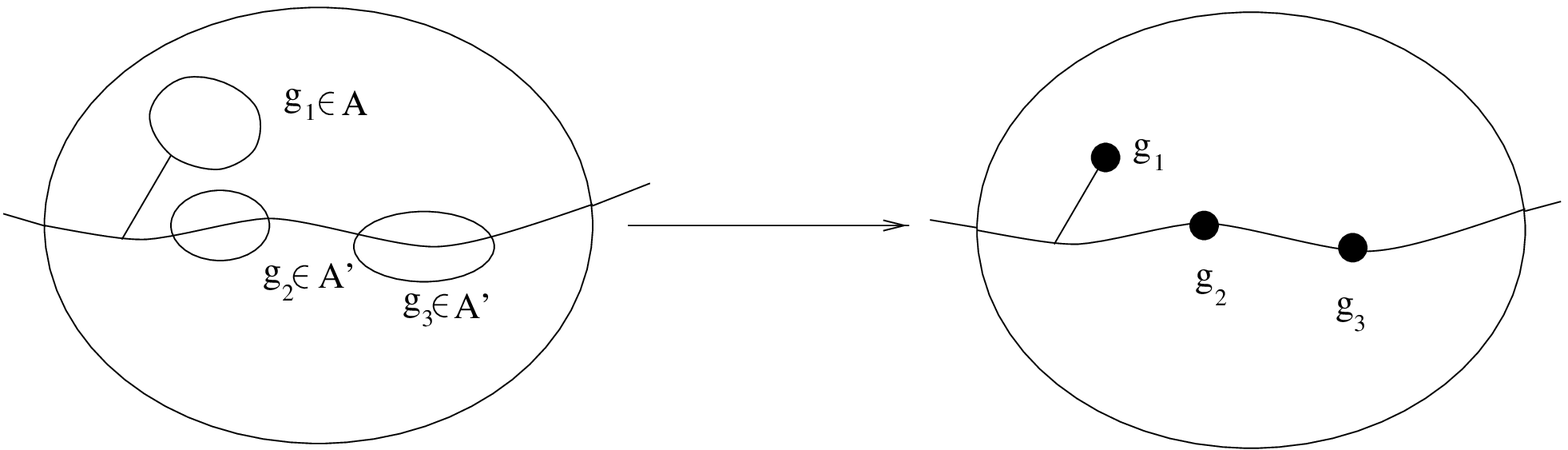,height=3cm,width=11cm}}}
\medskip
\centerline{Figure 19}
\medskip

By induction one can bound the number of choices in $J^1$ by
\be |J^1|\le \prod_{g \in D_{\cal C}^{1c}} 
 \sum_{j_{g}=1}^{L_{g}+ \sum_{g'\in {\cal A}'(g) }k_{g'}}
\sum_{k_{g}=1}^{j_{g}}  1 
\ee 
where this product is again ordered from the smallest to the largest
graphs.  
Now for any positive increasing function $f$ we have $\sum_{j=1}^L f(j)
\le\int_{1}^{1+L} f(x)dx $ so that 
\bqa |J^1|&\le& \prod_{g \in D_{\cal C}^{1c}} 
 \int_{1}^{1+L_{g}+ \sum_{g'\in {\cal A}'(g) }y_{g'}} dx_{g}
\int_{1}^{1+x_{g}}  dy_{g} 
\\
&\le& \prod_{g \in D_{\cal C}^{1c}} 
 \int_{0}^{1+L_{g}+ \sum_{g'\in {\cal A}'(g) }y_{g'}} dx_{g}
\int_{0}^{1+x_{g}}  dy_{g} \le e^{2r_{1}} 
 \prod_{g \in D_{\cal C}^{1c}} e^{L_{g}} \le e^{2\bar n}\no
\eqa
where in the last inequality, we bounded the last integral 
$\int_{0}^{1+x_{g_{r_{1}}}}  dy_{g_{r_{1}}}$ by 
$e^{1+x_{g_{r_{1}}}} $ and then effectuate each integral
exactly and bound each difference $e^x -1$ by $e^x$. Finally we used
the fact that every subgraph $g'$ is in ${\cal A}'(g)$ for at most one $g$,
and the fact that $2r_{1}\le \bar n -1$ (again [CR, Lemma C1]).
\hfill${\Box}$
\medskip

\noindent{\bf Remark}: 
This lemma does not apply to $J^0$. For a counter-example, the reader
can look at the following graph and tree, for which $J^0$ can be of order
$K^n (n/5)!$.

\medskip
\centerline{\hbox{\psfig{figure=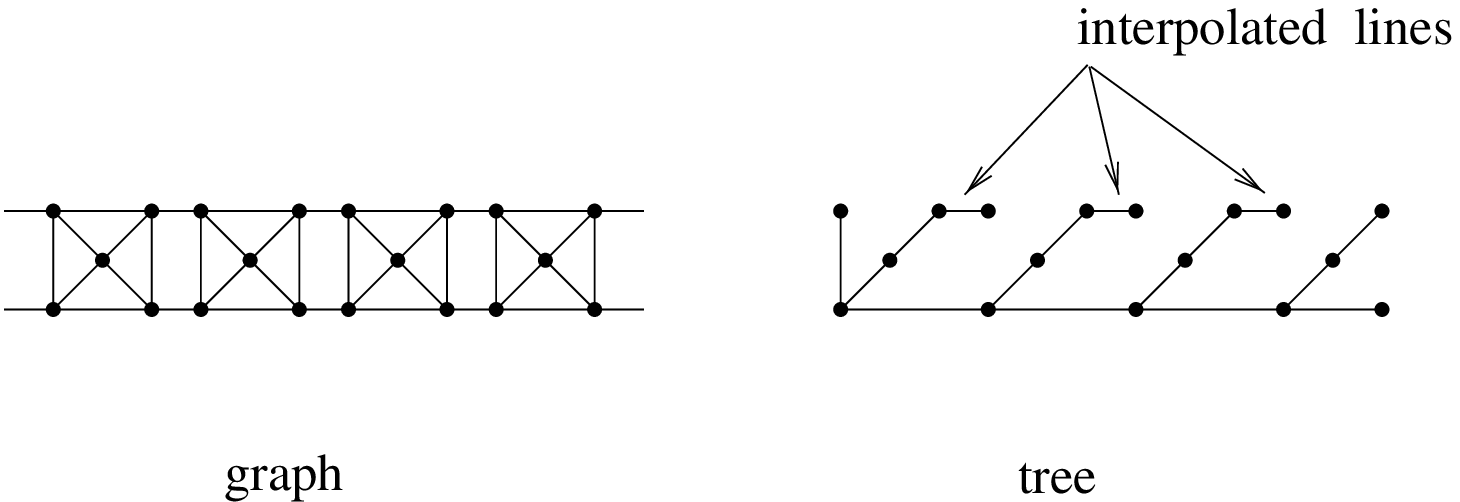,height=3cm,width=11cm}}}
\medskip
\centerline{Figure 20}
\medskip

\subsubsection{Bound on the sum over $J^0$, on the 
associated distance factors, and spatial integration of the vertices} 
It remains now to 
perform the sum over $J^0$ and the integral over the position of internal 
vertices, using the remaining tree decay in $B$ which is
$\prod_{l\in\tree(P,J) }e^{-(\e/4) 
|{\bar x}_l-x_l|\La^m_0(w_l)}$, and to check that the result,
as announced, is independent of $P$, $J$ and of the interpolation parameters
$t$. We recall that
$\La^m_0(w)$ was defined in (\ref{hello3}).

\begin{lemma}
\be
\sum_{J^0}\int dx\; B(x,J,P,\tree) \le K^{\bar n} 
\prod_{g_i\in{\cal D}_{\cal C}^{c0}(P)}
[\Lazero(w_{t(i)})-\Lazero(w_{{\cal A}(i)})]^{-1}
\prod_{q=1}^{\bar n-1} (\Lazero^m(w_q))^{-2}
\ee
where $K$ is some $\e$ dependent constant.
\end{lemma}

\noindent{\bf Proof}
First we 
divide one half of our remaining tree decay as in \ref{morceaux}.
This half will be used to  
bound each distance factor in
$\prod_{h_a\in L^{0}(P)}|x_a-x_{v}|$  
$\prod_{h_{i_e}\in E^{0}(P)}|x_{i_e}-x_{v}|$ as in Lemma 6, 
the sum over $J^0$ and the  distance factors
$\prod_{f_i,\bar f_i\in T^{0}(P)}|x_j-x_{j-1}|$.

As in Lemma 6, each distance factor in
$\prod_{h_a\in L^{0}(P)}|x_a-x_{v}|$  
 $\prod_{h_{i_e}\in E^{0}(P)}|x_{i_e}-x_{v}|$ leads to a bound
$K[\Lazero(w_{t(i)})-\Lazero(w_{{\cal A}(i)})]^{-1}$.

Then we perform the spatial integrals from the leaves of the tree 
$\tree(P,J)$ towards the root $x_{1}$, using the 
other half of the tree decay. In this inductive process
 when we meet an interpolated line
hooked at some interpolated point $x_{j}(t)$ or $x_{k}(t)$
two different
situations can occur, as pictured in Figure 21. The second situation
(interpolated point not towards the root) can occur only
for interpolations of two point subgraphs, hence associated
to the $J^1$ indices.  The first situation
(interpolated point towards the root) must occur 
for all interpolations associated to four point subgraphs
plus possibly some interpolation associated to two point subgraphs.
This is the consequence of our rule for the preferred vertices
of four point subgraphs (the interpolations associated to four point subgraphs
always bring nearer to the root).
 
The sum over $J^0$ is performed in pieces; the sum over each index
$j_{g}$ in $J^0$ is performed right after the spatial integration
which has used the corresponding interpolated line. By the remark above,
each sum over $j_{g}$ in $J^0$ occurs in the first situation of Figure 21.  

\medskip   
\centerline{\hbox{\psfig{figure=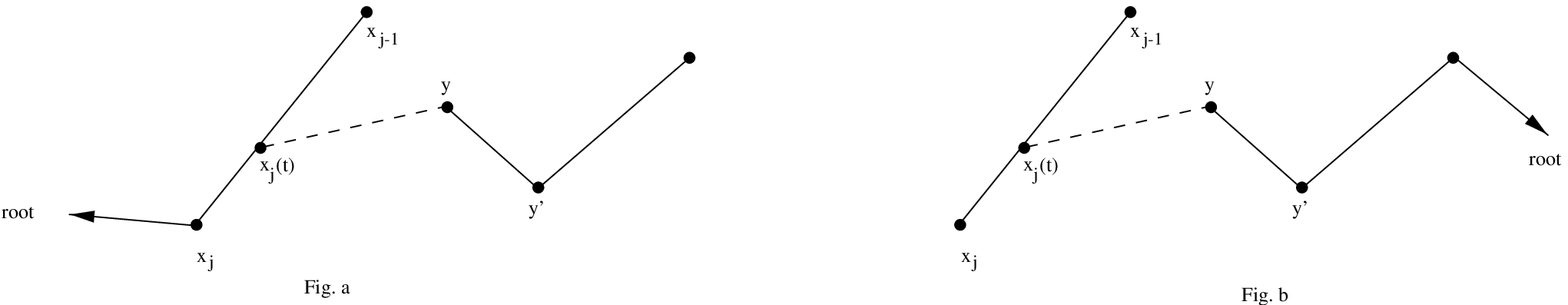,height=4cm,width=12cm}}}

\medskip\centerline{Figure 21}
\medskip

The decay of any tree line $l_i=[x,y]$ with both ends
$f_i,{\bar f}_i\in T^u(P)\cup T^{r0}(P)$ gives,  by  translation
invariance, a factor:
\be
\int e^{-(\e/8)|x-y|\La^m_0(w_i)} d^{2}x = 128\pi\e^{-2}[ \La^m_0(w_i)]^{-2} 
\label{A1}\ee
Surprisingly, the same result holds when one or both ends 
of the line have been 
interpolated, as we explain now.

- In the first situation of Figure 21, $y$, 
the other end of the interpolated line, 
is connected to the root through the interpolated
point (see Figure 21a). We can integrate over
$y$ before integrating over $x_{j}$ and $x_{j-1}$.
We use translation invariance to cancel the 
dependence from $x_{j}(t)$ in the interpolated covariance. 
The integral over the
variable $t$ is bounded by 1. The spatial integral over $y$
then gives the same factor as in (\ref{A1}). Then we perform the 
corresponding sum over $j$ in $J^0$ using
\be
\sum_{j=1}^{q} |x_j-x_{j-1}|
e^{-\frac{\e}{8}[\sum_{j=1}^{q}|x_j-x_{j-1}|][\Lazero(w_{t(i)})-
\Lazero(w_{{\cal A}(i)})]} \le K [\Lazero(w_{t(i)})-
\Lazero(w_{{\cal A}(i)})]^{-1}
\ee

-When the point $x(t)$ is connected to the root through 
 $y$ (see Figure  21b) 
we have to compute the integral 
\be
I= \int_{0}^1 dt
\int d^2x_j\; d^2x_{j-1}\; e^{-(\e/8)|x_j-x_{j-1}|\Lazero^m(w_j)}  
e^{-(\e/8)|x_j(t)-y|\Lazero^m(w_k)}
\ee
where 
$x_j(t)= x_{j-1}+t(x_j-x_{j-1})$.
Performing, for $t\ne 0$ the change of variables  
\be
v_1 = x_j(t) = t x_j  + (1-t)x_{j-1}\quad 
v_2 = \frac{1}{t} (x_{j-1}-x_j(t)) =  (x_{j-1}-x_j)
\ee
the integral becomes
\bqa
I &=& 
 \int_{0}^1 dt \int d^2v_1\; e^{- (\e/8)|v_1|\Lazero^m(w_k)}
\int  d^2v_2 \;  e^{-(\e/8)|v_2|\Lazero^m(w_j)}
\no\\ & =&[ 128\pi\e^{-2}]^2 
[ \La^m_0(w_j)]^{-2}[ \La^m_0(w_k)]^{-2}  
\eqa
This is again the same contribution as (\ref{A1}), hence the same 
contribution as if the line 
had not been interpolated!   

Following the tree from the leaves to the root we can perform 
the integrals over all positions in this way, except for $x_{i_1}$. 
This last point is
integrated using the test function $\phi_1$, which gives a factor 
$||\phi_1||_1$. 
Hence, the result of all spatial integrations is 
$K^n\prod_{q=1}^{\bar n-1} (\Lazero^{m}(w_q))^{-2}$
and the sum over $J^0$ has been performed at a cost of $K^n$
(although $|J^0|$ can be very large).
This completes the proof of Lemma 8.
\hfill${\Box}$\medskip

Since the result of Lemma 8 is independent of $J^1$, as announced, we can
apply Lemma 7.
Let us collect the result of Lemmas 6-8, together with the other
factors remaining after spatial integration.
The product over tree lines propagators, 
is bounded by
\bqa 
&& \quad K^{\bar n}(\Lainv-\Lazeroinv)^{\bar n-1}\;\; 
e^{-(\e/4)m^2\Lazero^{-2}(w_1) }\cdot  \no\\
&&\cdot  \prod_{q=1}^{\bar n-1} \Lazero^{3}(w_q) \ 
\prod_{f_q,\bar f_q\in T'\cup T^{0}(P)}  \Lazero(w_q)\;
\prod_{f_q,\bar f_q\in T^1} \Lazero^2(w_q) 
\label{infrared0}\eqa
where $K$ is some $\e$ dependent constant,and we included the
scaling prefactors in (\ref{hello1})-(\ref{hello2}). These factors
are

- a factor $\La_0(w)^3$ for each tree line,

- a factor $\La_0(w)$ for each half-line in $T^0(P)$,

- a factor $\La_0(w)^2$ for each half-line in $T^1$,

- a factor  $\La_0(w)$ for each half-line in $T'$,
the set of half-lines hooked to a $\de\ze$ vertex
which bears a derivative, hence has covariance
$C'$.

Remark that we kept for only one line, the lowest one, the massive
decay $ e^{-(\e/4)m^2\Lazero^{-2}(w_1)}$ from bounds 
(\ref{hello1})-(\ref{hello2}).
It will be useful only to conclude
the bound in the massive case
when $\La < m$. In this case the $n'$ vertices of type $\de m$ 
may create some infrared
difficulties if we were to replace directly for them 
the factor $ (\Lazero^m(w_q))^{-2}$ by $ (\Lazero(w_q))^{-2}$.
We introduce the set $\tree '$ of the tree lines used for integration
of the $\de m$ vertices. There are $n'$ of them (or $n'-1$ if the root
is of type $\de m$, a case we will exclude for simplicity).
Recall that by (\ref{hello3}) we have 
\be (\Lazero^m(w_q))^{-2} \le (\Lazero(w_q))^{-2}
\label{infrared1} 
\ee
\be (\Lazero^m(w_q))^{-2} \le m^{-1}(\Lazero(w_q))^{-1} 
\label{infrared2}\ee  
We use the bound (\ref{infrared2}) only for the lines of $\tree '$ 
when $\La < m$. For all other cases we use the bound (\ref{infrared1}).

\subsubsection{Integration over the parameters $w_i$}
Now, putting everything together, we describe first
the bound when $\La > m$, hence  $\La^m =\La $. Equation (\ref{sviluppo5}) 
is bounded by
\bqa
\lefteqn{|\Ga_{2p}^\Lambda(\phi_1,...\phi_{2p})| 
\leq 
||\phi_1||_1 \prod_{i=2}^{2p} 
||\phi_i||_{\infty,2} } \no\\
&&  N^{1-p} e^{-(1-\epsilon )\La^m 
d_T(\Om_1,...\Om_{2p})}
(\Lainv-\Lazeroinv)^{{\bar n}-1} \no\\
&& \sum_{n,n',n''=0}^\infty  (cK)^{\bar n} \frac{1}{n!n'!n''!}
 \sum_{o-\tree}\sum_{E,{\cal C}}  
\int_{0\le  w_{1} \le ...\le w_{{\bar n}-1}\le 1}\prod_{q=1}^{\bar n-1} dw_q
\no\\                       
&&\;\; \prod_{q=1}^{\bar n-1} \Lazero(w_q)  \;\;
\prod_{f_q,\bar f_q\in T'\cup T^{0}(P)}  \Lazero(w_q)\;\;
\prod_{f_q,\bar f_q\in T^1} \Lazero^2(w_q) \no\\
&&\;\; 
\prod_{a\in L^u(P)\cup L^{r0}(P)}  \Lazero^{\frac{1}{2}}(w_{M(a,{\cal C})})
\left [1-{\Lazero(w_{m(a,{\cal C})-1}) \over 
\Lazero(w_{M(a,{\cal C})}) }\right ]^{\frac{1}{2}}\no\\
&&\;\; \prod_{a\in L^0(P)}   \Lazero^{\frac{3}{2}}(w_{M(a,{\cal C})})
\left [1 - {\La_0^3(w_{m(a,{\cal C})-1}) \over
\Lazero^3(w_{M(a,{\cal C}) }) }\right ]^{\frac{1}{2}}\no\\
&&\;\; \prod_{a\in L^1}    \Lazero^{\frac{5}{2}}(w_{M(a,{\cal C})})
\left [1- { \Lazero^5(w_{m(a,{\cal C})-1}) 
\over \Lazero^5(w_{M(a,{\cal C}) }) 
}\right ]^{\frac{1}{2}} \no\\
&&\;\; \prod_{g_i\in {\cal D}_{\cal C}^{0c}} \Lazero^{-1}(w_{t(i)})
\left [ 1-{\Lazero(w_{{\cal A}(i)}) \over 
\Lazero(w_{t(i)})}\right ]^{-1}  \no\\
&&\;\;\prod_{g_i\in {\cal D}_{\cal C}^{1c}} \Lazero^{-2}(w_{t(i)})
\left [ 1-{\Lazero(w_{{\cal A}(i)}) \over 
\Lazero(w_{t(i)})}\right ]^{-2}  \label{sviluppo7}
\eqa
The differences $\left [1-\Lazero(w_{{\cal A}(i)})/ 
\Lazero(w_{t(i)})\right ]$ are dangerous 
as they appear with a negative exponent. They are the price to pay for
implementing continuous renormalization group. Indeed, in this continuous
formalism one has to perform renormalization even when the differences
between internal and external energies of subgraphs are arbitrarily small. 
However, there is a natural solution to this problem: 
each subgraph to renormalize has necessarily loop lines and these
 loop lines, when evaluated in the continuous formalism 
by Gram's inequality, give small factors precisely when the differences
between internal and external energies of subgraphs become arbitrarily small.
   
In other words, we can cancel the 
dangerous differences with a negative exponent
against the analogous differences 
with a positive exponent given by the loop lines.
This is the purpose of the next lemma.

\begin{lemma}
If $g_{i}\in{\cal D}_{\cal C}^{c0}(P) $  there is at least one loop line 
internal to $g_{i}$ which satisfies  $\Lazero(w_{M(a,{\cal C})})\leq 
\Lazero(w_{t(i)})$ and  $\Lazero(w_{m(a,{\cal C})-1}) \geq 
\Lazero(w_{{\cal A}(i)})$.
If $g_{i}\in{\cal D}_{\cal C}^{c1} $  there are at least two loop lines
internal to $g_{i}$ and which satisfy  $\Lazero(w_{M(a,{\cal C})})\leq 
\Lazero(w_{t(i)})$ and  $\Lazero(w_{m(a,{\cal C})-1}) \geq 
\Lazero(w_{{\cal A}(i)})$.
\end{lemma}

Assuming the lemma true, and using the relations
$
\sqrt{\frac{1-x^3}{1-x}}\leq \sqrt{3}$ and 
$ \sqrt{\frac{1-x^5}{1-x}}\leq \sqrt{5}$,
one obtains 
\bqa
&& \prod_{a\in L^u(P)\cup L^{r0}(P)} 
\left [1-{\Lazero(w_{m(a,{\cal C})-1}) \over 
\Lazero(w_{M(a,{\cal C})}) }\right ]^{\frac{1}{2}}  \prod_{a\in L^0(P)}   
\left [1 - {\Lazero^3(w_{m(a,{\cal C})-1}) \over
\Lazero^3(w_{M(a,{\cal C}) }) }\right ]^{\frac{1}{2}}\no\\
&& \prod_{a\in L^1}   
\left [1- { \Lazero^5(w_{m(a,{\cal C})-1}) 
\over \Lazero^5(w_{M(a,{\cal C}) }) 
}\right ]^{\frac{1}{2}}  \prod_{g_i\in {\cal D}_{\cal C}^{0c}}
\left [ 1-{\Lazero(w_{{\cal A}(i)}) \over 
\Lazero(w_{t(i)})}\right ]^{-1}  \;\no\\
&&
\prod_{g_i\in {\cal D}_{\cal C}^{1c}} 
\left [ 1-{\Lazero(w_{{\cal A}(i)}) \over 
\Lazero(w_{t(i)})}\right ]^{-2} \le \  \ 5^{n-1} 
\label{svilup}
\eqa
where we bounded by 1 
the loop lines differences that were not used to compensate 
some  $\left [1-\Lazero(w_{{\cal A}(i)})/ 
\Lazero(w_{t(i)})\right ]^{-1}$ factor.

\medskip
\noindent{\bf Proof of Lemma 9}\ \ 
We observe that the lowest tree line  $l_{t(i)}$ in $\tree_i(J,P)$ joining     
the interpolated line and the reference vertex  is external line
for the two subgraphs of $g_i$, $g_{t(i)1}$ and  $g_{t(i)2}$. One of these
two subgraphs has for external line the reference external line of $g_{i}$
and the other  has for external line the interpolated
line moved by the renormalization
$R^*_{g_{i}}$. But
$g_{t(i)1}$ and $g_{t(i)2}$  must both have at least some 
additional external lines, otherwise $g_{i}$ would be D1PR. By parity 
$g_{t(i)1}$ and $g_{t(i)2}$  must both have at least  {\em two} such
additional external lines.

We distinguish two cases:

- If  $g_{i}\in{\cal D}_{\cal C}^{c0}(P)$,
 since there are at most two 
additional external lines
of $g_{i}$, we find that there must be at least  {\em two}
external half-lines of $g_{t(i)1}\cup g_{t(i)2}$ different from $l_{t(i)}$
which are internal in $g_{i}$.
 If they are both loop 
half-lines we are done. If some of them is a tree half-line, 
the other half is external line for another 
subgraph of $g_i$, $g'$. Repeating the argument for $g'$
(as $|eg|=1$ is forbidden) we finally must find an associated loop half-line
(see Figure 12).

-If  $g_{i}\in{\cal D}_{\cal C}^{c1}$, since there was no 
additional external line
of $g_{i}$, then both 
 $g_{t(i)1}$ and $ g_{t(i)2}$ must have at least  {\em two}
external half-lines different from $l_{t(i)}$
which are internal in $g_{i}$. Either these four half-lines are loop half-lines
 and we are done, or some of them are tree lines, which we follow as above
until we find the corresponding loop half-lines.
\hfill${\Box}$
\medskip

After applying the bound (\ref{svilup}) 
we can take the limit $\Lazero\rightarrow \infty$. Performing the
change of variable   
$u_i=1-w_i$    equation (\ref{sviluppo7}) becomes:
\bqa
\lefteqn{|\Ga_{2p}^{\La}(\phi_1,...\phi_{2p})|\leq
 ||\phi_1||_1 \prod_{i=2}^{2p} ||\phi_i||_{\infty,2} e^{-(1-\epsilon )\La^m 
d_T(\Om_1,...\Om_{2p})}}\no\\
&&  N^{1-p}
\sum_{n,n',n''=0}^\infty   \La^{2-p-n'}(cK)^{\bar n} \frac{1}{n!n'!n''!}
\sum_{o-\tree}\sum_{E,{\cal C}} 
\int_{0\le  u_{{\bar n}-1} \le ...\le u_{1}\le 1}\prod_{q=1}^{\bar n-1} du_q  
\no\\                       
&&\left[ \prod_{q=1}^{\bar n-1} 
\frac{1}{\sqrt{u_q}}\right ]\left[
\prod_{f_q,\bar f_q\in T'\cup T^{0}(P)} \frac{1}{\sqrt{u_q}}\right]
\left[ \prod_{f_q,\bar f_q\in T^1}  
\frac{1}{u_q}\right]\no\\
&& \left[\prod_{a\in L^u(P)\cup L^{r0}(P)}  
(u_{M(a,{\cal C})})^{-\frac{1}{4}}\right]\left[
 \prod_{a\in L^0(P)}  (u_{M(a,{\cal C})})^{-\frac{3}{4}}\right] 
\left[ \prod_{a\in L^1}    (u_{M(a,{\cal C})})^{-\frac{5}{4}}\right] \no\\
&&\left[\prod_{g_i\in {\cal D}_{\cal C}^{0c}} (u_{t(i)})^{\frac{1}{2}}\right]
\left[\prod_{g_i\in {\cal D}_{\cal C}^{1c}} (u_{t(i)})\right]
\label{sviluppo8}
\eqa
To factorize the integrals we perform the  change of variable:
\be
u_i= \bt_i u_{i-1} \qquad \bt_i\in [0,1]
\ee
where by convention $u_{0}=1$.
The Jacobian of this transformation is the determinant of a
triangular  matrix hence it is given by:
\be
J=\bt_{1}(\bt_{1}\bt_{2})...(\bt_{1}\bt_{2}...\bt_{\bar n-2})=
\prod_{i=1}^{\bar n-1} \bt_i^{\bar n-1-i}.
\ee
We absorb $\La^{-n'}$ into the term $K^{\bar n}$
since we recall that $\La>m$ hence that $\La^{-n'}=(\La^m)^{-n'} $,
and that in Theorem 3
$\La^m$ remains in the compact $X$, hence is bounded away from 0.
Then the integral (\ref{sviluppo8}) becomes 
\bqa
\lefteqn{|\Ga_{2p}^\Lambda(\phi_1,...\phi_{2p})|\leq  
||\phi_1||_1 \prod_{i=2}^{2p} ||\phi_i||_{\infty,2} 
 e^{-(1-\epsilon )\La^m d_T(\Om_1,...\Om_{2p})}}\label{sviluppo9}\\
&& (\La^m)^{2-p}\; N^{1-p} \;
\sum_{n,n',n''=0}^\infty  (cK)^{\bar n} \frac{1}{n!n'!n''!}
\sum_{o-\tree}\sum_{E,{\cal C}}  \int_0^1 ...
\int_0^1 \prod_{i=1}^{\bar n-1}  d\bt_i 
\no\\                       
&& \prod_{i=1}^{\bar n-1}   
 \bt_i^{-1+\frac{1}{2}(\bar n-i)-\frac{1}{2}|N''_i|}
\left[
\prod_{f_q,\bar f_q\in T^{0}(P)} \frac{1}{\sqrt{\bt_{q}...\bt_1}}\right]
\left[\prod_{f_q,\bar f_q\in T^1}  
\frac{1}{\bt_q...\bt_1}\right]\no\\
&&\left[ \prod_{a\in L^u(P)\cup L^{r0}(P)}  
(\bt_{M(a,{\cal C})}...\bt_1)^{-\frac{1}{4}}\right]
 \left[\prod_{a\in L^0(P)}  (\bt_{M(a,{\cal C})}...\bt_1)^{-\frac{3}{4}}) 
\right] \no\\
&&
\left[ \prod_{a\in L^1} (\bt_{M(a,{\cal C})}...\bt_1)^{-\frac{5}{4}}\right]
\left[\prod_{g_i\in {\cal D}_{\cal C}^{0c}} 
(\bt_{t(i)}..\bt_1)^{\frac{1}{2}}\right]
\left[\prod_{g_i\in {\cal D}_{\cal C}^{1c}} 
(\bt_{t(i)}...\bt_1)\right]\no
\eqa
Each $\bt_i$ appears with the exponent $-1 + x_i$.
\bqa
x_i& =& 
\frac{1}{2}[{\bar n}-i]-\frac{1}{2}|N''_i|- \frac{1}{4} |IL_i({\cal C})|
\no\\
&&+\frac{1}{2}[|S_i^0({\cal C})|-|IT^0_i|- |IL^0_i({\cal C})|] +
[|S_i^{1}({\cal C})|
- |IT^1_i|-  |IL^1_i({\cal C})|]  \no\\
\eqa
where we defined
\bqa
IT^0_i &:=& \{f_j, {\bar f}_j\in T^0(P) | j\geq i \}\no\\
IT^1_i &:=&  \{f_j, {\bar f}_j\in T^1 | j\geq i \}\no\\
IL_i({\cal C})& :=& \{a\in L| M(a,{\cal C})\geq i\}\no\\
IL_{i}^0({\cal C})& :=& \{a\in L^0(P)| M(a,{\cal C})\geq i\}\no\\
IL_{i}^1({\cal C})& :=& \{a\in L^1| M(a,{\cal C})\geq i\}\no\\
S_i^0({\cal C}) &:=& \{ g_j\in{\cal D}^{0c}_{\cal C}(P)| t(j)\geq i\}\no\\
S_i^1({\cal C}) &:=& \{ g_j\in{\cal D}^{1c}_{\cal C}| t(j)\geq i\}.
\eqa 
$c(i)$ is the number of connected components in $T_i(P)$. 
All these definitions can be restricted to the connected components:
$IT^{0k}_i$,$IT^{1k}_i$ , $IL_i^{k}({\cal C})$, $IL_{i}^{0k}({\cal C})$ 
 $IL_{i}^{1k}({\cal C})$, $S_i^{0k}({\cal C})$ and  $S_i^{1k}({\cal C})$. 
We observe that 
$IL_i({\cal C})$ corresponds to the set of half-lines that could have,
in the class ${\cal C}$, $\mu(a)\geq i$ and it is the equivalent  of
$IL_i$ defined in (\ref{definizioni}).
$IT^0_i$ (respectively $IT^1_i$) and $IL_{i}^0({\cal C})$ 
(respectively
$IL_{i}^1({\cal C})$) are the set of tree half-lines and loop
half-lines at a level higher or equal to $i$,  which are the interpolated
external lines for some divergent subgraph in 
$ {\cal D}^{0c}_{\cal C}(P)$ (respectively in ${\cal D}^{1c}_{\cal C}$), 
$S_i^0({\cal C})$ (respectively $S_i^1({\cal C})$) 
is the set of subgraphs in
$ {\cal D}^{0c}_{\cal C}(P)$ (respectively in $ {\cal D}^{1c}_{\cal C}$) 
that have the internal tree line $l_{t(j)}$ of a level
higher or equal to $i$. In the same way,
we can define the equivalent of $EL_i$ and $E_i$ as
\be
EL_i({\cal C}) := \cup_{k=1}^{c(i)} el_i^k({\cal C})
\ee
which is the set of loop 
half-lines that are forced to have $\mu(a)\leq i$, and  
\be
E_i({\cal C}):= \cup_{k=1}^{c(i)} eg_i^k({\cal C}).
\ee
The integral in the variable $d\bt_i$ can be performed only if 
the exponent of $\bt_i$ is bigger than $-1$. Using the relations
\bqa
{\bar n}-i &=& \sum_{k=1}^{c(i)} [|N_i^k|+|N^{'k}_i|+|N''^{k}_i|-1]\no\\
 |E_i({\cal C})| &= &  |EL_i({\cal C})|+ |ET_i| +|EE_i|\no\\
&&= 2|N_i| +2  - |IL_i({\cal C})| \ ,
\eqa  
the exponent of $\bt_i$ can be written as $-1 + \sum_{k=1}^{c(i)} x_i^k $,
where
\bqa
x_i^k &:= &\frac{1}{2}
\biggr [\frac{1}{2}(|E^k_i({\cal C})|+ 2|N_i^{'k}|-4)\label{powerc}\\ 
&&+ [ |S_i^{0k}({\cal C})|-|IT^{0k}_i|-|IL^{0k}_{i}({\cal C})|] +2 
[|S_i^{1k}({\cal C})|-|IT^{1k}_i|-|IL^{1k}_{i}({\cal C})|]\biggr]\no
\eqa
Remark that for any level $i$ we have
\be
[|S_i^{0}({\cal C})|-|IT^0_i|-|IL^0_{i}({\cal C})|]\geq 0
\quad 
[|S_i^{1}({\cal C})|-|IT^1_i|-|IL^1_{i}({\cal C})|]\geq 0
\ee
as each half-line in $IT^0_{i}$ ($IT^1_{i}$) or 
$IL^0_{i}({\cal C})$ ($IL^1_{i}({\cal C})$) is the external interpolated  line
for a subgraph $g_j$. This subgraph 
$g_{j}$ must have $j>i$ hence have $t(j)>i$. 
Therefore for each half-line in one of these sets
there is always at least one corresponding  
half-line in $S_i^{0}({\cal C})$  ($S_i^{1}({\cal C})$). 

\medskip
\begin{lemma}
For any connected component in $\tree_i^k$ we have $x_i^k\geq 1/2$. 
\end{lemma}
\medskip

\noindent{\bf Proof} We distinguish three situations.
\paragraph{-} If $|E^k_i({\cal C})|\ge 5$, in fact, by parity
of the number of external half-lines of any subgraph,
$|E^k_i({\cal C})|\ge 6$ and
then 
\be
x_i^k
\geq (1/4)(|E^k_i({\cal C})| -4 ) \geq 1/2 .
\ee
\paragraph{-}  If $|E^k_i({\cal C})|=4$, then there must be a subgraph
$g_j\in{\cal D}^{0c}_{\cal C}(P)$ with $j\geq i$ ($j=i$ only if $l_i$ 
belongs to the connected component $\tree_{i}^k (J,P)$) and 
${\cal A}({j})<i$. Hence the interpolated line for $g_{j}$
does not belong to  
$IT^{0k}_i$ or $IL^{0k}_{i}({\cal C})$, but the corresponding internal
line $l_{t(j)}$ belongs to $S_i^{0k}$. Then 
\be
 |S_i^{0k}|-|IT_i^{0k}|-|IL^{0k}_{i}({\cal C})|\geq 1 
\ee
and 
\be
x_i^k = 
\frac{1}{2}[ |N'^{k}_i| + [ |S_i^{0k}|-|IT^{0k}_i|-|IL^{0k}_{i}({\cal C})|]+
2 [|S_i^{1k}|-|IT^{1k}_i|-|IL^{1k}_{i}({\cal C})|]\geq \frac{1}{2}.
\ee
\paragraph{-} Finally if $|E^k_i({\cal C})|=2$ 
one can  see, by the same arguments, that  
\be
 |S_i^{1k}|-|IT_i^{1k}|-|IL^{1k}_{i}({\cal C})|\geq 1 
\ee
and
\[
x^k_i = 
\frac{1}{2}[-1+ |N_i^{'k}| + [ |S_i^{0k}|-|IT_i^{0k}|-|IL^{0k}_{i}({\cal C})|]
+2 [|S_i^{1k}|-|IT^{1k}_i|-|IL^{1k}_{i}({\cal C}|)
\]
\be
\geq   [-1 + 2 [|S_i^{1k}|-|IT_i^{1k}|-|IL^{1k}_{i}({\cal C})|]
\geq  1/2.
\ee
\hfill${\Box}$

Now we can perform the integrals in equation (\ref{sviluppo9}) and we obtain
\bqa
\lefteqn{|\Ga_{2p}^\Lambda(\phi_1,...\phi_{2p})|\leq 
||\phi_1||_1 \prod_{i=2}^{2p} ||\phi_i||_{\infty,2}
 e^{-(1-\epsilon )\La^m 
d_T(\Om_1,...\Om_{2p})}}\label{sviluppo10} \\
&& (\La^m)^{2-p} N^{1-p}  
\sum_{n,n',n''=0}^\infty (cK)^{\bar n} \frac{1}{n!n'!n''!}
\sum_{u-\tree}\sum_{\si}\sum_{E,{\cal C}} 
  \prod_{i=1}^{\bar n-1}  
\frac{1}{\sum_{k=1}^{c(i)}x_i^k}
\no\eqa
where we wrote the sum over ordered trees as sum over unordered trees and sum
over all possible orderings $\si$ of the tree.
The sum $\sum_{{\cal C}}$ is over a 
set whose cardinal is bounded
by  $K^{\bar n}$ so it's sufficient to bound them with the supremum 
over this set, 
as  we are interested in a theorem at weak coupling $\la$. 
However the sum over $E$ to attribute the $2p$ external lines to particular
vertices runs over a set of at most $\bar n ^{2p}$ (this is an overestimate!).
This will lead to the factorial in Theorem 3. We remark however that
a better bound on the behaviour of the vertex functions at large $p$
can presumably be obtained when the external points are sufficiently
spread (not too closely packed), but we leave this improved estimate
to a future study.
 
Moreover,  we bound $\frac{1}{({\bar n})!}\sum_{u-\tree} f(\tree)$ by 
$\frac{{\bar n}^{{\bar n}-2}}{{\bar n}!}
\sup_{u-\tree}|f(\tree)|$ using Cayley's theorem. 
Therefore,  by Stirling's formula, it's  enough to consider the 
unordered  tree $\tree$ which gives the $\max_{u-\tree}|f(\tree)|$. 
The sum that could still give some factorial  is $\sum_{\si}$. To bound it 
we use the product of fractions obtained after integration on the $\bt_i$.
\begin{itemize}
\item{} if $|ET_i^k|\geq 5$ we have
\be
(|ET_i^k|+|EE_i^k|+|EL_i^k({\cal C})|-4)/4\geq (|ET_i^k|-4)/4\geq 
\frac{|ET^k_i|+1}{24}
\ee
\item{} if  $|ET_i^k|< 5$ we have
\be
x_i^k \geq 1/2 \geq \frac{|ET^k_i|+1}{24}
\ee
\end{itemize}
Now  $|ET_i|$ depends on the (now unordered) tree $\tree$
{\em and on its ordering} $\sigma$. Therefore we call it from now
on $|ET_i^{\si}|$. Recall that it is
the total number of external tree half-lines
of the subset $\tree_i^{\si}$  of $\tree$ made of the $\bar n - i$ 
highest lines in the permutation $\si$.
Since $\sum_{k}(|ET_i^k |+1)\ge |ET_i^{\si}|+1$, equation
(\ref{sviluppo10}) becomes
\bqa
\lefteqn{|\Ga_{2p}^\Lambda(\phi_1,...\phi_{2p})|\leq 
||\phi_1||_1 \prod_{i=2}^{2p} ||\phi_i||_{\infty,2}e^{-(1-\epsilon )\La^m 
d_T(\Om_1,...\Om_{2p})}  }\no\\
&& (\La^m)^{2-p} N^{1-p} 
\sum_{n,n',n''=0}^\infty \bar n ^{2p} (cK)^{n+n'+n''} 
 \sum_{\si} \prod_{i=1}^{\bar n-1}  
\frac{1}{|ET_i^{\si}|+1}\label{sviluppo11}
\eqa

At this point we can apply a result of [CR] (Lemma A,1, B.3, B.4) which
states that for any tree we have
\be
 \sum_{\si}\prod_{i=1}^{{\bar n}-1}\frac{1}{|ET_i^{\si}|+1}\leq 4^{\bar n}.
\ee
For completeness let us recall the proof of this result.
For each tree $\tree$ we can define a mapping $\xi$ of   
$\tree$ on a chain-tree with the
same number of vertices: 
\be
\xi: \tree \rightarrow \xi_{\tree}
\ee
To define $\xi$, we turn around $\tree$ starting from an arbitrary end
line, and we number the lines in the order we meet them for
the first time. The
lines of $\xi_{\tree}$ are numbered in the same way and $\xi_\tree$ associates
the lines with the same number.

\centerline{\hbox{\psfig{figure=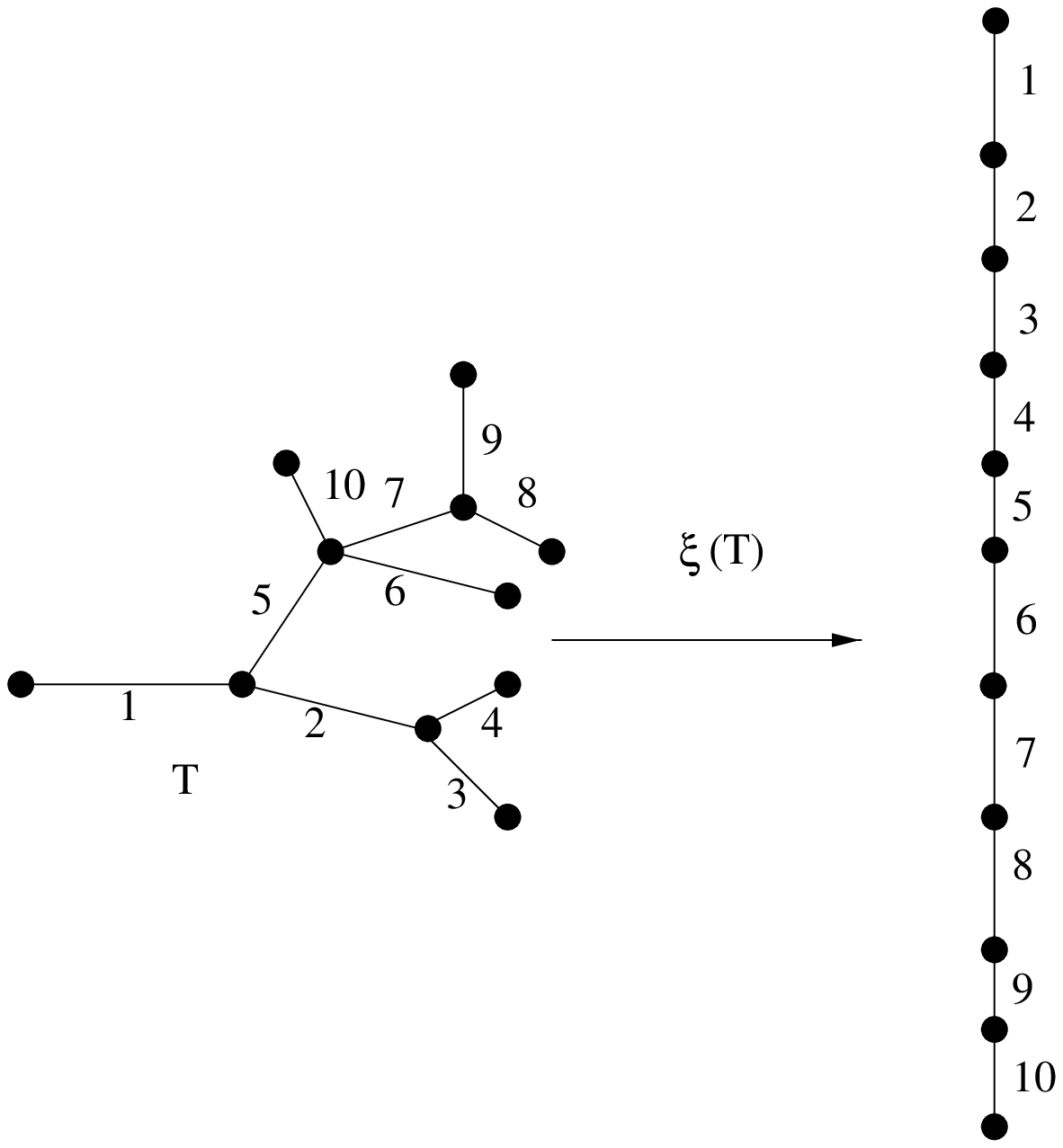,height=6cm,width=6cm}}}
\nobreak\centerline{Figure 22}
\goodbreak

Now we observe that the sum over the orders on $\tree$ corresponds to the
sum over all permutations of the indices in $\xi(\tree)$. Moreover Lemma B.3
in [CR] proves that for any connected or disconnected subgraph $R$ of
$\tree$, 
we have
\be
E\tree(R)+1\geq c(\xi_{\tree}(R))
\ee
where $c(\xi_\tree(R))$ is the number of connected components of the image of
$R$ $\xi_\tree(R)$ and $E\tree(R)$ is the number of
external half-lines of $R$ in $\tree$. 
Finally we note that $\xi(T_i)$ is the set of lines 
with number $j\geq \bar n - i$ so we can write
\be
 \sum_{\si}\prod_{i=1}^{{\bar n}-1}\frac{1}{|ET_i^\si|+1}
= \sum_\si \prod_{i=1}^{{\bar n}-1}\frac{1}{c(D_i^\si)}\ :=\ \Delta_{\bar n} \ ,
\ee
where $D_i^\si$ is the set of lines in the chain-tree $\xi(\tree)$, 
that have $\si(j)\geq \bar n - i$ (after the permutation $\si$).
Now, applying Lemma B.4. in [CR], we obtain
\be
\De_{\bar n}\leq 4^{\bar n}
\ee
We recall that  this can be proved by remarking that $\De_{\bar n}$
satisfies the inductive equation 
\be
\De_{\bar n}= \sum_{k=1}^{\bar n -1}
\De_{p}\De_{\bar n -k} \ ,
\ee
so that equation (\ref{sviluppo11}) becomes
\bqa
\lefteqn{
|\Ga_{2p}^\Lambda(\phi_1,...\phi_{2p})|\leq   ||\phi_1||_1 \prod_{i=2}^{2p} 
||\phi_i||_{\infty,2}e^{-(1-\epsilon )\La^m 
d_T(\Om_1,...\Om_{2p})} }\no\\ 
&& (\La^m)^{2-p} N^{1-p}\; \sum_{n,n',n''=0}^\infty  
\bar n ^{2p}(4cK)^{\bar n} 
\eqa
where $K$ depends only on $\e$.
Taking $c$ small enough completes the proof of the theorem
in the case $\La >m$, since $\sum_{\bar n }\bar n ^{2p}e^{-\bar n}\le 
K^p (p!)^2 $.

In the case $\La < m$, we have a few changes to perform.
Replacing the lines of $\tree '$ in (\ref{infrared0})
by the bound (\ref{infrared2}), keeping the massive
decay factor $e^{-(\e/4)m^2\Lazero^{-2}(w_1) }$  in (\ref{infrared0}) 
and passing to the limit
$\Lazero \to \infty$
we have the following changes: in (\ref{sviluppo8})
we add the factors
\be
(\La/m)^{n'}\bigl[ \prod_{l_{q}\in\tree'} (u_{q})^{-1/2} \bigr]
e^{-(\e/4) u_1 m^2\La^{-2} }
\ee
The factor $(\La/m)^{n'}$ exactly changes $\La^{2-p-n'}$ into 
$\La^{2-p} m^{-n'}= \La^{2-p}(\La^m)^{-n'}$. The factor $(\La^m)^{-n'}$
is absorbed in $K^{\bar n}$ since $\La^m$ in the hypothesis of Theorem 3
remains in the compact $X$.
Passing to the variables $\bt_i$,
the factor $\bigl[ \prod_{l_{q}\in\tree'} (u_{q})^{-1/2} \bigr]$
is bounded by the factor $\prod_{i} \bt_{i}^{|N'^k_i|}$
in (\ref{powerc}), which was previously bounded by 1, hence not used at all.
Finally the last integral over $\bt_{1}$ becomes bounded,
for $p>2$ by:

\be \La^{2-p}\int_{0}^1 \bt_{1}^{(p -2)/2} {d\bt_{1}\over \bt_{1}}
e^{-(\e/4)m^2\bt_1\La^{-2}} 
\ee
Changing to the variable $v=(\e/4)m^2\bt_1\La^{-2} $ we obtain 
for the final bound a factor
\be (4/\e m^2)^{(p -2)/2}\int_{0}^{\e m^2\La^{-2}/4} 
v^{(p -2)/2} {dv\over v}
e^{-v} \le (\La^m)^{2-p} K^p  \sqrt{p!}
\ee
The case $p=2$ is easy and left to the reader.
Hence Theorem 3 holds in every case, by combining the factor $ \sqrt{p!}$
with the factor $ (p!)^2$ coming from the sum over $E$. Remark that
in the case $m=0$ we have never $\La < m$, hence the factor $(p!)^{5/2}$
can be replaced by $(p!)^2$.

\section{The Renormalization Group Equations}

In this section we establish the renormalization group equations
obtained when varying $\La$ and we check that for a fixed and
small renormalized coupling constant, the effective constants
remain bounded and small as predicted by the well known
perturbative analysis of the model, which is asymptotically
free in the ultraviolet regime [MW].

The derivative $\p{\La}\Ga_{2p}^{\La\Lazero}(\phi_1,...\phi_{2p})$ 
can be written, using the expression 
(\ref{sviluppo4}), as:
\be
\p{\La } \Ga_{2p}^{\La\Lazero}(\phi_{1},...,\phi_{2p}) =  
T\Ga_{2p}^{\La\Lazero} (\phi_{1},...,\phi_{2p}) + 
L\Ga_{2p}^{\La\Lazero} (\phi_{1},...,\phi_{2p}).
\label{GE0}\ee
The first term $T\Ga_{2p}^{\La\Lazero} (\phi_{1},...,\phi_{2p}) $ is the 
series  obtained when the derivative falls on a tree line propagator
(see Figure 23a):
\bqa
\lefteqn{T\Ga_{2p}^{\La\Lazero}(\phi_1,...\phi_{2p})= 
\sum_{n,n',n''=0}^\infty  \frac{1}{n!n'!n''!}
\sum_{o-\tree} 
\sum_{E,\mu}\sum_{{\cal C}ol,\Om}  \e(\tree, \Om)\int  
d^2x_1...d^2x_{\bar n} }\no\\ 
&& 
\int_{0\le w_{1} \le ...\le w_{\bar n-1}\le 1}\prod_{q=1}^{\bar n-1} 
dw_q \;  
\left [\prod_{v} \lp \frac{\la_{w(v)}}{N}\rp \right ] 
\left [\prod_{v'}\de m_{w(v')}\right ] 
\left [\prod_{v''}\de \ze_{w(v'')}\right ] \no\\
&& \prod_{G_i^k\in D_\mu} R_{G_i^k} \;  
\biggr[ \sum_{q'=1}^{\bar n-1} \p{\La} D_{\La}^{\La_{0},  w_q}
(\bar x_{l_{q'}}, x_{l_{q'}})\no\\
&& \prod_{q\neq q'}  D_{\La}^{\La_{0},  w_q}
(\bar x_{l_{q}}, x_{l_{q}}) \;\; \det {\cal M}(\mu)
\;\;\phi_1(x_{i_{1}})...\phi_{2p}(x_{j_{p}})\biggr] 
\label{GE1}\eqa

\centerline{\hbox{\psfig{figure=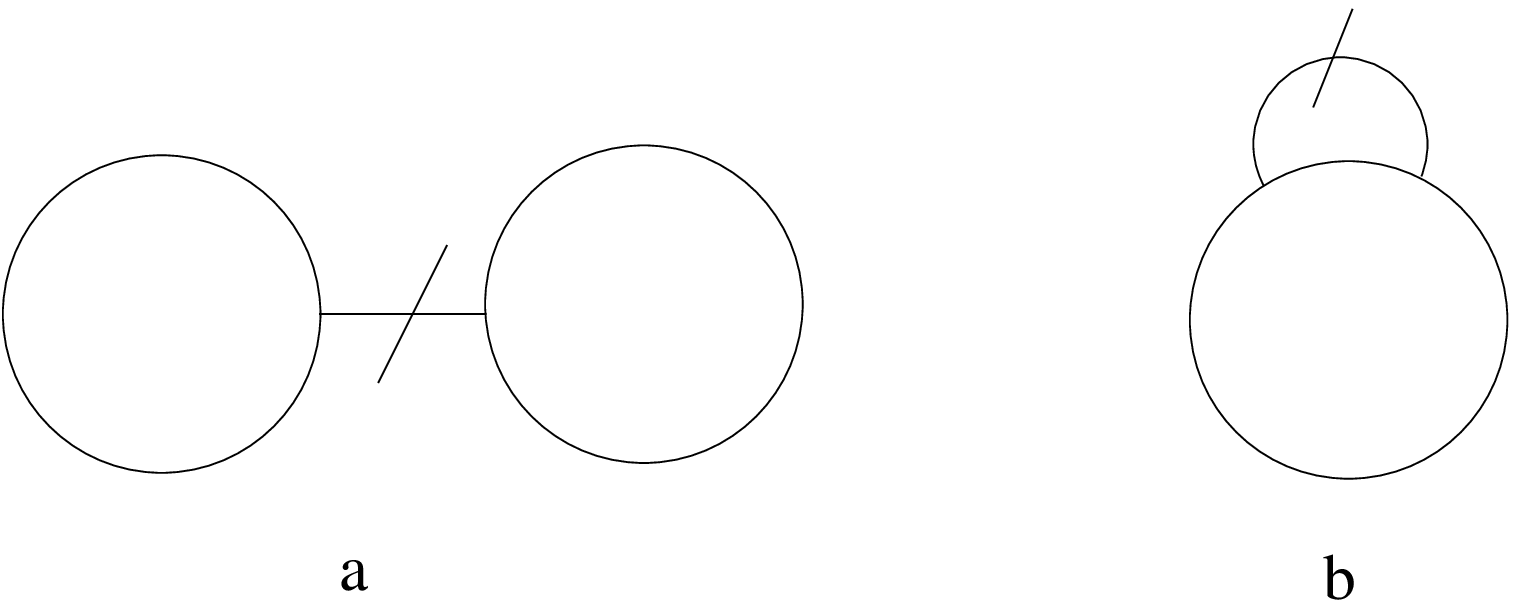,height=3cm,width=7cm}}}
\centerline{Figure 23}

The second term $L\Ga_{2p}^{\La\Lazero} (\phi_{1},...,\phi_{2p}) $ is the 
series  obtained when the derivative falls on a loop line in the determinant
(see Figure 23b):
\bqa
\lefteqn{L\Ga_{2p}^{\La\Lazero}(\phi_1,...\phi_{2p})= 
\sum_{n,n',n''=0}^\infty  \frac{1}{n!n'!n''!}
\sum_{o-\tree} 
\sum_{E,\mu}\sum_{{\cal C}ol,\Om}  \e(\tree, \Om)\int  
d^2x_1...d^2x_{\bar n} }\no\\ 
&& 
\int_{0\le w_{1} \le ...\le w_{\bar n-1}\le 1}\prod_{q=1}^{\bar n-1} 
dw_q \;  
\left [\prod_{v} \lp \frac{\la_{w(v)}}{N}\rp \right ] 
\left [\prod_{v'}\de m_{w(v')}\right ] 
\left [\prod_{v''}\de \ze_{w(v'')}\right ] \no\\
&& \prod_{G_i^k\in D_\mu} R_{G_i^k} \;  
\biggr[ \prod_{q=1}^{\bar n-1}  
D_{\La}^{\La_{0},w_q}(\bar x_{l_{q}}, x_{l_{q}}) 
\phi_1(x_{i_{1}})...\phi_{2p}(x_{j_{p}})\no\\
&&  \sum_{h_f,h_g| \mu(f)= \mu(g)}
(-1)^{\e(f,g)}\p{\La} C_{\La_0(w_{\mu(f)})}^{\La_{0}(w_{\mu(f)-1})}(x_f, x_g)
  {\det}_{left}{\cal M}(\mu)
\biggr]\label{GE2} 
\eqa
where $\e(f,g)$ is a sign coming from the development of the determinant.
The convergence proofs of course extend to both terms of equation
(\ref{GE0}). Indeed, in the first one, the sum 
over the tree lines is bounded by a factor $\bar n$, and in the second one
the sum is over the set of loop half-lines which is 
bounded by a factor
$\bar n^2$. Therefore these sums cannot generate any factorial. Then we obtain
the same bound as in (\ref{sviluppo11}), with an additional factor
$1/\La$. This factor disappears when, as usual, the renormalization group
equations are written as derivatives with respect to $\log \Lambda$
rather than $\Lambda$.

From these equations
one can derive also equations for the flow of the effective constants
defined in (\ref{eff-const}).  For instance to obtain
the flow of the effective coupling constant $\la$ 
which is the four-point vertex function at zero external momenta,
we can use equations (\ref{GE0})-(\ref{GE2}) 
in which we let $\phi_{1}\to \de(0)$,  $\phi_{2}$,
$\phi_{3} $, $\phi_{4}\to 1$. This is compatible with our 
$L_{1}$-$L_{\infty}$ bounds, so that everything remains bounded. We obtain
in this way the famous {\em continuous} flow
equation which gives the derivative of the coupling constant
with respect to $\log \La$:
\be
\p{\log \La } N \widehat\Ga^{\La}_{4} (0,0,0,0)= \p{\log \La }  \la_{\La}
=  \beta_{2} \la^2_{\La} + O(c^3) +  \la^2_{\La}O(\La^{-\alpha})
\label{GE3}
\ee
where 
\be \beta_{2} =  -2(N-1)/\pi 
\ee
is the first non trivial term corresponding to the four-point graph with one
tree line and one loop line, and the last term $\la^2_{\La}O(\La^{-\alpha})$
is an infrared correction to the asymptotic flow (see [FMRS]).
 The negative sign of $\beta_{2}$
is responsible for the asymptotic freedom of the model.
Similar equations hold
for the flow of $\de m$ and $\de \ze$ (which remain bounded).
For these equations up to one loop, see [MW] [GN] [GK] [FMRS]. 
For the computation up to two loops, we refer to [W]. 

From these renormalization group equations one can control the behavior of
the effective constants and
check that they remained bounded (until now this was assumed). The reader
might be afraid that there is something circular in this argument.
In fact this is not the case. Let us discuss for simplicity the massless
case where  the renormalized coupling
$\widehat
\Ga^{\La\Lazero}_{4} (0,0,0,0)$ is only a function of $\Lazero/\La$
and of the bare coupling $\la$.
We know  that it
is analytic at the origin as function of the bare coupling $\la$ [AR2].
Therefore from (\ref{GE0})-(\ref{GE3})
it is for small bare $\la$ and $\Lazero/\La$
a monotone increasing function of the ratio $\Lazero/\La$ (although
this function might blow up in finite time).

% One can use a
%continuous induction, like in the proof of 
%completeness of flows of vector fields on compact manifolds.

Inverting the map from bare to renormalized couplings,
one can prove that conversely for small renormalized coupling
$\widehat
\Ga^{\La\Lazero}_{4} (0,0,0,0)$ all the effective
constants $\la_{w}$ remain bounded by the
renormalized one. Therefore one can pass to the ultraviolet
limit $\Lazero \to \infty $, in analogy with the 
completeness of flows of vector fields on compact manifolds.
Furthermore one can compute the asymptotic behavior
of the bare coupling which tends to 0
as $1/(|\beta_{2}| \log(   \Lazero/\La ))$). Similar arguments hold for the 
mass and wave function effective constants and achieve the proof of Theorem 1.

We recall for completeness  that it is easy to build the Schwinger functions
from the vertex functions and 
that the Osterwalder-Schrader axioms of continuous Euclidean
Fermionic theories hold for the massive Gross-Neveu
model at $\La=0$. The simplest proof is to remark that being the Borel
sum of the renormalized expansion, the Schwinger functions we build are unique.
Building them as limits of theories with different kinds of cutoffs
prove the axioms since different sets of cutoffs violate different axioms
[FMRS]. 
 
\medskip
\noindent{\bf Acknowledgements}
\medskip

It is a pleasure to thank D. Brydges,
C. de Calan, H. Kn{\"o}rrer, C. Kopper, G. Poirot and 
M. Salmhofer for discussions related to this
paper. In particular we thank C. Kopper for his careful reading 
of this preprint, which lead us to correct the initial version
of subsections IV.3.4 and IV.3.5.

\medskip 
\noindent{\bf References}
\medskip

[AR1] A. Abdesselam and V.  Rivasseau, Trees, forests and jungles: a
botanical garden for cluster expansions, in Constructive Physics, ed by
V. Rivasseau, Lecture Notes in Physics 446, Springer Verlag, 1995.

[AR2] A. Abdesselam and V. Rivasseau, Explicit Fermionic Cluster
Expansion,
preprint: cond-mat/9712055.

[CR] C. de Calan and V. Rivasseau, Local existence of the Borel
transform in Euclidean $\phi^{4}_{4}$,  Commun. Math. Phys. 82, 69 (1981).

[FMRS] J. Feldman, J. Magnen, V. Rivasseau and R.
S{\'e}n{\'e}or, A renormalizable field theory: the massive Gross-Neveu
model in two dimensions, Commun. Math. Phys. 103, 67 (1986).

[GK] K. Gawedzki and A. Kupiainen, Commun. Math. Phys. 102, 1 (1985). 

[GN] D. Gross and A. Neveu, Dynamical Symmetry breaking in asymptotically
free field theories, Phys. Rev. D10, 3235 (1974).

[KKS] G. Keller, Ch. Kopper and M. Salmhofer,
Helv. Phys. Acta 65 32 (1991).

[KMR] C. Kopper, J. Magnen and V. Rivasseau,
Mass Generation in the Large N
Gross-Neveu Model, Commun. Math. Phys. 169, 121 (1995).

[MW] P.K. Mitter and P.H. Weisz, Asymptotic scale invariance in a massive
Thirring model with U(n) symmetry, Phys. Rev. D8, 4410 (1973).

[P] J. Polchinski, Nucl. Phys. B 231 269 (1984).

[R] V. Rivasseau, From perturbative to constructive renormalization,
Princeton University Press (1991).

[S] M. Salmhofer,
Continuous renormalization for Fermions and Fermi liquid theory,
ETH preprint, cond-mat/9706188.

[W] W. Wetzel, Two-loop $\beta$-function for the Gross-Neveu model,
Phys. Lett. 153B, 297 (1985).

\end{document}